\documentclass[10pt]{iopart}

\usepackage{graphicx}

\usepackage{iopams}

\expandafter\let\csname equation*\endcsname\relax

\expandafter\let\csname endequation*\endcsname\relax

\usepackage{amsmath}

\usepackage[hyphens]{url}
\usepackage{bbm}
\usepackage[]{units}
\usepackage{hyperref}
\usepackage{braket}

\newcommand{\nnb}{\nonumber \\}

\newcommand{\bv}{\left( \begin{array}{c}}
\newcommand{\ev}{\end{array} \right)}
\newcommand{\E}{\mathrm{e}}

\newcommand{\st}[1]{_{\text{#1}}}
\newcommand{\bo}[1]{\boldsymbol{#1}}
\newcommand{\I}{\mathrm{i}}

\newcommand{\tun}{\, U}

\begin{document}
\topical{Three-electron spin qubits}
\author{Maximilian Russ and Guido Burkard}
\address{Department of Physics, University of Konstanz, D-78457 Konstanz, Germany}
\begin{abstract}
The goal of this article is to review the progress of three-electron spin qubits from their inception to the state of the art. We direct the main focus towards the resonant exchange (RX) qubit and the exchange-only qubit, but we also discuss other qubit implementations using three electron spins. The RX qubit is a qubit implementation in a triple quantum dot with the exchange interaction always turned on, hence, a modified version of the exchange-only qubit\cite{Bacon2000,Nature2000}. For each three-spin qubit we describe the qubit model, the physical realization, the implementations of single-qubit operations, as well as the read-out and initialization schemes. Two-qubit gates and decoherence properties are discussed for the RX qubit and the exchange-only qubit, thereby, completing the list of requirements for a viable candidate qubit implementation for quantum computation. We start with describing the full system of three electrons in a triple quantum dot, then discuss the charge-stability diagram and restrict ourselves to the relevant subsystem, introduce the qubit states, and discuss important transitions to other charge states\cite{Russ2016}. Introducing the various qubit implementations, we begin with the exchange-only qubit\cite{Nature2000,Laird2010}, followed by the spin-charge qubit\cite{Kyriakidis2007}, the hybrid qubit\cite{Shi2012,Koh2012,Cao2016}, and the RX qubit\cite{Medford2013,Taylor2013}, discussing for each the single-qubit operations, read-out, and initialization methods, whereas the main focus will be on the RX qubit, whose single-qubit operations are realized by driving the qubit at its resonant frequency in the microwave range similar to electron spin resonance. Two different types of two-qubit operations are presented for the exchange-only and the RX qubit which can be divided into short-ranged and long-ranged interactions. Both of these interaction types can be expected to be necessary in a large-scale quantum computer. The short-ranged interactions use the exchange coupling by placing qubits next to each other and applying exchange-pulses\cite{Nature2000,Fong2011,Setiawan2014,Zeuch2014,Doherty2013,Shim2016}, while the long-ranged interactions use the photons of a superconducting microwave cavity as a mediator in order to couple two qubits over long distances\cite{Russ2015b,Srinivasa2016}. The nature of the three-electron qubit states each having the same total spin and total spin in $z$-direction (same Zeeman energy) provides a natural protection of the qubit to several sources of noise\cite{Nature2000,Kempe2001,Taylor2013,Russ2015}. The price to pay for this improvement is an increase in gate complexity. We also take into account the decoherence of the qubit through the influence of magnetic noise\cite{Ladd2012,Mehl2013,Hung2014}, in particular dephasing due to the presence of nuclear spins, as well as dephasing due to charge noise\cite{Medford2013,Taylor2013,Fei2015,Russ2015,Shim2016}, fluctuations of the energy levels on each dot due to noisy gate voltages or the environment. Several techniques are discussed which partly decouple the qubit from magnetic noise\cite{West2012,Setiawan2014,Rohling2016} while for charge noise it is shown that it is favorable to operate the qubit on the so-called ``sweet spots''\cite{Taylor2013,Russ2015} or better ``double sweet spots''\cite{Fei2015,Russ2015,Shim2016}, which are least susceptible to noise, thus providing a longer lifetime of the qubit.

\end{abstract}
\submitto{\JPCM}
\maketitle
\ioptwocol
\section{Introduction}
\subsection{Quantum computation}
\label{ssec:QC}

Since the early beginnings, certain aspects of quantum mechanics such as entanglement and non-locality have fascinated those who studied it due to their counterintuitive behavior in comparison with everyday life, thus fueling heated debates. 
With the rise of the information processing technology the question arose whether these quantum properties could be exploited and used also for information processing, thus, opening the research field of quantum information processing (see e.g. Ref.~\cite{nielsen2000} for a historical account). These quantum computers which follow a logic based on quantum mechanics can be seen as superior to classical computers since they have the ability to solve certain problems which classical computers cannot solve in any reasonable time. It does not seem that quantum computers surpass classical computers in performing arbitrary tasks but since they operate according to quantum mechanical laws, they are certainly better in simulating other quantum systems\cite{Feynman1982,Lloyd1996}, a problem classical computers are unable to tackle efficiently. Furthermore, quantum computers can use certain quantum aspects to solve some very specialized problems. The most well-known application is the Shor algorithm with is able to efficiently factorize integers\cite{Shor1997} and thereby subvert the ability to encode encrypted information using prime factors as a private key\cite{Rivest1978}. Other known applications are the Deutsch algorithm, its generalization, the Deutsch-Josza algorithm\cite{nielsen2000}, and the Grover algorithm for efficient search in unsorted databases\cite{Grover1996,Grover1998}.

Following directly after its potential applications, concepts for a physical realization of such a quantum computer were proposed. These concepts, among others, are based on the charge of confined electrons\cite{Lloyd1993}, the spin of electrons in quantum dots\cite{Kloeffel2013}, nuclear spin\cite{Vandersypen2005}, photons in resonators and cavities\cite{Miller2005}, trapped ions\cite{Cirac1995}, low-capacitance Josephson junctions\cite{Nakamura1999,Makhlin2001}, donor atoms\cite{Kane1998}, colored centers in diamond\cite{Doherty2013b} or silicon-carbide\cite{Dobrovitski2013}, or linear optics\cite{Kok2007}.

\begin{figure}
\begin{center}
\includegraphics[width=0.9\columnwidth]{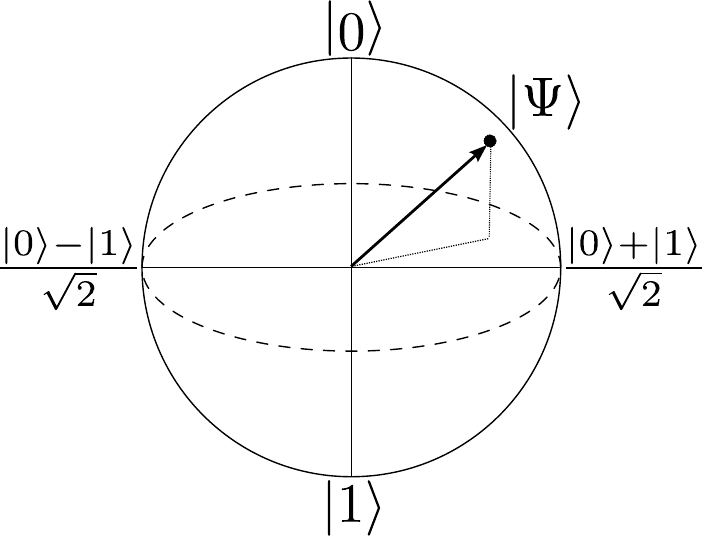}
\caption{Visualization of a qubit on the Bloch sphere. The basis states $\ket{0}$ and $\ket{1}$ are located on the poles and arbitrary pure qubit states $\ket{\Psi}$ are represented by vectors pointing to an arbitrary point on the surface of the sphere.}
\label{fig:bloch}
\end{center}
\end{figure}

The five DiVincenzo criteria for a fully functioning quantum computer\cite{DiVincenzo2000,Delgado2011,Martinis2015,Bejanin2016} help to decide which concepts are reliable which, in a few words, comprise the scalability of the system, initialization and read-out schemes, the durability of the stored information, and concepts for precise controlling of the quantum bits (qubits)\cite{DiVincenzo2000}. These qubits are the fundamental building block of each quantum computer and are not limited to two states, 0 and 1, but can be in any superposition of these, $\ket{\Psi}=\alpha\ket{0}+\beta\ket{1}$ with $|\alpha|^{2}+|\beta|^{2}=1$\cite{nielsen2000}, thus forming a quantum two-level system. Each qubit can be visualized on the Bloch sphere (see Fig.~\ref{fig:bloch}) where the basis states of the qubit, $\ket{0}$ and $\ket{1}$, are located on the poles while vectors pointing to the surface of a unity sphere represent all possible pure states of the qubit. Thus, complete control of a qubit requires interactions corresponding to two independent axes of rotations on the Bloch sphere\cite{nielsen2000}. The physical realization of these rotations depends on the chosen qubit system. 

The aim of this topical review is to review the recent progress and advances of three-spin qubit systems which are considered as a possible and promising candidate for a functioning quantum computer. The organization of the paper is as follows. In Section~\ref{ssec:QC}, we start with an introduction in which we briefly cover the fundamental concepts needed for three-spin qubits such as the exchange interaction (see subsection~\ref{ssec:exchange}), the single-spin qubit (see subsection~\ref{ssec:LDQ}), and the two-spin singlet-triplet qubit (see subsection~\ref{ssec:STQ}). Subsequently in Section~\ref{sec:3spin}, the three-spin qubits are introduced. We start with explaining the experimental realization (see subsection~\ref{ssec:physReal}), and the electrical (see subsection~\ref{ssec:chargestab}), and spin properties of three-spin qubits (see subsection~\ref{ssec:spinprop}) constructing a framework for the investigation. Afterwards, we introduce and discuss the different physical implementations of three-spin qubits in the light of the DiVincenzo criteria, the exchange-only qubit (see subsection~\ref{ssec:EOQ}), the spin-charge qubit (see subsection~\ref{ssec:SCQ}), the hybrid qubit (see subsection~\ref{ssec:HQ}), the resonant-exchange qubit (see subsection~\ref{ssec:RXQ}), and the always-on exchange-only qubit (see subsection~\ref{ssec:AEON}). In Section~\ref{sec:two}, two-qubit gates are discussed for the three-spin qubit with the focus on exchange-based short-ranged (see subsection~\ref{ssec:shortTwo}) and cavity-mediated long-ranged operations (see subsection~\ref{ssec:longTwo}). Finally, in Section~\ref{sec:dec}, we investigate the behavior of the qubit under the influence of the two main sources of decoherence, magnetic noise (see subsection~\ref{ssec:magDec}) and charge noise (see subsection~\ref{ssec:chargeDec}), and provide concepts for reduction of the decoherence effects. We conclude in Section~\ref{sec:pers} with a summary and future perspectives.

\subsection{The exchange interaction}
\label{ssec:exchange}

The most important tool for spin quantum computation with electrons in quantum dots (QDs) is the exchange interaction, originating from the sign change under exchange of fermionic particles, since it can be electrically controlled both very precisely and fast by detuning of the externally applied electrostatic gate voltages\cite{Petta2005,Hanson2007}.

A sufficient explanation for exchange interaction between $N_{e}$ electrons in $N$~QDs is provided by the Hubbard model\cite{Hubbard1963,Burkard1999} with symmetric spin-conserving nearest neighbor hopping
\begin{align}
H\st{Hub}=&\sum_{i=1}^{N} \left[\frac{\tilde{U}}{2}n_i(n_i-1) + V_in_i\right] \nnb
&+ \sum_{\braket{i,j}} U_{C} n_{i}n_{j} + \sum_{\sigma=\uparrow,\downarrow}\left(t_{ij}c_{i,\sigma}^\dagger c_{j,\sigma}+\text{h.c.}\right),
\label{eq:Hub}
\end{align}
where the operator $c_{i,\sigma}^{\dagger}$ ($c_{i,\sigma}$) creates (annihilates) an electron in QD $i$ with spin $\sigma=\uparrow,\downarrow$, $V_{i}$ is the gate energy in QD~$i$, $\tilde{U}$ is the energy penalty for doubly occupying a single QD due to the Coulomb repulsion, and $U_{C}$ is the energy to pay for two electrons in neighboring QDs. We define the number operator $n_i\equiv\sum_\sigma c_{i,\sigma}^\dagger c_{i,\sigma}$ and the gate-controlled hopping matrix elements $t_{ij}=t_{ji}=t$. For a  number of electrons $N_{e}$ matching the number $N$ of QDs, $N=N_{e}$, the low-energy Hamiltonian for suitable adjusted gate potentials $V_{i}$, in the limit $2t_{ij}\ll |2U\pm (V_{1})-V_{3}|,\,|2U\pm 2V_{2}- (V_{1})+V_{3}|$, can be approximated by a Schrieffer-Wolff transformation\cite{Schrieffer1966,Bravyi2011} yielding an Heisenberg spin chain
\begin{align}
H\st{Heis}=\sum\limits_{\braket{i,j}} J_{ij} \boldsymbol{S}_{i}\,\cdot\,\boldsymbol{S}_{j},
\end{align}
where $\boldsymbol{S}_{i}$ is the spin operator of the $i$-th electron in QD~$i$ and the sum runs over neighboring pairs of electrons. While the general case is quite interesting, in this review we are only interested in small systems with $N\leq 3$.

We explicitly demonstrate the simplest case for exchange, two electrons in $N=2$ QDs. Considering a single (valley-) orbital for each QD and restricting ourselves to the $S_{z}=0$ subspace there are four relevant states
\begin{align}
	\ket{s}&\equiv \frac{1}{\sqrt{2}}\left(c_{1,\uparrow}^\dagger c_{2,\downarrow}^\dagger-c_{1,\downarrow}^\dagger c_{2,\uparrow}^\dagger\right)\ket{\text{vac}},   \\
	\ket{t_{0}}&\equiv \frac{1}{\sqrt{2}}\left(c_{1,\uparrow}^\dagger c_{2,\downarrow}^\dagger+c_{1,\downarrow}^\dagger c_{2,\uparrow}^\dagger\right)\ket{\text{vac}},   \\
	\ket{s}_{11}&\equiv  c_{1,\uparrow}^\dagger c_{1,\downarrow}^\dagger\ket{\text{vac}},\\
	\ket{s}_{22}&\equiv  c_{2,\uparrow}^\dagger c_{2,\downarrow}^\dagger\ket{\text{vac}}.
	\label{eq:STstates}
\end{align}
States with $S_{z}=\pm 1$ and $\ket{t_{0}}$ are pure triplet states, thus, not coupled to any intermediate states with (2,0) or (0,2) charge configurations and, therefore, not affected by the exchange interaction.
Introducing the dipolar detuning parameter $\varepsilon\equiv(V_{1}-V_{2})/2$, the charging energy $U=\tilde{U}-U_{C}$ and assuming real valued tunneling parameters, the matrix representation of Eq.~\eqref{eq:Hub} is given by
\begin{align}
H\st{ST}=\left(
	\begin{matrix}
	 0 & 0 & \sqrt{2}t &\sqrt{2} t\\
	 0 & 0 & 0 & 0 \\
	\sqrt{2} t &  0 & U + \varepsilon \\
	\sqrt{2} t & 0 & 0 & U-\varepsilon
	\end{matrix}
	\right).
	\label{eq:hubmatrix}
\end{align}

\begin{figure}
\begin{center}
\includegraphics[width=0.9\columnwidth]{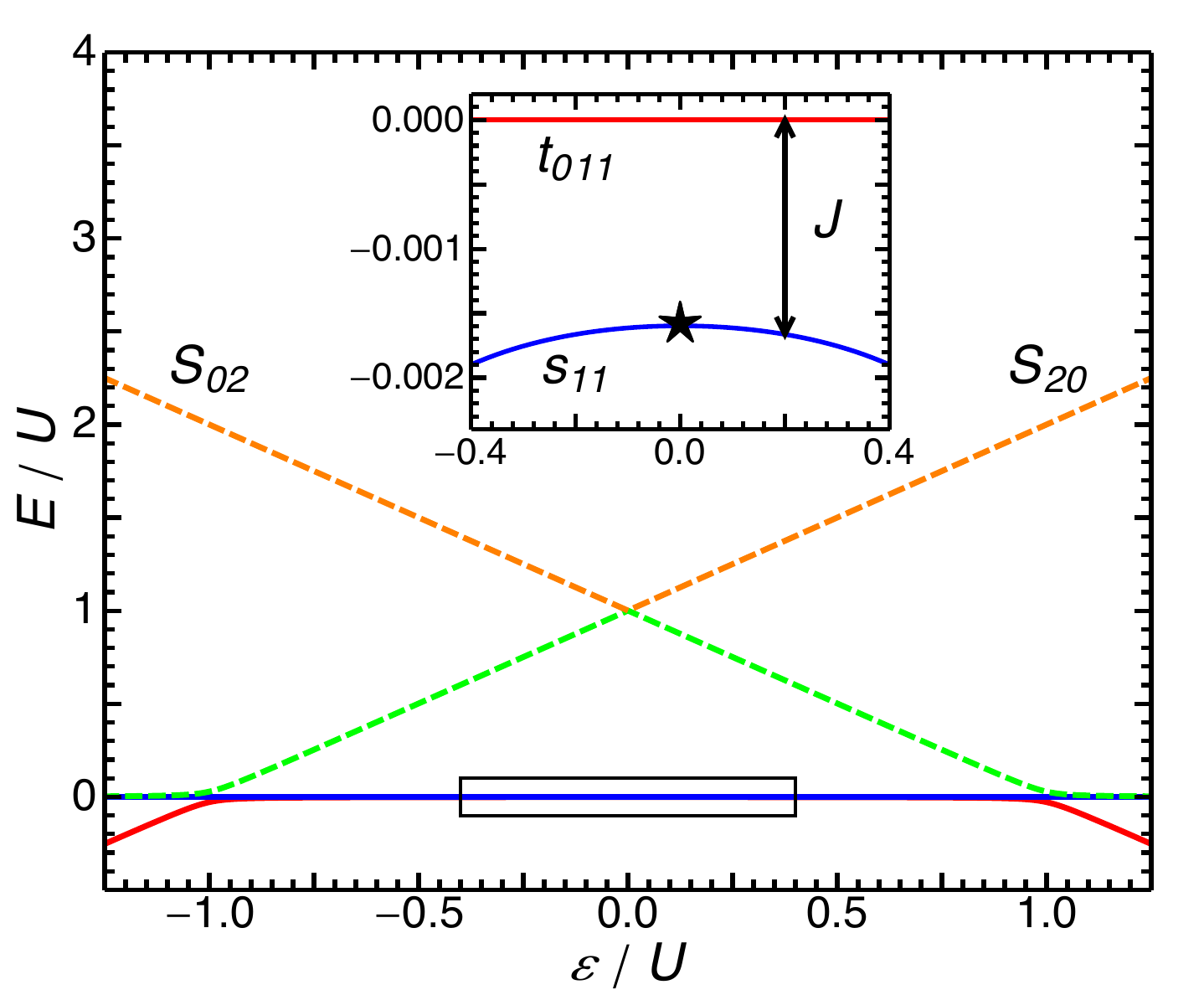}
\caption{Eigenenergies (in units of the charging energy $U$) of the singlet-triplet (ST) qubit as a function of detuning $\varepsilon$ for weak tunneling, $t=0.02\tun$. The inset magnifies the energy splitting between the $\ket{s}$ and $\ket{t_{0}}$ state due to the exchange interaction $J$. Inside the (1,1) charge configuration regime $J\approx4t^{2}U/(U^{2}-\varepsilon^{2})$\cite{Burkard1999}. The eigenenergies are labeled with their dominating charge configuration. There is a sweet spot (black star in inset) at zero detuning, $\varepsilon=0$.}
\label{fig:eigenenergiesST}
\end{center}
\end{figure}

In Fig.~\ref{fig:eigenenergiesST} the eigenenergies of this Hamiltonian are plotted as a function of the detuning $\varepsilon$. Inside the (1,1)-charge configuration regime the singlet qubit state is hybridized, $\ket{s}\rightarrow\ket{\tilde{s}}$, by the admixture of the other charge states, thereby, splitting the qubit by the exchange interaction $J\approx4t^{2}U/(U^{2}-\varepsilon^{2})$\cite{Burkard1999} which can be used either for entangling two-qubit gates or single qubit rotations depending on the implementation of the logical qubit.

\subsection{Spin-$\frac{1}{2}$ qubit}
\label{ssec:LDQ}

The original idea for a semiconductor electron spin qubit was proposed by Loss and DiVincenzo\cite{Loss1998} two decades ago. As the simplest choice, the spin-$\frac{1}{2}$ qubit is encoded in the two-level system associated with the spin-degree of freedom, i.e., the $\ket{0}\equiv\ket{\uparrow}$ and $\ket{1}\equiv\ket{\downarrow}$ states, of a single electron confined in the lowest orbital of a single QD. Since the qubit states have opposite spin projections the qubit is susceptible to magnetic fields. An external magnetic field $\boldsymbol{B}(t)$ lifts the degeneracy between the qubit states by the Zeeman energy $E=g\mu_{B}\boldsymbol{B}(\tau)\cdot\boldsymbol{S}$\cite{Hanson2007,Kloeffel2013} and fixes the quantization axis. 

Considering $\boldsymbol{B}(\tau)=(B_{x}(\tau),0,B_{z})^{T}$ with a large time-independent magnetic field in $z$-direction, $B_{z}=B$, and a small oscillatory driving field in $x$-direction, $B_{x}(\tau)=B_{D}\cos(\omega\tau)$. In qubit space the Zeeman term takes the expression for electron spin resonance (ESR)
\begin{align}
H\st{ESR}=\hbar\omega_{z}\sigma_{z} + \hbar\omega_{x} \cos(\omega\tau)\sigma_{x}
\label{eq:ESR}
\end{align}
with the Zeeman energies $\hbar\omega_{z}=g\mu_{B}B$ and $\hbar\omega_{x}=g\mu_{B}B_{D}$. Turning on the oscillating field, therefore, causes Rabi transitions between the spin states which together with the energy splitting of the qubit states provide full control of the qubit\cite{Kloeffel2013}. As an alternative for oscillating magnetic fields one can also modulate the g-factor of the material instead which yields the same expression\cite{Loss1998}. The speed of the gates depends on the strength of the oscillatory driving field and the weak magnetic dipole interaction resulting in typical gate times $\tau_{g}\approx\unit[100]{ns}$\cite{Koppens2006}.

Electric dipole spin resonance (EDSR) can be seen as an improvement of ESR which allows for electric driving of the qubit instead of magnetic driving. In the presence of spin-orbit interaction an electric field $E(\tau)=E_{0}\cos(\omega\tau)$ induces, in general, non-zero components of a pseudo magnetic field $b(\tau)$ perpendicular to the static magnetic field\cite{Golovach2006,Kloeffel2013}. This perpendicular (pseudo) magnetic field yields the same dynamics in the system as a real magnetic field due to $b(\tau)\propto E(\tau)$ with a coupling strength $\omega_{\tilde{x}}$ depending on the spin-orbit parameters\cite{Golovach2006}. Experiments demonstrate successful qubit rotations including spin flips achieving gate times on the order of $\tau_{g}\approx\unit[100]{ns}$ in GaAs devices ($f\st{Rabi}\approx\unit[3]{MHz}$\cite{Brunner2011}, $f\st{Rabi}\approx\unit[9]{MHz}$\cite{Obata2010}) and in silicon devices ($f\st{Rabi}\approx\unit[5]{MHz}$\cite{Kawakami2014}). However, the proposed $\tau_{g}\approx\unit[10]{ns}$\cite{Golovach2006} are not reached yet due to problems occurring at high electric fields , e.g., incomplete spin flips,\cite{Khomitsky2012}. Since the Rabi oscillations depend on the strength of the spin-orbit interaction, materials with strong spin-orbit coupling, such as InAs nanowires, can increase spin-flip frequency correspondingly\cite{Li2013}. Rabi oscillation as fast as $f\st{Rabi}\approx\unit[58]{MHz}$ were demonstrated\cite{Nadj2010}. However, the qubit fidelity of these fast gates is quite poor ($\approx$ 50\%) due to strong dephasing from nuclear spins\cite{Nadj2010}. 

Alternatively, one can use an oscillating magnetic field combined with a gradient in the magnetic field (slanting magnetic field) which does not rely on spin-orbit interaction \cite{Tokura2006} allowing for the use of materials with weak spin-orbit coupling such as silicon. The electric field $E(\tau)$ induces an oscillation of the electron position such that the electron experiences an oscillating magnetic field of the same frequency $B(\tau)\propto E(\tau)$. Experiments in a GaAs device using the Overhauser fields demonstrate a spin-flip time comparable to standard ESR ($\tau_{g}\approx\unit[110]{ns}$)\cite{Nowack2007,Laird2007} while the use of an integrated magnet yield a spin-flip time as fast as $\tau_{g}\approx\unit[20]{ns}$\cite{Pioro2008} due to larger field gradients. Reaching a high gate fidelity should be feasible for silicon devices due to the large amount of spin free nuclei.

Initialization and read-out schemes, among others, require a nearby auxiliary QD in order to enable a spin-to-charge conversion which is detectable by a quantum point contact (QCP) or the coupling to the lead\cite{Hanson2007}. Since this technique is of a more general kind and applicable for other qubit implementations we postpone its description to subsection~\ref{ssec:physReal}.

The common implementation of two-qubit gates for spin-$\frac{1}{2}$ qubits makes use of the exchange interaction\cite{Petta2005,Hanson2007} between two electrons in neighboring QDs (see previous subsection~\ref{ssec:exchange}) which induces the universal $\sqrt{\textsc{swap}}$-gate as described in the original proposal\cite{Loss1998}. Experiments demonstrate  two-qubit gates with a gate time $\tau_{g}\approx\unit[180]{ps}$\cite{Petta2005,Hanson2007,Foletti2009} with gate fidelities exceeding 99\%. A full demonstration of universal quantum control in two spin-$\frac{1}{2}$ qubits was demonstrated about a decade after the original proposal$\cite{Brunner2011}$.

The main advantage of single spin-$\frac{1}{2}$ qubits is their immunity to electric charge noise from background fluctuations; however, two or more coupled spin-$\frac{1}{2}$ qubits are not immune, since the exchange interaction needed for two-qubit gates is very sensitive to these, limiting the gate fidelity\cite{Culcer2009b}. Improvements use, among others, dynamical decoupling techniques or operate the qubits at a sweet spot\cite{Vion2002,Paladino2014,Gong2016}, working points least susceptible to the noise. In this review we do not focus on decoherence in single-spin qubits which is already extensively studied in related reviews (see Ref.~\cite{Chirolli2008} and Ref.~\cite{Kloeffel2013}).

It is disadvantageous, that electric control of the spin-$\frac{1}{2}$ qubits is challenging to implement and realize due to their dependence on either slanting magnetic fields or spin-orbit effects whereby oscillating magnetic fields are as an alternative not very reliable due to the weak coupling to the spins.
Another drawback of the spin-$\frac{1}{2}$ qubit is the rather strong susceptibility to (global) magnetic fields which are on one hand required for fast single qubit operations while on the other hand they couple the qubit to magnetic noise giving rise to strong decoherence. The strongest source of decoherence is magnetic noise due to nuclear spins\cite{Merkulov2002,Coish2004,Fischer2009,Coish2009,Kloeffel2013} while relaxation processes are dominated by the spin-orbit interaction\cite{Khaetskii2001,Golovach2004,Bulaev2005,Falko2005,Kloeffel2013,Prada2016}. Theoretical studies\cite{Merkulov2002} and experimental demonstration\cite{Petta2005,Hanson2007} show typical dephasing times on the order of $T_{2}^{\star}\approx\unit[10]{ns}$ in GaAs devices. However, due to the slow dynamic of the nuclear field the coherence time of the qubit can significantly increased by polarizing the nuclear spin\cite{Coish2009} which, however, has not been successfully demonstrated yet. Relaxation processes on the other hand scale with an external magnetic field and are several orders of magnitudes larger\cite{Hanson2007,Amasha2008}. Both main sources for decoherence are reduced significantly in silicon devices due to the smaller number of nuclear spins and weaker spin-orbit interaction\cite{Zwanenburg2013}. 

\subsection{Singlet-triplet (ST) qubit}
\label{ssec:STQ}

One idea to achieve (partial) electrical control of the qubit gates and to counteract the sensitivity of spin-$\frac{1}{2}$ qubits to fluctuations in global magnetic field (here labelled global magnetic noise) is to encode the quantum information in the $S_{z}=0$ subspace of two electrons in a double quantum dot (DQD). One state is the $S_{z}=0$ triplet state $\ket{t_{0}}\equiv(\ket{\uparrow,\downarrow}+\ket{\downarrow,\uparrow})/\sqrt{2}$ and the other state is the singlet state $\ket{s}\equiv(\ket{\uparrow,\downarrow}-\ket{\downarrow,\uparrow})/\sqrt{2}$. Since both states have the same $S_{z}=0$ quantum number, global magnetic noise pointing along the quantization axis has no effect on these two states, thus, the singlet-triplet (S-T) qubit is protected against such noise and a simple example of a decoherence free subspace (DFS) qubit. 

One axis of qubit control is provided by the electrically controllable exchange interaction between the electrons in the DQD due to the hybridization of the singlet energy given by admixture of charge states with doubly occupied dots giving rise to a splitting of the singlet and triplet energy. It is the same mechanism that provides the two-qubit gates for the spin-$\frac{1}{2}$ qubit which can be controlled to a very high degree yielding gate fidelities exceeding 99\%\cite{Gong2016} with gate times in below one nanoseconds\cite{Petta2005,Gong2016}.

A second axis of control is provided by a gradient of the magnetic field in the DQD which lifts the degeneracy between the states $(\ket{t_{0}}+\ket{s})/\sqrt{2}=\ket{\uparrow,\downarrow}$ and $(\ket{t_{0}}-\ket{s})/\sqrt{2}=\ket{\downarrow,\uparrow}$ due to the difference in magnetic fields in the two QDs. This leads to rotations of the qubit around an axis orthogonal to the quantization axis\cite{Levy2002}. Experimentally, these gradients can either be implemented by the Overhauser fields of the nuclear spins in the host material, typical for GaAs, or using the artificial magnetic field gradient by placing a micromagnet in the vicinity of the DQD QD\cite{Obata2010}. The latter implementation is needed for materials without (with a low density of) nuclear spins such as silicon or optionally for better control of the interaction strength.

Read-out and state preparation can be achieved in the same way as for spin-$\frac{1}{2}$ qubits via ``spin-to-charge'' conversion, where the gates are adiabatically detuned in such a way that one of the doubly occupied states is energetically highly favored\cite{Hanson2007}. Due to the Pauli exclusion principle only the anti-symmetric singlet state $\ket{s}$ is coupled via tunneling $t$ to the doubly occupied state while the triplet state transition is forbidden giving rise to a read-out technique with high fidelity exceeding $99\%$. The requirements are spin conserving hopping and a single non-degenerate ground state in the QD with a sufficient energy gap to the excited states.

Two-qubit gates can be implemented by the short-ranged exchange interaction together with spin-orbit interaction (exchange interaction alone is insufficient due to its high symmetry)\cite{Barenco1995,Klinovaja2012}, magnetic field gradients\cite{Wardrop2014}, by capacitively coupled DQDs\cite{Taylor2005,Coish2005,Stepanenko2007,Calderon2015} or an auxiliary dot\cite{Mehl2014}. Long-ranged two-qubit gates can either use the electrostatic coupling between the DQDs\cite{Hanson2007,Taylor2007,Shulman2012,Srinivasa2015}, and/or the coupling of two DQDs to the same microwave cavity\cite{Imamoglu1999,Childress2004,Burkard2006,Taylor2006,Hu2012,Petersson2012,Viennot2015}. The later two approaches only differ by the use of a the microwave cavity as a mediator and have been under intense investigation recently due to the access to high quality and high conductance microwave cavities\cite{Guilherme2014,Samkharadze2016}. The requirements for such an interaction, reaching strong-coupling regime where the transfer of information is faster than its loss, was very recently indicated\cite{Kontos2016,Viennot2016,Mi2016}.

The downside of all the advantages provided by the ST qubit is the opening of a channel which couples the qubit to charge noise, electric fluctuations of the environment or the gate potentials. Electric fields can couple to the qubit through the detuning parameter $\varepsilon$ and in this way give rise to an exponential decay of coherence due to dephasing. The exact decay depends on the spectral density of the noise $S_{q}(\tilde{\omega})=A |\tilde{\omega}|^{-\gamma}$ where $A$ is the strength of the noise, $\tilde{\omega}$ is the noise frequency, and $\gamma$ is the spectral density exponent which usually has to be set phenomenologically or needs to be measured in experiments and strongly depends on the host material and device fabrication. Typical values for $\gamma$ range from $0.7$ to $2.3$\cite{Jung2004,Mueller2006,Dutta1981,Weissman1988,Paladino2014}, but higher values are not unusual\cite{Beaudoin2015}. Protection against such charge noise can be obtained by operating the qubit system at a high symmetry point, where the transition to both asymmetric charge states is equal. At such a sweet spot, the ground state energy gap as a function of $\varepsilon$ has an extremum, thus, the qubit is immune to energy fluctuations in $\varepsilon$ due to charge noise in first order\cite{Ramon2010,Ramon2011,Hiltunen2015,Gong2016,Reed2016,Martins2016}. However, second or higher order effects still limit the dephasing time.

\section{Three-electron spin qubits}
\label{sec:3spin}

Taking the idea of electrical control and protection of the qubit against noise one step further leads ultimately to the three-spin qubit which can be controlled fully electrically. Some three-spin qubits also form a decoherence-free subspace (DFS) qubit implying that they are immune to all collective decoherence, decoherence which affects all spins in one qubit, many nearby qubits, or ideally in the full quantum computer in the same way\cite{Lidar2012}. There are many different ways of implementing such a three-spin qubit. In this review we cover the exchange-only (EO) qubit\cite{Nature2000}, the spin-charge qubit\cite{Kyriakidis2007}, the hybrid qubit\cite{Shi2012,Koh2012,Cao2016}, the resonant exchange (RX) qubit\cite{Medford2013,Taylor2013}, and on the always-on exchange-only (AEON) qubit\cite{Shim2016}. All of these qubit implementations are realized using three electrons in either a single quantum dot, double quantum dot (DQD), or triple quantum dot (TQD) depending on the qubit implementation. The full spin-space is spanned by $\mathcal{H}\st{3spin}=\mathcal{H}_{1/2}\otimes \mathcal{H}_{1/2}\otimes \mathcal{H}_{1/2}$ which can be divided into two spin-$\frac{1}{2}$ and one spin-$\frac{3}{2}$ subspace, thus, $\mathcal{H}\st{3spin}=\mathcal{H}_{1/2}\oplus \mathcal{H}_{1/2}\oplus \mathcal{H}_{3/2}$, where $\mathcal{H}_{\sigma}$ denotes the Hilbert space with total spin $\sigma$. In other words the Hilbert space can be separated into a $ S=3/2 $ quadruplet and a degenerate $ S=1/2 $ doublet\cite{Buchachenko2002,Laird2010} which can further be split into a high and low energy doublet by an external magnetic field along the $z$-axis. The qubit states for these qubits are chosen in such a way that they have identical spin quantum numbers, both the total spin $S=1/2$ and the total spin projection along the quantization axis $S_{z}=1/2$ giving rise to global immunity against magnetic fluctuations. Different qubit realizations are introduced and discussed in the following subsections in detail and we postpone a more detailed discussion about DFSs and further dynamical (noise) decoupling schemes in three spin qubits to section~\ref{sec:dec} and refer to a related review\cite{Lidar2012} for more details. However, before delving into the qubit implementations, some basic properties of electrons in TQDs need to be introduced.

For the description of TQDs, the extended Hubbard Hamiltonian (see Eq.~\eqref{eq:Hub} for $N=3$ quantum dots) is an appropriate choice throughout almost the full review since it combines all key features of the three-spin qubits while the expressions are still succinct. Therefore, in this review we skip a realistic and comprehensive discussion of the exact energy levels and their microscopic dependence on gate voltages, the geometry of the TQD, the number of electrons, and the magnetic field\cite{Waugh1995,Usukura2005,Korkusinski2007,Delgado2007,Li2007,Hsieh2010,Hsieh2012} and only briefly introduce the key points while sticking to the Hubbard model in the remainder. A comprehensive study of these can be found in the related review \cite{Hsieh2012}. 

\subsection{Physical realization and measurement techniques}
\label{ssec:physReal}

Although three-spin qubits were already proposed in 2000\cite{Nature2000} it took several years for the appearance of devices capable to operate three-spin qubits due to technical and engineering challenges. The main challenges for a functioning three-spin (here three electrons in a TQD) device is the complexity of the system since several gate electrodes are needed in order to address each electron individually. The first device, which has neither linearly or evenly triangular geometry, was realized around 2006 \cite{Gaudreau2006} and was soon followed by other realizations\cite{Schroeer2007,Ihn2007,Rogge2008, Gaudreau2009,Yamahata2009,Pierre2009,Laird2010,Takakura2010,Granger2010,Gaudreau2012,Amaha2012,Pan2012,Busl2013} which improve the positioning of the electrons. A functioning three-spin qubit device capable of quantum computation is demonstrated in a DQD\cite{Kim2014,Shi2014,Kim2015b,Cao2016} and in a linear TQD\cite{Gaudreau2009,Laird2010,Gaudreau2012,Aers2012,Amaha2012,Medford2013,Medford2013b,Sanchez2014,Eng2015,Poulin2015}. While a triangular shape provides some interesting new features, e.g., chirality\cite{Scarola2004,Scarola2005,Hsieh2012b,Luczak2012,Urbaniak2013,Tooski2014,Luczak2014} and faster qubit operations\cite{Nature2000,Setiawan2014}, the advantages currently do not seem to outweigh the experimental drawbacks and difficulties. Therefore and since almost all experiments and most theoretical studies use the linear geometry we also mostly stick in this review to the linear geometry and implicitly consider that each TQD is linearly arranged, unless otherwise mentioned.

\begin{figure}[t]
\begin{center}
\includegraphics[width=1.0\columnwidth]{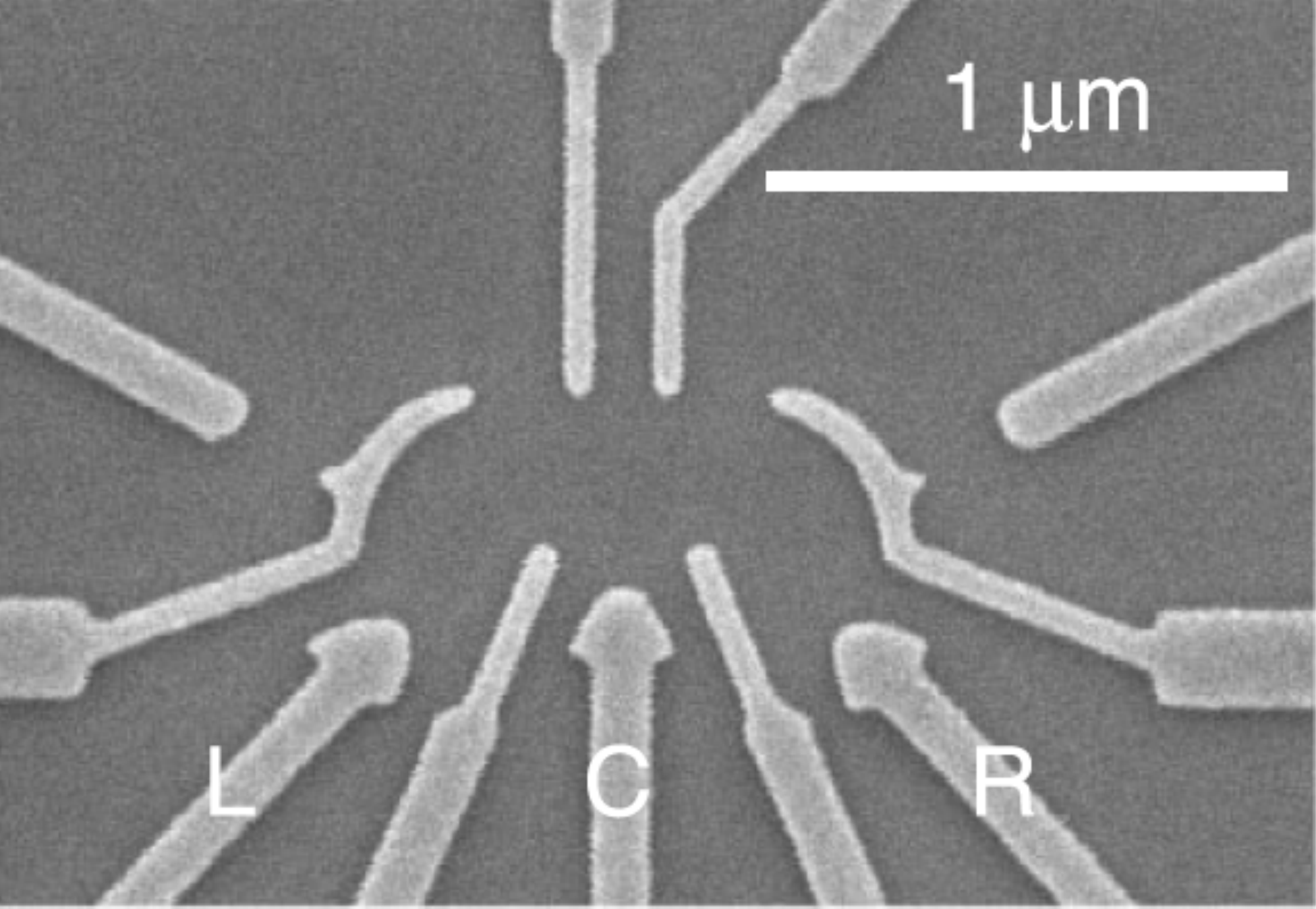}
\caption{Scanning electron micrograph of a depletion triple quantum dot (TQD) device in GaAs. Metallic gates (light gray) are deposited on top of the heterostructure (dark gray) to deplete the 2DEG underneath in order to form isolated structures. The number of electrons and the energy in each QD is controlled by the gates labeled as L, C, and R. Tunneling between the electrons is controlled by the thinner gates in between. Figure taken from Ref.~\cite{Sanchez2014}.}
\label{fig:model}
\end{center}
\end{figure}

The most common technique implementing semiconductor QDs are lateral QDs where a two dimensional electron gas (2DEG) is further confined by electrostatic potentials provided by the gate electrodes forming a (approximately) zero dimensional structure (see Fig.~\ref{fig:model}). While GaAs\cite{Hanson2007} and silicon (Si)\cite{Zwanenburg2013} are the typical choice of material, more exotic semiconductors\cite{Recher2009,Churchill2009,Reynoso2011,Reynoso2012,Song2015,Song2015b,Laird2015,Wang2016} are also possible whereby rare. Crucial for the implementation is the 2DEG that usually is realized by the fabrication of a heterostructure which accumulates the electrons at the interface; the GaAs layer is sorrounded by a AlGaAs layer\cite{Hanson2007} and Si by either a SiGe\cite{Zwanenburg2013} layer or a $\text{SiO}\st{x}$\cite{Angus2007,Jones2016} layer. Advances in the fabrication process allow for layer interfaces at atomistic scales and gate structures with a very high precision giving rise to scalable and controllable quantum dot devices (selected examples see Refs.\cite{Yang2013,Veldhorst2014,Muhonen2014,Eng2015,Morello2015,Borselli2015,Zajac2015,Reed2016,Ward2016,Knapp2016}) with the record of nine individually addressable QDs\cite{Zajac2016}. Regarding the choice for material there is a clear tendency towards favoring Si since Si consists of $\approx 95\%$ nuclear free isotopes which can be increased with isotopic purification\cite{Itoh2014}. For further confinement of the electrons, metallic gates are placed on top and/or underneath the heterostructure which locally deplete the 2DEG forming an isolated island. In Si/SiGe the 2DEG is typically empty in the beginning and the gates accumulate electrons in the 2DEG\cite{Zwanenburg2013}. Tuning of the gate voltages giving rise to the few electron regime\cite{Hanson2007,Zwanenburg2013}. Due to the higher complexity of the multiple-dot nano-devices recent devices use a stacking gate architecture useful for scaling up\cite{Kawakami2014,Zajac2015,Borselli2015,Zajac2016,Jones2016}. A typical setup for a TQD consists of at least five gates (see Fig.~\ref{fig:model}), three gates (L-, C-, R-gate in Fig.~\ref{fig:model}) on top of each QD in order to control the energy and two gates in between for control of the tunneling coupling. Additional gates (number depends on the material and fabrication, e.g., four in the device seen in Fig.~\ref{fig:model}) are required to form the dots as well as to control the coupling to the lead which allows for initialization. 

Common techniques for measurement of the QD devices require additional QDs or quantum point contacts (QPC)\cite{Field1993,Barthel2009}, single electron transistors\cite{Devoret2000,Barthel2010,House2016}, or tunnel junctions\cite{House2014} in order to sense the number and movement of the electrons in each dot in a time resolved measurement\cite{Hanson2007,Zwanenburg2013}. Due to the finite range of the charge sensors, large arrays of QDs have multiple sensors, e.g., a nine-dot device has three charge sensors\cite{Zajac2016}. Another measurement technique involves photons that carry the information out of the device. Connecting the device to a microwave cavity allows for read-out of device parameters using cavity quantum electrodynamics (cQED) without directly interfering with the device\cite{Kerman2008,Delbecq2011,Kerman2013,Gonzalez2014,Didier2015,Burkard2016,Mi2016,Beaudoin2016a}. Both measurements also allow for read-out of the qubit state; cavity read-out additionally requires strong coupling.
The measurement techniques can be grouped as either invasive, e.g., emptying the QD\cite{Hanson2007,Zwanenburg2013}, or noninvasive, sensing the charge or spin of the electrons in the QD without changing the electron number\cite{Hanson2007,Zwanenburg2013}.

\subsection{Electric properties of electrons in a TQD}
\label{ssec:chargestab}

\begin{figure}
\begin{center}
\includegraphics[width=0.9\columnwidth]{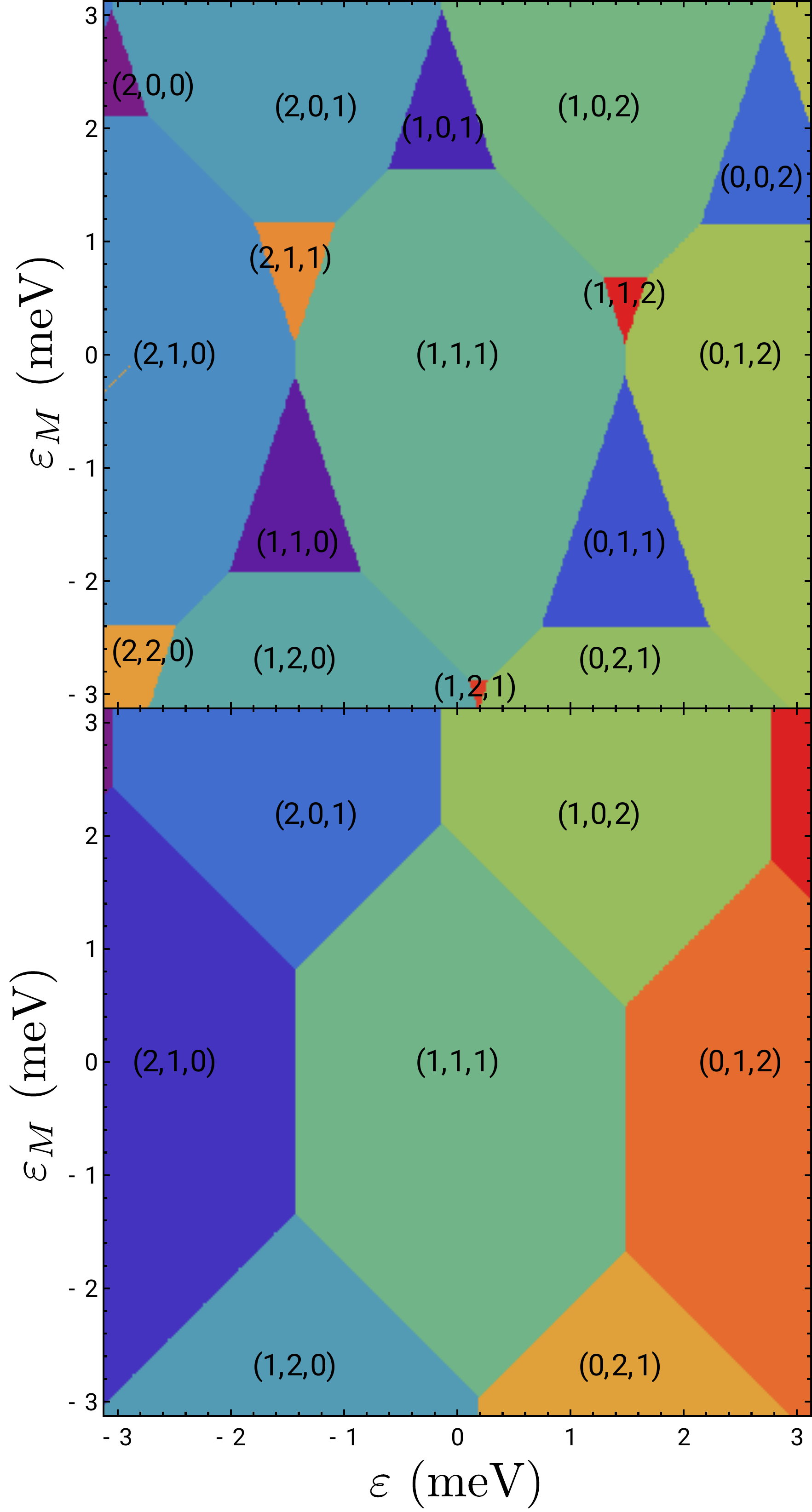}
\caption{Charge stability diagram of a triple quantum dot (TQD) with realistic parameter settings as a function of detuning parameters $\varepsilon$ and $\varepsilon_{M}$ from Eq.~\eqref{eq:defDet} (a) for an arbitrary number of electrons and (b) for a fixed number $n=3$ of electrons. Both diagrams were obtained using a capacitance model of the TQD adapted from Refs.~\cite{Schroeer2007,Rogge2009}.}
\label{fig:csTQD}
\end{center}
\end{figure}

As a first step to visualize, navigate, and find relevant states in the large Hilbert space of multi-electron states in a TQD, the charge stability diagram of the TQD is helpful as it highlights the charge transitions between different occupancies of multiple QDs\cite{Wiel2002,Vidan2005,Schroeer2007,Rogge2009,Granger2010,Hsieh2012,Seo2013,Broome2016} and neglects all spin related effects. To generate the charge stability diagram, we use here a modified version of the algorithm used in Refs.~\cite{Khoi2013,Powell2014} with a maximum number of (four) electrons in the TQD and a fixed rate for tunneling of the electrons. 
Fig.~\ref{fig:csTQD} shows the low electron occupancy part of the charge stability diagram as a function of the two detuning parameters defined as
\begin{align}
\varepsilon =& (\mu_{1}-\mu_{3})/2, \\
\varepsilon_{M} =& \mu_{2} - (\mu_{1}+\mu_{3})/2,
\label{eq:defDet}
\end{align}
for a fixed value of the avarage voltage $eV\st{av} = (\mu_{1}+\mu_{2}+\mu_{3})/3$. The parameter $\mu_{i}=\sum_{j=1}^{3}v_{i,j}V_{j}$ with $i,j=1,2,3$ is the chemical potential of QD~$i$ given by the gate voltages $V_{i}$ (see Fig.~\ref{fig:model}) underneath each QD and  $v_{i,j}$ which describes the electrostatic interaction between QD~$i$ and the gate underneath QD~$j$ and depend of the charging energies\cite{Wiel2002,Schroeer2007}. In brief words, the $v_{i,j}$ describe how each gate has to be adjusted in order to change the chemical potential in each dot.

Taking into account a finite coupling between the QDs due to cross capacitance effects, i.e., adding an electron in one QD changes the potential of the neighboring QD, the typical honeycomb structured diagrams shown in Fig.~\ref{fig:csTQD} are obtained. 

In the center of the charge stability diagram for a fixed value of $V\st{tot}$ lies the (1,1,1) charge configuration regime with one electron in each QD surrounded by the six asymmetric charge configurations, (2,0,1), (1,0,2), (1,2,0), (0,2,1), (2,1,0), and (0,1,2) with the same number of electrons. Here, $(l,m,n)$ labels a charge configuration with $l$ electrons in the left~QD, $m$ electrons in the center~QD, and $n$ electrons in the right~QD. Each of these asymmetric states except the last two are interlinked with the (1,1,1) charge configuration through the motion of a single electron, the last two states require the motion of two electrons. States with triple occupation of a single QD are located at more extreme values of the detuning parameters. Note that the average voltage $V\st{av}$ roughly sets the total number of electrons in the TQD due to the finite coupling to the leads. 

Special points of interest for quantum computation and qubit implementations are typically centered inside a charge configuration regime or located at the charge transition points where multiple charge configurations intersect since these points provide a high symmetry with respect to charge configurations. 

\subsection{Spin properties of three-spin qubits}
\label{ssec:spinprop}

In a second step, spin and orbital effects are reintroduced which in general further subdivides the stability diagram. 
The Hilbert space of three electron-spins with spin $\frac{1}{2}$ in a TQD is $\mathcal{H}\st{3spin}=\mathcal{H}_{1/2}\otimes \mathcal{H}_{1/2}\otimes \mathcal{H}_{1/2}$ and combined with only a single available orbital in each QD contains in total 20 possible states (220 possible states for a second available orbital, e.g., additional valley). There are eight states with a symmetric charge configuration (1,1,1), and two states with asymmetric charge configurations (2,0,1), (1,0,2), (1,2,0), (0,2,1), (2,1,0), and (0,1,2) each. States with a triply occupied QD (3,0,0), (0,3,0), and (0,0,3) are excluded due to the restriction to a single available orbit in each dot. 

\begin{figure*}
\begin{center}
\includegraphics[width=0.9\textwidth]{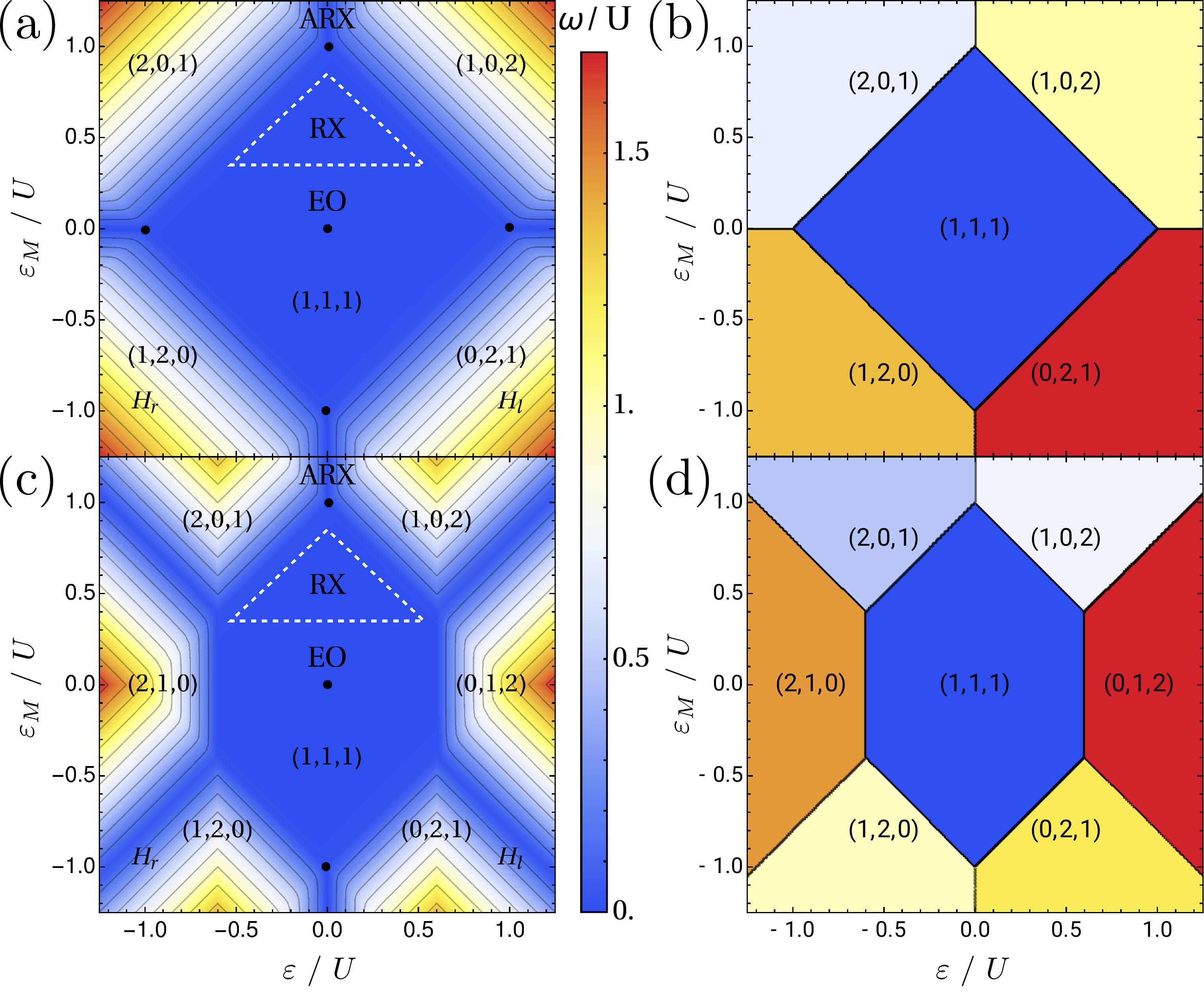}
\caption{(left column) Energy landscape of the ground-state energy gap $\omega$ of a three-spin qubit as a function of the detuning parameters $\varepsilon$ and $\varepsilon_{M} $ in a triple quantum dot in units of the charging energy $U$ (see also Refs.~\cite{Russ2016,Russ2016b}). For the tunneling parameters the ratios $t_{l}=0.022\tun$ and $t_{r}=0.015\tun$ and for the mutual charging energy $U_{C}=0.2\tun$  are used. Maneuvering through the ($\varepsilon$, $\varepsilon_{M} $)~plane one can access various parameter regimes that allow the use of different qubit implementations in different charge configurations $(l,m,n)$, where $l$ electrons are in the left, $m$ electrons in the center, and $n$ electrons in the right QD. We indicated the double sweet spots (DSS) (black dots), the location of the exchange-only (EO) qubit, the resonant exchange (RX) qubit (dashed triangle), the asymmetric resonant exchange (ARX) qubit, and the left and right hybrid (H$_{l,r}$) qubit highlighted.
(right column) Energetic optimal charge configuration of the ground state of a three-spin qubit in the absence of tunneling, $t_{l}=t_{r}=0$ as a function of the detuning parameters $\varepsilon$ and $\varepsilon_{M} $ in a triple quantum dot in units of the charging energy $U$. For plots~(a) and (b) the states with (2,1,0) and (0,1,2) charge configurations are neglected corresponding to large values of $U_{C}\lesssim U$ while for plots~(c) and (d) small values $U_{C}=0.2\tun$ are considered.
}
\label{fig:energygap}
\end{center}
\end{figure*}

The corresponding spin Hilbert space $\mathcal{H}\st{3spin}=\mathcal{H}_{3/2}\oplus \mathcal{H}_{1/2}\oplus \mathcal{H}_{1/2}$ can be divided into a quadruplet $\mathcal{H}_{3/2}$ with effective spin-$3/2$ and two degenerate doublets $\mathcal{H}_{1/2}$ which combined with different orbits and restricted to the total spin $S=1/2$ subspace gives rise to a two-fold degenerate subspace $\mathcal{H}\st{+1/2}\oplus \mathcal{H}\st{-1/2}$. This subspace is effectively decoupled from the $S=3/2$ subspace considering weak magnetic field gradients\cite{Hung2014} and weak spin orbit interaction\cite{Mehl2013}. Leakage into the $S=3/2$ and $S_{z}=\pm 1/2$ states is suppressed by exchange\cite{Hung2014}. These two subspaces, distinguished by $S_{z}=\pm 1/2$, are interchangeable with respect to the exchange interaction, thus an external magnetic field allows us to focus on only one of them, e.g., $S=1/2$, $S_{z}=+1/2$. Without loss of generality, the $S=S_{z}=+1/2$ subspace is spanned by the basis states
\begin{align}
	\ket{0}&\equiv\ket{s}_{13}\ket{\uparrow}_{2},\label{eq:basisstatesBeg}\\
	\label{state:0}
	\ket{1}&\equiv\sqrt{\frac{2}{3}}\ket{t_{+}}_{13}\ket{\downarrow}_{2}-\frac{1}{\sqrt{3}}\ket{t_{0}}_{13}\ket{\uparrow}_{2},\\
	\label{state:1}
	\ket{2}&\equiv \ket{s}_{11}\ket{\uparrow}_{3},   \\
	\ket{3}&\equiv \ket{\uparrow}_{1}\ket{s}_{33},\\
	\ket{4}&\equiv \ket{\uparrow}_{1}\ket{s}_{22},   \\ 
	\ket{5}&\equiv \ket{s}_{22}\ket{\uparrow}_{3}, \\
	\ket{6}&\equiv \ket{s}_{11}\ket{\uparrow}_{2},\\
	\ket{7}&\equiv \ket{\uparrow}_{2}\ket{s}_{33}.
	\label{eq:basisstates}
\end{align}
with the two-electron singlet state $\ket{s}_{ij}\equiv(\ket{\uparrow}\ket{\downarrow}-\ket{\downarrow}\ket{\uparrow})/\sqrt{2}$ and the two-electron triplet states $\ket{t_{0}}_{ij}\equiv(\ket{\uparrow}\ket{\downarrow}+\ket{\downarrow}\ket{\uparrow})/\sqrt{2}$ and $\ket{t_{+}}_{ij}\equiv\ket{\uparrow}\ket{\uparrow}$ occupying QD~$i$ and QD~$j$.
Since the doubly occupied states $\ket{2}$, $\ket{3}$, $\ket{4}$, and $\ket{5}$ are obtained from $\ket{0}$ and $\ket{1}$ via the motion of a single electron and the states $\ket{6}$ and $\ket{7}$ require the motion of two electrons, the latter two states are neglected in most studies.\cite{Medford2013,Taylor2013,Mehl2013,Fei2015,Russ2015,Shim2016,Russ2016} The resulting matrix representation of the Hamiltonian Eq.~\eqref{eq:Hub} in this basis up to a global energy shift, is\cite{Russ2016} 
\begin{align}
H=\left(
\begin{smallmatrix}
 0 & 0 & t_{l}/2 & t_{r}/2 & t_{r}/2 &t_{l}/2 \\
 0 & 0 & \sqrt{3}t_{l}/2& -\sqrt{3}t_{r}/2 & -\sqrt{3}t_{r}/2 &\sqrt{3} t_{l}/2 \\
 t_{l}/2 & \sqrt{3}t_{l}/2 & E_{2} & 0 & 0 & 0 \\
t_{r}/2 & -\sqrt{3} t_{r}/2 & 0 & E_{3} & 0 & 0 \\
t_{r}/2 & -\sqrt{3} t_{r}/2 & 0 & 0 & E_{4}& 0 \\
t_{l}/2 & \sqrt{3}t_{l}/2 & 0 & 0 & 0 & E_{5}
\end{smallmatrix}
\right).
\label{eq:EOmatrix8}
\end{align}
The symmetric tunneling parameters are $ t_{12}=t_{21}\equiv t_{l}/ \sqrt{2}$, $ t_{23}=t_{32}\equiv t_{r}/ \sqrt{2}$, and $t_{13}=t_{31}=0$ and the simplified expressions for the charging energies of the states are
\begin{align}
E_{2} &= \varepsilon-\varepsilon_{M}+U,\\
E_{3} &= -\varepsilon-\varepsilon_{M}+U,\\
E_{4} &= \varepsilon+\varepsilon_{M}+U,\\
E_{5} &= -\varepsilon+\varepsilon_{M}+U.
\label{eq:chargeEnergieEO6}
\end{align}
In this case, all of the charging energies $E_{i}$ depend only on the two detuning parameters 
\begin{align}
\varepsilon =& (V_{1}-V_{3})/2, \\
\varepsilon_{M} =& V_{2} - (V_{1}+V_{3})/2+U_{C},
\label{eq:defDet}
\end{align}
and the charging energy $U=\tilde{U}-U_{C}$. For the general expressions this is not true and one needs more charging energies for a full description\cite{Shim2016}. 

The logical choice for a qubit is the two-level system consisting of the ground state and the first excited state, energetically split by the ground-state energy gap $\omega$. For $t_{l,r}\ll E_{i}$, these are essentially the states $\ket{0}$ and $\ket{1}$ with corrections $\propto t_{l,r}/E_{i}$. In Figs.~\ref{fig:energygap}~(a) and \ref{fig:energygap}~(c) the ground-state energy-gap is plotted as a function of the two detuning parameters, $\varepsilon$ and $\varepsilon_{M}$, with labels indicating the dominant charge configuration (a) excluding and (c) including the states $\ket{6}$ and $\ket{7}$. Fig.~\ref{fig:energygap}~(b) shows the dominant contribution of the ground state from Eq.~\eqref{eq:basisstates} whereas in Fig.~\ref{fig:energygap}~(d) the states $\ket{6}$ and $\ket{7}$ are included. Therefore, the qubit states have different charge configurations depending on the exact location in the detuning space. In the (1,1,1) charge configuration regime the spin qubit states are $\ket{0}$ and $\ket{1}$
hybridized by the admixture of the asymmetric states $\ket{2},\,\ket{3},\,\ket{4},\,\ket{5}$ (to a less degree also by $\ket{6}$ and $\ket{7}$) giving rise to a finite energy gap between the states $\omega$.

\subsection{Exchange-only (EO) qubit}
\label{ssec:EOQ}

The idea of all-electric qubit control ultimately leads to the exchange-only qubit which, as the name suggests, provides the possibility for full qubit control with only the exchange interaction\cite{Nature2000}. Analogously to the ST qubit (see subsection~\ref{ssec:STQ}), the exchange interaction originates from the hybridization of the logical qubit states with asymmetric charge states and can be precisely controlled by electrostatic control of the gates underneath and in between the QDs. In this subsection we try to provide an overview of preceding experimental and theoretical developments of the exchange-only qubit. The organization is as follows. We start with the model and subsequently follow with the single-qubit operations, where we discuss the two main types of experimental realizations. We discuss two-qubit operations and the decoherence of our qubit due to environment separately in the next sections~\ref{sec:two} and~\ref{sec:dec}.

\subsubsection{Model}
\label{sssec:qubit}

For the EO~qubit the focus is on the eight-dimensional subspace with a symmetric (1,1,1) charge configuration which can be separated into a $ S=3/2 $ quadruplet and in a degenerate $ S=1/2 $ doublet\cite{Buchachenko2002,Mizel2004,Laird2010} which is lifted by an external magnetic field alined along the $z$-axis. We are interested in these doublets since each provides a two level system with two identical quantum numbers, one being the total spin $S=1/2$ the other being the projection of the total spin along the quantization axis $S_{z}=\pm 1/2$. For the $S_{z}=+ 1/2$ doublet an appropriate basis is given by 
\begin{align}
	\ket{0_+}&\equiv\ket{s}_{13}\ket{\uparrow}_2=\frac{1}{\sqrt{2}}\left(\ket{\uparrow,\uparrow,\downarrow}-\ket{\downarrow,\uparrow,\uparrow}\right) ,\\
	\ket{1_+}&\equiv\sqrt{\frac{2}{3}}\ket{t_{+}}_{13}\ket{\downarrow}_{2}-\frac{1}{\sqrt{3}}\ket{t_{0}}_{13}\ket{\uparrow}_{2}\nnb
	&=\frac{1}{\sqrt{6}}\left(2\ket{\uparrow,\downarrow,\uparrow}-\ket{\uparrow,\uparrow,\downarrow}-\ket{\downarrow,\uparrow,\uparrow}\right),
	\label{eq:EOplusStates}
\end{align}
while for the $S_{z}=- 1/2$ doublet all spins are flipped,
\begin{align}
	\ket{0_-}&\equiv\ket{s}_{13}\ket{\downarrow}_2=\frac{1}{\sqrt{2}}\left(\ket{\downarrow,\downarrow,\uparrow}-\ket{\uparrow,\downarrow,\downarrow}\right) ,\\
	\ket{1_-}&\equiv\sqrt{\frac{2}{3}}\ket{t_{-}}_{13}\ket{\uparrow}_{2}-\frac{1}{\sqrt{3}}\ket{t_{0}}_{13}\ket{\downarrow}_{2}\nnb
	&=\frac{1}{\sqrt{6}}\left(2\ket{\downarrow,\uparrow,\downarrow}-\ket{\downarrow,\downarrow,\uparrow}-\ket{\uparrow,\downarrow,\downarrow}\right).
	\label{eq:EOminusStates}
\end{align}
A special feature of the EO qubit is the possibility for two different qubit encodings using either the ``subspace'' or the ``subsystem'' encoding. For the subspace, as the name suggests, the qubit states are encoded in a real subspace of the total Hilbertspace, either in the positive doublet, $\ket{0}=\ket{0_{+}}$ or $\ket{1}=\ket{1_{+}}$, or in the negative doublet, $\ket{0}=\ket{0_{-}}$ or $\ket{1}=\ket{1_{-}}$. This implementation needs a sufficiently strong magnetic field along the quantization axis to break the degeneracy of the doublets and energetically favor one of the doublets depending on the sign of the magnetic field. Here, we use the convention that the positive ($S_{z}=+1/2$) subspace qubit is energetically favorable. Particularly, fine-tuning of the confinement potentials and the strength of the magnetic field energetically separates the doublet states from the quadruplet states\cite{Hawrylak2005} in Si\cite{Hsieh2012,Ren2014} and GaAs\cite{Hawrylak2005,Hsieh2012}. For a finite magnetic field, this pushes one of the doublets down in energy, such that it forms the ground and first excited state, hence, eliminating orbital relaxation. 

The second type of encoding is the ``subsystem'' qubit which utilizes both doublets, thus, $\ket{0}=\ket{0_{\pm}}$ and $\ket{1}=\ket{1_{\pm}}$ giving rise to a qubit implementations with one leftover degree of freedom. In the absence of a magnetic field there are parameter regimes where the orbital energies dominate pushing the quadruplet up in energy and the doublets down in energy\cite{Hsieh2012,Ren2014} allowing for the implementation of such a subsystem qubit\cite{Hsieh2012,Ren2014}. It is crucial for this implementation that there are no interactions which couple the $\ket{0_{\pm}}$ states differently than the $\ket{1_{\pm}}$ states. Under this condition, the doublets are not entangled and the additional degree of freedom can be rewritten into a global degree of freedom allowing for a well-defined qubit\cite{Kempe2001}. In realistic systems, the exchange interaction fulfills these conditions while local magnetic field gradients and spin-orbit coupling violate it.

In the low energy subspace in the (1,1,1) charge configuration regime a Schrieffer-Wolff transformation yields (analogously to the ST qubit) an effective Heisenberg Hamiltonian for the hybridized states, however, with three exchange coupling parameters
\begin{align}
	H\st{eff,TQD}=\frac{J_{12}}{4}\bo{\sigma_1}\cdot\bo{\sigma_2}+\frac{J_{23}}{4}\bo{\sigma_2}\cdot\bo{\sigma_3}+\frac{J_{13}}{4}\bo{\sigma_1}\cdot\bo{\sigma_3}.
	\label{eq:heisTQD}
\end{align}
For a linear arrangement and neglecting superexchange $J_{13}=0$.
The expressions for the exchange couplings depend on the choice of the subspace taken into account. Since the states with (2,1,0) and (0,1,2) charge configurations are not directly coupled to the (1,1,1) charge states they are usually neglected for the derivation of the exchange couplings. In this case the exchange couplings are given by\cite{Russ2016}
\begin{align}
J_{12}=J_{l}=& 2t_{l}^{2}U/\left[U^{2}-(\varepsilon-\varepsilon_{M})^{2}\right],\\
J_{23}=J_{r}=&2t_{r}^{2}U/\left[U^{2}-(\varepsilon+\varepsilon_{M})^{2}\right].
\label{eq:exchangeEO}
\end{align}
The more general expressions which include different Coulomb terms $U_{i}$ in each QD are given in Ref.~\cite{Shim2016} and are not shown here. 
Note that the resulting expressions for the exchange interactions are identical for both subspaces, $S_{z}=\pm1/2$. The resulting energy splitting between the qubit states is given by\cite{Taylor2013}
\begin{align}
\omega=\sqrt{J_{l}^{2}+J_{r}^{2}-J_{l}J_{r}}.
\end{align}
 
\subsubsection{Single-qubit operations}
\label{sssec:sqo}

The exchange-only qubit allows for all-electrical control of the qubit rotations with only the exchange interactions allowing for $J_{l}$ and $J_{r}$ two independent axes of control.
In the hybridized qubit basis, $\ket{0}$ and $\ket{1}$, the Heisenberg Hamiltonian from Eq.~\eqref{eq:heisTQD} can be expressed as
\begin{align}
H\st{qubit}=E_{0}\mathbbm{1}_{2}-\frac{J}{2}\sigma_{z}-\frac{\sqrt{3}\,j}{2}\sigma_{x}
\label{eq:EOqubit}
\end{align}
with the qubit Pauli matrices, $\sigma_{z}\equiv\ket{0}\bra{0}-\ket{1}\bra{1}$ and $\sigma_{x}\equiv\ket{0}\bra{1}+\ket{1}\bra{0}$, and the exchange energies $J\equiv (J_{l}+J_{r})/2$ and $j\equiv (J_{l}-J_{r})/2$. The first term $\propto E_{0}$ only contributes to a global phase of the qubit, thus, can be ignored. Note, that the rotation axes are provided by the sum and difference of the exchange interaction between the dots (see Eq.~\eqref{eq:exchangeEO}), thus, the rotation axes corresponding to an exchange pulse of $J_{l,r}$ are not perpendicular on the Bloch sphere. To be exact, the angle between the rotation axis by the pure exchange interactions $J_{12}$ and $J_{23}$ is $120^{\circ}$; a symmetric pulse $J_{12}=J_{23}$ provides rotations around the $z$-axis while the rotation around the $x$-axis is given by a three step pulse due to the exchange interaction always being positive. This can be visualized using the classic Euler angle construction as rotations around three axis can simulate a rotation around any axis. Hence, in total four exchange pulses (for $J_{13}\neq0$ three pulses) are always sufficient to create arbitrary single qubit operations\cite{Nature2000}.
In experiments there are two ways of controlling the exchange interaction which differentiate in the choice of the operating gates.

\begin{figure}
\begin{center}
\includegraphics[width=1.\columnwidth]{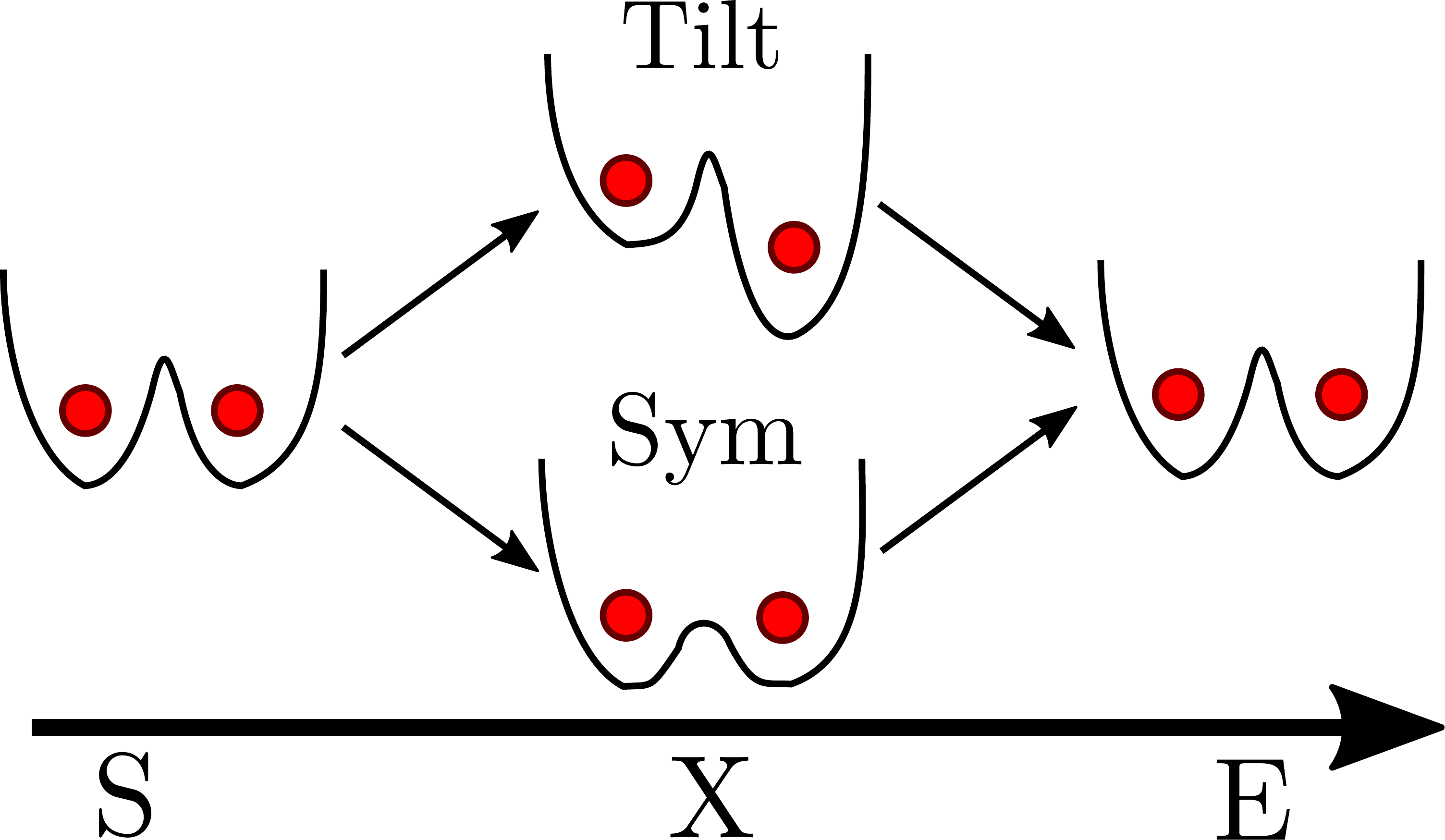}
\caption{Schematic illustration of the two methods for operating the qubit described in the main text (here a singlet-triplet qubit in a double quantum dot (DQD)). The black lines represent the energy potential of the DQD which is filled with two electrons (red dots). The qubit is in some initial state at time S. Then, a single-qubit operation, tilt or symmetric, is performed at time X, leaving the qubit in some final state at time E. For the tilt method, the detuning, the energy potential difference between the quantum dots, is changed to operate the single-qubit gate and for the symmetric method the tunnel-barrier, the height of the energy potential separating the quantum dots, is lowered for the gate operation. Figure is inspired by Ref.~\cite{Reed2016}.}
\label{fig:SOP}
\end{center}
\end{figure}

\paragraph{Tilting based operations}
\label{par:EO:det}

The usual way to control the exchange interaction in a TQD is by varying the gate potentials underneath each QD adapted from DQDs\cite{Petta2005,Hanson2007} and successfully demonstrated for TQDs\cite{Gaudreau2012,Aers2012,Medford2013b,Eng2015,Poulin2015}. The exchange interactions $J_{l,r}(\varepsilon,\varepsilon_{M})$ are controlled by adjusting the gate potentials, maneuvering through the detuning space spanned by the two detuning parameters $\varepsilon$ and $\varepsilon_{M}$. An exchange-pulse, thus, requires the movement of the point of operation to the correct spot at which $J_{l}$ and $J_{r}$ take the desired values for the single qubit operation. Visualized in parameter space, this corresponds to maneuvering to a region where either $J_{l}$ or $J_{r}$ dominates the exchange interaction. In a ST qubit this corresponds to a tilting of the QD potential (see schematically in Fig.~\ref{fig:SOP}) while for the EO qubit both detuning parameters play a role. Precisely, a pure $J_{l}$-pulse requires $|\varepsilon-\varepsilon_{M}|\gg|\varepsilon+\varepsilon_{M}|$ and a pure $J_{r}$-pulse requires $|\varepsilon-\varepsilon_{M}|\ll|\varepsilon+\varepsilon_{M}|$, while the requirement for the (1,1,1) charge configuration regime $|\varepsilon\pm\varepsilon_{M}|<U$ must still hold. Since the detuning parameters have to be operated adiabatically and are located far away in detuning space, this limits the speed of arbitrary qubit rotations since they require a sequence of $J_{l}$ and $J_{r}$ pulses. To speed up gate operations, optimized pulse sequences can be used. Universal control is demonstrated experimentally in the two most common materials, Si\cite{Eng2015} and GaAs\cite{Gaudreau2009,Laird2010,Gaudreau2012,Amaha2012,Aers2012,Medford2013,Medford2013b,Sanchez2014}, yielding control over two independent rotation axes with both exchange couplings exceeding $J\approx\unit[100]{MHz}\approx\unit[40]{\mu eV}$\cite{Eng2015}. Strong dephasing from hyperfine interactions in GaAs devices\cite{Gaudreau2012,Medford2013b} and charge noise in Si devices\cite{Eng2015}, however, limits the fidelity of the qubit rotations. A significant improvement is to be expected by operating the qubit at charge noise sweet spots\cite{Taylor2013,Fei2015,Russ2015}, using dynamical decoupling sequences\cite{Kawano2006,Lee2008,Lidar2012,Hickman2013}, and using devices with nuclear-spin-free isotops\cite{Eng2015}.

\paragraph{Symmetric operations}
\label{par:EO:sop}

Another concept for improved single qubit rotations is the symmetric operation point (SOP)\cite{Reed2016}. Looking back to the expression for the exchange couplings in Eq.~\eqref{eq:exchangeEO} one finds that controlling the tunneling amplitudes $t_{l,(r)}$ also leads to control over the exchange couplings due to $J_{l,(r)}\propto t_{l,(r)}^{2}$. To be exact, this way to control exchange was already proposed in the original paper by Loss and DiVincenzo\cite{Loss1998}. In a ST qubit this corresponds to a lowering of the interdot potentials (see schematically in Fig.~\ref{fig:SOP}) while for the EO qubit both interdot potentials have to be lowered accordingly. Since recent architectures for quantum dot devices\cite{Hanson2007,Zwanenburg2013} always include an additional (static) gate to set the tunnel coupling between the dots the symmetric operation point does not require new quantum dot architectures\cite{Reed2016,Martins2016}. However, SOP allows for heavy filtering of the detuning gates together with a symmetric way of control; both points decreasing the effects of the charge noise to the qubit. In this sense, the symmetric point of operations resembles a charge noise sweet spot which we discuss in more detail in section~\ref{ssec:chargeDec}. However, time-dependent control of the tunneling parameters also opens another channel for coupling noise to the qubit, via the tunnel couplings, due to the absence of heavy filtering\cite{Russ2016}. Experiments in DQDs demonstrate a significant improvement of the qubit properties and dephasing times compared of the standard implementation using detuning as control\cite{Reed2016,Martins2016}, therefore, indicating that noise coupled to the qubit via detuning dominates over noise coupled via tunneling. Up to date there is no experimental demonstration of symmetric operation of three-spin qubits, however, experiments successfully demonstrated control over various QDs\cite{Takakura2014}.

\paragraph{Other methods}
\label{par:EO:other}


Very recently another type for entanglement of multiple spin-$\frac{1}{2}$ qubits was proposed which uses magnetic field gradients in order to implement a phase gate instead of a CNOT gate between two spin-$\frac{1}{2}$ qubits in different quantum dots. The experiment demonstrates a successful entanglement of three spins in a TQD\cite{Noiri2016}. This can hypothetically also be adapted for single-qubit rotations of the three-spin qubit which are independent of the exchange interaction, thus orthogonal. However, one has to be careful not to leak out of the qubit subspace\cite{Hung2014,Poulin2015}.

Up to this point we have only considered a linearly aligned TQD which is used in most experimental setups. In the following, we briefly introduce triangularly arranged TQD systems (TQD molecules) where we mainly focus on the implementation of qubit rotations in such a system which differ from the linear case. For more details about the energy structure and properties we refer to the review by Chan-Yu Hsieh (see Ref.~\cite{Hsieh2012}) or the original works\cite{Scarola2004,Scarola2005}. In addition to the exchange interaction $J_{13}$ between the first and the last dot an (equilateral) triangular shape adds another feature, the chirality, to the system. This allows for a new set of qubit states in the same $S=1/2$ and $S_{z}=\pm1/2$ subspace 
\begin{align}
\ket{+}&=\frac{1}{\sqrt{3}}\left(\ket{\uparrow\uparrow\downarrow}+\E^{+\I\frac{2\pi}{3}}\ket{\uparrow\downarrow\uparrow}+\E^{+\I\frac{4\pi}{3}}\ket{\downarrow\uparrow\uparrow}\right),\\
\ket{-}&=\frac{1}{\sqrt{3}}\left(\ket{\uparrow\uparrow\downarrow}+\E^{-\I\frac{2\pi}{3}}\ket{\uparrow\downarrow\uparrow}+\E^{-\I\frac{4\pi}{3}}\ket{\downarrow\uparrow\uparrow}\right),
\label{eq:chiralBasis}
\end{align}
which are the eigenstates of the chirality operator\cite{Scarola2004} with $\chi=\pm1$. A unitary transformation connects them with the conventional eigenstates from Eq.~\eqref{eq:EOplusStates}. The low-energy subspace can also be approximated by a Heisenberg exchange Hamiltonian\cite{Scarola2005}, however, the exchange couplings include additional terms arising from the circular structure and chirality\cite{Anderson1964,Stoehr2006,Kostyrko2011,Urbaniak2013}. Applying an electric field breaks the symmetry of the system and gives rise to terms $\propto\sigma_{y}$ in the qubit space, corresponding to rotations around the $y$-axis on the Bloch sphere\cite{Luczak2012,Urbaniak2013}. Combining the in-plane electric field with spin-orbit effects, very fast Rabi oscillations between the chiral qubit states are proposed with $\tau\st{Rabi}=\unit[0.1-10^{3}]{ps}$ depending on the realization of the device\cite{Trif2008}. Additionally, the ring structure allows for the application of topologically protected quantum computation due to the non-trivial phase an electron acquires when traveling around a circle\cite{Hsieh2012,Hsieh2012b}. Since this is beyond the scope of this review, we end the excursion of triangular shaped TQDs and continue with further qubit implementations of three-spin qubits.

\subsection{Spin-charge qubit}
\label{ssec:SCQ}

The spin-charge qubit is a unique implementation for a three-spin qubit since all three electrons are located in a single quantum dot occupying the three lowest orbitals\cite{Kyriakidis2007}. The qubit states are
\begin{align}
	\ket{1}&\equiv\ket{s}_{01}\ket{\downarrow}_{2},\\
	\ket{0}&\equiv\sqrt{\frac{2}{3}}\ket{t_{-}}_{01}\ket{\uparrow}_{2}-\frac{1}{\sqrt{3}}\ket{t_{0}}_{01}\ket{\downarrow}_{2},	\label{eq:qubitstatesSC}
\end{align}
where each orbital, $0,1,2$, is occupied by a single electron which corresponds to the $S=1/2$ and $S_{z}=-1/2$ subspace (see subsection~\ref{ssec:spinprop}). Orbital relaxation processes can be suppressed by designing the confinement potential in such a way that the $S=1/2$ and $S_{z}=-1/2$ doublet form the ground and the first excited state\cite{Usukura2005,Korkusinski2007,Delgado2007,Li2007,Hsieh2010,Hsieh2012}. In this sense the qubit implementation is very similar to the exchange-only qubit where the quantum dot (position degree of freedom) is interchanged with the orbital (orbital degree of freedom). Therefore, single qubit rotations are not possible anymore through conventional electric control, i.e., control over the exchange interaction through biasing of the gate voltages underneath or in-between the QDs. Instead of controlling the detuning or the barrier between the QDs one can acquire single-qubit rotations by controlling the confinement potential, particularly, the eccentricity of the confinement potential. Going beyond the Hubbard Hamiltonian and considering electrons in a elliptic confinement potential with eccentricities $\omega_{x}$ and $\omega_{y}$ the Hamiltonian in the qubit space can be written as\cite{Kyriakidis2005,Kyriakidis2007}
\begin{align}
H\st{qubit}=b_{x}\sigma_{x}+b_{z}\sigma_{z}+b_{0}\mathbbm{1}_{2}.
\label{eq:SCqubit}
\end{align}
The parameters are $b_{x}=\sqrt{3}(V_{0220}-V_{1221})/2$, $b_{z}=-V_{0110}+(V_{1221}+V_{0221})$, and $b_{0}=V_{0101}+V_{1212}+V_{0202}$ with the usual matrix elements $V_{o_{1}o_{2}o_{3}o_{4}}=\bra{o_{1},o_{2}}\mathcal{V}\st{Coulomb}\ket{o_{3},o_{4}}$ and $o_{i}\in\lbrace 0,1,2\rbrace$ originating from the long-range Coulomb interaction. A direct comparison of Eq.~\eqref{eq:SCqubit} and Eq.~\eqref{eq:EOqubit} show that the matrix elements of the form $V_{o_{1}o_{2}o_{2}o_{1}}$ resemble an orbital exchange interaction, thus, $V_{0110}\sim J_{l}$, $V_{1221}\sim J_{r}$, and $V_{0220}\sim J_{13}$ (omitted in Eq.~\eqref{eq:EOqubit}). While the explicit expressions can be found in Ref.~\cite{Kyriakidis2005}, the main result is a different dependence of $b_{x}$ and $b_{z}$ with respect to the eccentricity ratio $r\equiv\omega_{x}/\omega_{y}$ which both can be electrically adjusted by the gates. This allows for fast and all-electric single qubit gates with sub-nanosecond gate times ($\tau_{g}\approx\unit[1-10]{ps}$) in GaAs and faster in silicon due to stronger confinement\cite{Kyriakidis2007}. 

The next requirement for quantum computation are feasible read-out and initialization schemes. In the case that the qubit states (see Eq.~\eqref{eq:SCqubit}) are the ground states, the initialization is trivial and just a matter of thermalization\cite{Kyriakidis2007}. In the general case initialization techniques may be adapted from the EO qubit or the ST qubit; a singlet state is initialized in an isolated QD and in a second step the adiabatic opening of the tunnel barriers allows for the tunneling of a third electron. For read-out, a destructive measurements is suggested that detects if a fourth electron is resonantly tunneling in the QD or not, following the same protocol as used for a single-spin qubit\cite{Kyriakidis2007,Elzerman2004}. Since the qubit states are not degenerate, read-out techniques using cavity quantum electrodynamics (cQED) should be adaptable\cite{Beaudoin2016a,Royer2016,Beaudoin2016b,Burkard2016}.

\begin{figure}
\begin{center}
\includegraphics[width=1.\columnwidth]{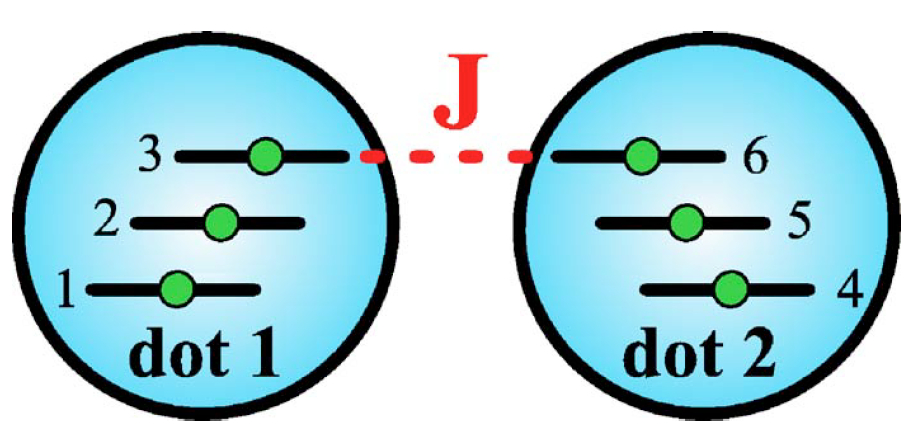}
\caption{Schematic illustration of the two-qubit coupling of two spin-charge qubits. The inter-dot exchange interaction $J$ can be controlled by the tunnel barriers. The shortest full universal pulse sequence consists of nine pulses. Figure taken from Ref.~\cite{Kyriakidis2007}.}
\label{fig:spincharge}
\end{center}
\end{figure}

Together with arbitrary single qubit gates a universal two-qubit gate is needed for universal quantum computation. In a minimal coupling approach only the highest orbitals of two neighboring spin-charge qubits are coupled via the next neighbor exchange interaction (see schematic illustration in Fig.~\ref{fig:spincharge})\cite{Kyriakidis2007}. Note that the resulting two-qubit coupling is identical to the EO qubit (see Fig.~\ref{fig:spincharge}) except the always-on intra-dot exchange interaction, thus, similar but not identical pulse sequences can be used. The shortest universal pulse sequence consists of a minimum of 9 pulses\cite{Kyriakidis2007}.

\subsection{Hybrid qubit}
\label{ssec:HQ}

The holy grail for quantum computation is claimed by the qubit implementation which allows most high-fidelity operations during its coherence time. There are basically two ways of winning the race, either the coherence time is increased or the gate time is decreased, i.e., making the qubit operations faster. The hybrid qubit (HQ) is a representative of  the latter approach, which in short, combines the longevity of spin qubits and the fast qubit operations of a charge qubit\cite{Shi2012}. Note, that this subsection is far from complete and only covers the core concepts and recent advances providing a first insight of the hybrid qubit and that the hybrid qubit deserves a review article on its own.

The HQ qubit is implemented in a double quantum dot (DQD) analogously to the ST-qubit, however, filled with three electrons. As usual for three-spin qubits, the qubit states are 
\begin{align}
	\ket{0}&\equiv\ket{s}_{L}\ket{\downarrow}_{R},\\
	\ket{1}&\equiv-\sqrt{\frac{2}{3}}\ket{t_{-}}_{L}\ket{\uparrow}_{R}+\frac{1}{\sqrt{3}}\ket{t_{0}}_{L}\ket{\downarrow}_{R},	\label{eq:qubitstatesHy}
\end{align}
where the left QD is doubly occupied while the right QD only singly which corresponds to the $S=1/2$ and $S_{z}=-1/2$ subspace (see subsection~\ref{ssec:spinprop}). While the lowest orbital allows the singlet state $\ket{s}_{L}$ the triplet states $\ket{t_{-}}_{L}$ and $\ket{t_{0}}_{L}$ are forbidden in the lowest orbital due to the Pauli exclusion principle, hence, occupy the first excited orbital\cite{Shi2012}. Assuming that the described singlet and triplet states are lowest in energy\cite{Borselli2011,Liu2012}, higher excited singlets and triplets can be neglected due to fast spin conserving orbital relaxation processes\cite{Tahan2014} which immediately relaxes the higher state into the ground state. The essential difference between the HQ and the EO qubit are the use of a DQD instead of a TQD and the the double occupation of the left QD which includes occupation of higher orbital states (only in the left QD). This leads to a total of three relevant states, the two qubit states, $\ket{0}$ and $\ket{1}$, and a virtually occupied state $\ket{v}=\ket{\downarrow}_{L}\ket{s}_{R}$ while other states are negligible. Analogously to the EO qubit, the low-energy subspace Hamiltonian is approximated by a Schrieffer-Wolff transformation (the formal derivation and the expressions can be found in the supplementary material of Ref.~\cite{Shi2012}). A more precise calculation using the projector method and taking additional states into consideration yields qualitatively the same results\cite{Ferraro2014}.

Arbitrary single qubit rotations consist of two independent axis of control, one axis is provided by changing the energy splitting between the qubit states $\hbar\omega$ while the other axis drives transitions between the qubit state, hence, 
\begin{align}
H\st{qubit}=\hbar\omega_{z}\sigma_{z}+j\sigma_{x}.
\label{eq:Hybridqubit}
\end{align}
For the hybrid qubit the energy gap between the qubit state is dominated by the orbital singlet-triplet splitting $E_{ST}$ in the doubly occupied QD, thus, $\hbar\omega=J+E\st{ST}\approx E\st{ST}$. In particular, $J\propto J_{S}+3J_{T}$, hence, also depends on the singlet exchange coupling $J_{S}$ between $\ket{0}$ and $\ket{v}$ and the triplet exchange coupling $J_{T}$ between $\ket{1}$ and $\ket{v}$, however, $|J|\ll|E\st{ST}|$. Since the singlet-triplet $E\st{ST}$ splitting can be controlled by changing the gate voltages in the QD\cite{Amasha2008,Shi2011,Boykin2004,Saraiva2009}, this provides a controllable qubit energy splitting\cite{Shi2012}. Transitions between the qubit states are induced by the off-diagonal terms of the qubit Hamiltonian which are the sum of the exchange couplings, $j\propto J_{S}+J_{T}$ with proportionalities $J_{S,(T)}\propto t^{2}_{S,(T)}/\Delta E_{S,(T)}$. Here, $t_{S,(T)}$ is the corresponding tunneling amplitude between and $\Delta E_{S,(T)}$ is the corresponding energy difference between the virtual state and the singlet (triplet) state. Therefore, either a modulation of the tunnelings $t_{S,(T)}$ or the modulation of the energy differences $\Delta E_{S,(T)}$ give rise to transitions between the qubit states. Considering Si/SiGe as the QD host material, sub-nanosecond ($f=\unit[10]{GHz}$) gate times have been predicted\cite{Shi2012} and experimentally demonstrated\cite{Kim2014}. Moreover, since both tunneling couplings $t_{S}$ and $t_{T}$ can be tuned independently (also independent of $E\st{ST}$) as well as the ration $r=t_{S}/t_{T}$ between them, a larger set of elementary single qubit rotations becomes accessible. This provides a more ``fine-grained'' control of the qubit which reduces the number of the pulses needed for two qubit gates\cite{Shi2012,Koh2012}. Experiments demonstrate $\pi$-rotations around two orthogonal axis with rotation times $t_{\pi}\approx\unit[100]{ps}$ and 86\% (transition between states) and 94\% (control over qubit splitting) gate fidelity\cite{Kim2014} which is further improved if resonantly modulated yielding 93\% and 96\% gate fidelity in experiments\cite{Kim2015b}. This allows for more than 100 coherent exchange oscillations within the dephasing time $T_{2}^{\star}$ in Si/SiGe quantum dot devices\cite{Shi2014}. Numerical results predict further improvement of the coherence time using a quantum point contact\cite{Lei2014}. A recent demonstration of a modified version of the hybrid qubit in GaAs which operates at the (2,3)-(1,4) charge transition yields over 10 coherent Rabi oscillations during the coherence time\cite{Cao2016}.

An initialization and read-out scheme requires the coupling of the doubly occupied QD to the lead with a significant difference in the tunneling rates between the qubit states. A large difference in the tunneling rates allows for a time-resolved measurement which yields information about the qubit state to be initialized or read-out. The crucial requirement, significantly  different tunneling rates, are experimentally demonstrated in GaAs\cite{Hanson2005} and Si/SiGe\cite{Simmons2011,Shi2012} devices.

\begin{figure}
\begin{center}
\includegraphics[width=1.\columnwidth]{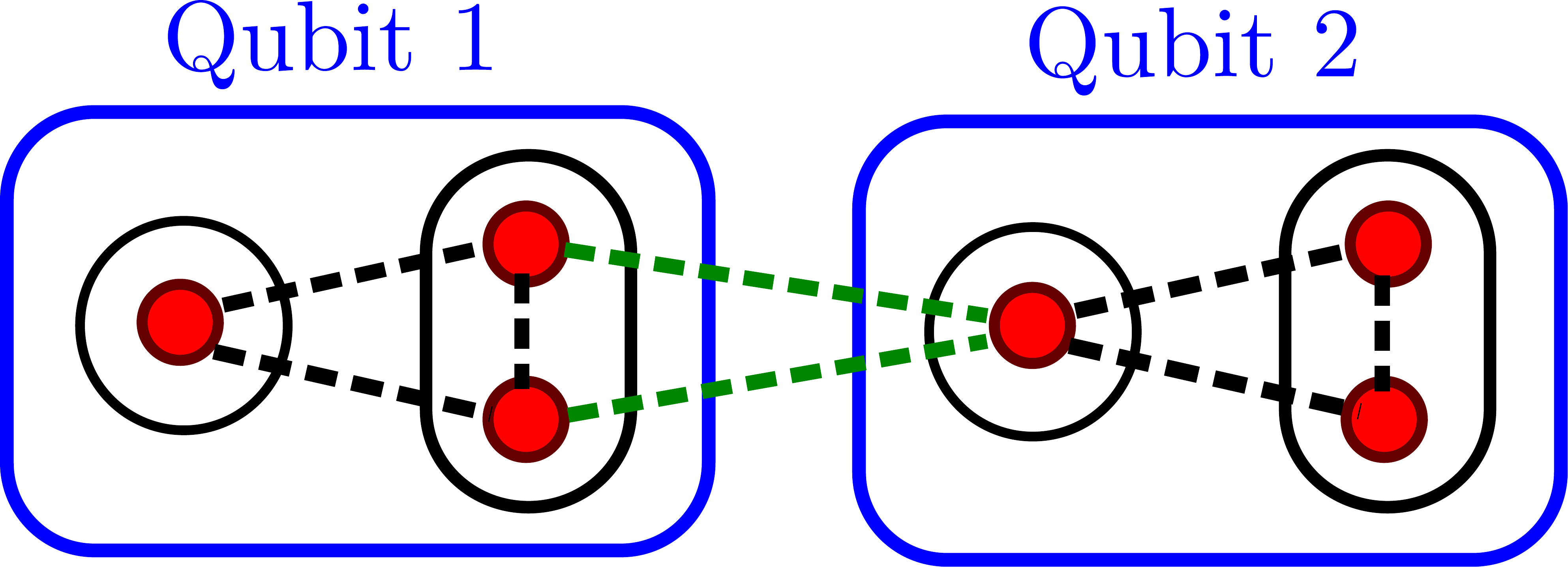}
\caption{Schematic illustration of the two-qubit coupling of two hybrid qubits (qubit 1 and qubit 2). The black circles represent the quantum dots of each qubit (blue box) which each consists of three electrons (red dots), the black dashed lines correspond to the intra-qubit exchange interactions, and the green dashed lines to the inter-qubit exchange interaction. The shortest full universal pulse sequence consists of 14 exchange pulses. Figure inspired from Ref.~\cite{Shi2012}.}
\label{fig:hybrid}
\end{center}
\end{figure}

The remaining issue is the implementation of two qubit gates which can either be performed by pulse sequences of the inter-qubit exchange interaction\cite{Shi2012,Ferraro2015} considering neighboring hybrid qubits or by capacitively coupling of the qubits\cite{Koh2012,Mehl2015,Serina2016}. Exchange based two qubit gates require complex pulse sequences identical to the EO qubit due to a difference in the operational and computational subspace (a detailed discussion can be found in subsection~\ref{ssec:shortTwo} in the next section). However, since the hybrid qubit has additional qubit control (schematically illustrated in Fig.~\ref{fig:hybrid}) shorter pulse sequences are possible consisting of only 14 exchange pulses\cite{Shi2012,Setiawan2014} summing up to an overall gate time on the order of nanoseconds. For the capacitative coupling the relevant interaction is the dipole-dipole coupling originating from the charge difference of the qubit states proposing a fast and feasible two qubit pulsed gate\cite{Koh2012} while a more realistic analysis hints possible problems due to charge noise\cite{Mehl2015}.

\subsection{Resonant exchange (RX) qubit}
\label{ssec:RXQ}

Going the opposite way as the hybrid qubit in the race for the holy grail of quantum computation, i.e., increasing the coherence time $T_{2}$ with still fast qubit gates brings us to the resonant exchange (RX) qubit. The RX qubit\cite{Medford2013,Taylor2013,Doherty2013} is a modified version of the EO qubit where the exchange interaction is always turned on while the qubit is operated (as the name suggests) through resonant driving of the qubit energy gap. As a first thought this may sound like a step backwards to the original spin-$\frac{1}{2}$ qubit which depends on the (slow) qubit rotation through ESR (electron spin resonance)\cite{Hanson2007} or EDSR (electric dipole spin resonance)\cite{Rashba2008}, however, due to the permanently turned on exchange interaction which induces a strong qubit splitting, the qubit operations are much faster, on the order of nanoseconds\cite{Medford2013}.

Analogously to the EO qubit, the qubit states are given by Eq.~\eqref{eq:EOplusStates}, and therefore, still located inside the (1,1,1) charge configuration regime. However, due to $\varepsilon_{M}\gg|\varepsilon|$ the qubit state are strongly hybridized by the admixture of the (2,0,1) and (1,0,2) charge configurations resulting in a large energy gap between the qubit states while the influence of the (1,2,0) and (0,2,1) charge configurations is negligible. Inside the ($\varepsilon,\varepsilon_{M}$)-landscape of the ground state energy gap the RX regime is located in the upper part of the diamond formed (1,1,1) charge regime (white triangle in Fig.~\ref{fig:energygap}). The RX qubit Hamiltonian in its eigenbasis with a modulated detuning $\varepsilon\rightarrow\varepsilon+\delta\varepsilon$ takes the form\cite{Taylor2013}
\begin{align}
H\st{qubit}=\frac{\hbar}{2}\omega\st{RX}\sigma_{z}+\delta\varepsilon\,\eta\sigma_{x}
\label{eq:RXqubit}
\end{align}
with the resonance frequency $\hbar\omega\st{RX}=\sqrt{J^{2}+3j^{2}}$, the modulation coupling $\eta = \sqrt{(\partial J/\partial\varepsilon)^{2}+3(\partial j/\partial\varepsilon)^{2}}$, and the exchange couplings $J=(J_{l}+J_{r})/2$ and $j=(J_{l}-J_{r})$. Due to the negligible influence of the (1,2,0) and (0,2,1) charge configurations the exchange coupling are approximated by $J_{l,r}=t_{l,r}^{2}/(U-\varepsilon_{M}\pm\varepsilon)$\cite{Taylor2013,Russ2015}.

Rabi oscillation corresponding to qubit rotations become accessible through resonant driving of the detuning $\varepsilon$ near the qubit's resonance frequency $\omega\st{RX}$, thus, $\delta\varepsilon=f(\tau)\cos(\nu\tau+\phi)$ with an adjustable phase $\phi$, while the modulation amplitude $f(\tau)$ varies slow (compared to $\omega\st{RX}$) in time $\tau$. Nearby resonance, one finds $\delta\ll \omega\st{RX}$ with $\delta=\nu-\omega\st{RX}$, the Rabi frequency is given within the rotating frame approximation by
\begin{align}
\Omega(\tau)\approx \frac{f(\tau)}{\hbar}\frac{\sqrt{3}\,t^{2}}{(U-\varepsilon_{M})^{2}},
\end{align}
while the axis of rotation is set by the adjustable phase $\phi$ of the driving. Experiments in a GaAs TQD device demonstrate $\pi$ rotations of the qubit around two axis of control on nanosecond time scales, $t\st{gate}=\unit[2.5]{ns}$\cite{Medford2013}. Combined with a coherence time $T_{2}\approx\unit[10]{\mu s}$, this allows for more than $10^{3}$ coherent gates\cite{Medford2013}. In this experiment the modulation amplitude $f(\tau)$ is given by the Overhauser field gradients which is also the limiting noise source for the coherence time. Therefore, an experimental realization in a silicon device (Si/SiGe or SiMos) significantly improves the RX qubit due better control (replacing Overhauser fields with a gradient from a micromagnet\cite{Obata2010}) combined with a longer decoherence time from isotopic purification\cite{Gordon1958,Chiba1972,Tezuka2010,Zwanenburg2013,Itoh2014}. 

The initialization techniques, read-out schemes, and physical implementation are identical to the conventional EO qubit. As a remark, both initialization and read-out schmes using either spin-to-charge conversion or cQED based techniques should be feasible due to the short distance in ($\varepsilon,\varepsilon_{M}$) parameter space with respect to the (2,0,1) and (1,0,2) charge configurations which strongly hybridize the qubit states\cite{Taylor2013,Russ2015b,Srinivasa2016}.

\subsection{Always-on exchange-only (AEON) qubit}
\label{ssec:AEON}

A recent addition to the list of three-spin qubits is the AEON qubit which has a favorable noise robustness combined with symmetric electrical gate control for the single qubit rotations. The always-on, exchange-only (AEON) qubit\cite{Shim2016} is also a modified version of the original EO qubit\cite{Nature2000} where the exchange interaction is either completely turned on or completely turned off, thus, implemented in a TQD filled with three electrons. Since we already introduced the pertinent model earlier we focus here on the difference in operating the single-qubit gates. We again postpone the noise properties and the realization of two qubit gates to the next sections which allows us for a direct comparison of the AEON qubit with the EO qubit and the RX qubit. 

Analogously to the EO and RX qubit, the qubit states are given by Eq.~\eqref{eq:EOplusStates}
Due the hybridization of the qubit states with states with the same quantum numbers, $S=1/2$ and $S_{z}=+1/2$, and different charge configurations, the effective low-energy qubit Hamiltonian takes the familiar form Eq.~\eqref{eq:EOqubit}.
Full control over the qubit is possible through the two exchange interactions $J=(J_{l}+J_{r})/2$ and $j=(J_{l}-J_{r})$ consisting of the left (right) exchange coupling $J_{l,(r)}$ with the approximated expression $J_{l,r}=2t_{l,r}^{2}U/\left[U^{2}-(\varepsilon\mp\varepsilon_{M})^{2}\right]$ (for the general expression see Ref.~\cite{Shim2016}). So far, there is no difference to the EO qubit except the detailed expressions for the exchange couplings. However, the specific expressions for the exchange coupling in the AEON qubit allow for the existence of a double sweet spot (DSS) which is insensitive to noise in lowest order (and additional small second order components). The DSS for the AEON qubit is located directly in the center in the energy landscape of the ground-state energy gap $\omega=\sqrt{J^{2}+3j^{2}}/2$ (see Fig.~\ref{fig:energygap}), thus, possessing the highest symmetry with respect to all (directly tunnel coupled) asymmetric charge configurations. Since the location of the DSS is provided by the geometry of the TQD, thus, independent of the tunneling parameters, it still exist even for less symmetric geometries albeit not located in the center\cite{Shim2016}. This allows for operating the qubit by tuning the tunneling parameters (symmetric operations) while staying permanently on the DSS. Setting the tunneling parameters to be symmetric $t_{l}=t_{r}$ (turning on both exchange coupling simultaneously) results in a rotation around the $z$-axis, while setting $t_{l}=(\sqrt{6}+\sqrt{2})t_{r}/2$ results in a rotation around the $\boldsymbol{n}=-(\boldsymbol{x}+\boldsymbol{z})/\sqrt{2}$-axis which together with a rotation around the $z$-axis causes a rotation around the orthogonal $x$-axis\cite{Shim2016}. Therefore, a three-pulse sequence is sufficient for arbitrary single qubit gates which is one pulse less than needed for the conventional EO qubit\cite{Nature2000}. Since the exchange couplings are either completely turned on or completely turned off, symmetric gate operations (see paragraph \ref{par:EO:sop}), which control the tunnel barriers $t_{l,r}$ directly, are the requirement. Simultaneously, this makes the AEON qubit robust against leakage induced by a magnetic field gradient\cite{Hung2014}, albeit to a lesser degree as the RX qubit due to smaller exchange couplings.

The initialization techniques, read-out schemes, and physical implementation are identical to the conventional EO qubit or the RX qubit. We note however that initialization and read-out using spin-to-charge conversion is not favored for this implementation since one needs to traverse the RX regime in parameter space\cite{Shim2016}.

\section{Two-qubit gates for three-spin qubits}
\label{sec:two}
\subsection{Using short-ranged exchange}
\label{ssec:shortTwo}

After the experimental demonstrations of arbitrary single qubit rotations\cite{Gaudreau2012,Eng2015} the remaining challenge  is the demonstration of a universal two qubit operations in order to achieve universal quantum computation according to the DiVincenzo criteria\cite{DiVincenzo1995} which should be achievable since the bulk of two qubit gates are universal\cite{Lloyd1995}. However, for the case of exchange coupled three-spin qubits the story is more complex. The main problem arises from fact, that the computational two-qubit space, $\mathcal{H}\st{2qubit}=\mathcal{H}\st{+1/2}\oplus\mathcal{H}\st{+1/2}$, represents only a subspace of the sector with spin quantum numbers $S=1$ and $S_{z}=+1$. The inter-qubit exchange coupling leads to excursions outside the computational space during the pulse sequences, and thus, the possibility of leakage into the non-computational space. There are two distinct approaches to counteract the leakage which we discuss in detail; in the first approach, complex pulse sequences are applied in order to make sure that the mapping between the non-computational space and the qubit subspace at the end of the sequence vanishes. In the second approach a (large) energy difference between the computational space and the non-computational space in combination with fast gates (approximately) prevents leakage into the non-computational subspace.

\paragraph{Exact gate sequences}

There are many different pulse sequences for implementing an exact entangling gate between two three-spin qubits. In order to keep the expressions simple, we consider the time steps $\tau$ of the exchange interaction in units of a full swap gate, $\tau_{\textsc{swap}}=2\hbar/\pi J$, between neighboring QDs in the remainder of this subsection. This justifies a consideration where all exchange couplings are identical, $J_{ij}=J$ since the resulting two-qubit gate $U_{ij}(\tau)=\exp[\I \int_{0}^{\tau}dt^\prime J_{ij}(t^\prime)\,\bo{\sigma_i}\cdot\bo{\sigma_j}/4\hbar]$ is independent of the pulse shape of $J_{ij}(t)$. In the original proposal, a minimal pulse sequence consisting of 19 exchange interactions between the QDs was found numerically yielding a \textsc{cnot}-gate up to local single qubit gates. The sequence can be implemented in 13 time steps since some exchange interactions can be run in parallel\cite{Nature2000}. However, this sequence yields a leakage-free entangling gate only for the subspace qubit while there is still leakage in the case of the subsystem qubit. As a brief reminder, the subspace qubit is encoded in either of the doublets, $S=1/2$ and $S_{z}=+1/2$ or $S=1/2$ and $S_{z}=-1/2$, whereas the subsystem simultaneously uses both doublets $S=1/2$ and $S_{z}=\pm1/2$ for the encoding\cite{Kempe2001}. An exact \textsc{cnot}-gate sequence for the subsystem qubit consists of 22 pulses in 13 time steps\cite{Fong2011} with all time steps being multiples of $\tau\st{swap}/4$, while the shortest pulse sequence for a \textsc{cnot} gate up to local single qubit gates consists of 18 pulses in 11 time steps\cite{Setiawan2014}. It should be noted that all sequences were discovered using a numerical minimization algorithm due to the very large Hilbert space for six spins-$\frac{1}{2}$ (dimension $2^{6}=64$). A full understanding and analytical derivation of the path through the Hilbert space associated with the exact \textsc{cnot} gate has subsequently been found\cite{Zeuch2014,Zeuch2016}.

Taking into consideration other geometries which have more connections (exchange couplings) between the two three-spin qubits, shorter and faster pulse sequences are possible. The shortest sequence consisting of 12 pulses in 9 time steps was found for a butterfly geometry where only the center QDs of each three-spin qubit are connected.

Nevertheless, the mutual feature that all sequences require more than 10 pulses makes the exact two-qubit gate somewhat vulnerable to a noisy exchange interaction, i.e., charge noise in the tunnel parameters and detuning parameters or hyperfine interaction due to nuclear spin. Treating the effects of nuclear noise requires a noise-correction scheme consisting of permutations which decouple the static effects of the noise\cite{Fong2011,Setiawan2014}. In simple words, a single pulse is divided into several pulses such that each electron ``feels'' the same nuclear fields at each given time step, thus, unavoidably increases the pulse sequences\cite{Fong2011,Setiawan2014}. The procedure is comparable to a spin echo where the effect of dephasing is reversed by a spin flip and can also be adapted for (quasi) static charge noise. However, due to the nature of charge noise which also consists of high frequency components the correction scheme is better suited for counteracting the effects of nuclear noise due to its slow dynamics. For charge noise other techniques are usually considered, i.e., operating on a charge noise sweet spot. At this point we postpone a detailed discussion regarding the exchange interaction under the influence of charge noise to Section~\ref{ssec:chargeDec}. 

\paragraph{Approximated gate sequences}

Instead of maneuvering on complex paths through the Hilbert space in order to minimize leakage into the non-computational space, one can use short-cuts, gate sequences consisting of a single exchange pulse\cite{Doherty2013,Wardrop2016,Shim2016}. However, these short-cuts are only feasible if there exists a favorably large energy gap between the the computational and non-computational subspaces. This energy gap is crucial since it reduces the amount of leakage during the operation depending on the size of the energy gap. It should also be noted that the amount of leakage can never reach zero for a finite energy gap. Practically speaking, this energy gap is provided by the energy splitting between the qubit states while it is reduced by the inter-qubit exchange interaction\cite{Doherty2013}, thus, making the RX qubit an ideal candidate for its two-qubit scheme due to the large and always turned-on exchange interaction. Another good candidate is the AEON qubit since there the exchange interaction is always turned on or off but has naturally a smaller qubit splitting than the RX qubit. We want to discuss two concrete methods for implementing two qubit gates, the first consisting of a DC pulse while in the second the exchange interaction is modulated by an RF signal. Both methods provide fast two-qubit gates with suppressed but still finite leakage.  
\begin{table}
\begin{center}
\begin{tabular}{|cc|}
\hline\hline
Two qubit state & Energy + $E\st{Zeeman}$\\
\hline 
$\ket{Q,Q}$, $\ket{Q_{3/2},Q_{-}}$, $\ket{Q_{-},Q_{3/2}}$ & 0\\
$\ket{0,Q}$, $\ket{0_{-},Q_{3/2}}$ & $-J_{z,A}/2$\\
$\ket{Q,0}$, $\ket{Q_{3/2},0_{-}}$ & $-J_{z,B}/2$\\
$\ket{0,0}$ & $-(J_{z,A}+J_{z,B})/2$\\
$\ket{1,Q}$, $\ket{1_{-},Q_{3/2}}$ & $-3J_{z,A}$\\
$\ket{Q,1}$, $\ket{Q_{3/2},1_{-}}$ & $-3J_{z,B}/2$\\
$\ket{1,0}$ & $-(3J_{z,A}+J_{z,B})/2$\\
$\ket{0,1}$ & $-(J_{z,A}+3J_{z,B})/2$\\
$\ket{1,1}$ & $-3(J_{z,A}+J_{z,B})/2$\\
\hline\hline
\end{tabular}
\caption{All 15 states in the $S_{z}=1$ subspace of two three-spin qubits with their respective eigenenergies, where $J_{z,A\,(B)}$ is the exchange splitting between the qubit states $\ket{0}$ and $\ket{1}$ in qubit A (qubit B). For the notation we use $\ket{A,B}=\ket{A}\ket{B}$, where the leakage states are defined as follows; $\ket{Q_{3/2}}=\ket{S=\frac{3}{2},S_{z}=+\frac{3}{2}}=\ket{\uparrow,\uparrow,\uparrow}$, $\ket{Q}=\ket{\frac{3}{2},S_{z}=+\frac{1}{2}}=(\ket{\uparrow,\uparrow,\downarrow}+\ket{\uparrow,\downarrow,\uparrow}+\ket{\downarrow\uparrow,\uparrow})/\sqrt{3}$, and $\ket{0_{-}}$ and $\ket{1_{-}}$ being the qubit states for $S_{z}=-1/2$ (see Eq.~\eqref{eq:EOminusStates}). Note, that all qubit states differ in energy from the leakage states. This table was adapted from Ref.~\cite{Doherty2013}.}
\label{fig:leakageEnergies}
\end{center}
\end{table}

Considering a Heisenberg type Hamiltonian for the interaction between the the electrons in the singly occupied QDs the system is described by $H=H\st{Q1} + H\st{Q2}+H\st{int}$ where $H\st{Q1,Q2}$ are the uncoupled single qubit Hamiltonians introduced in Eq.~\eqref{eq:EOqubit}. Focusing on the relevant subspace $S_{z}=1$ which has dimension $n=15$, there are 11 leakage states\cite{Doherty2013}, however, six states cannot be accessed by the exchange interaction alone since it conserves the total spin\cite{Nature2000}. In Table~\ref{fig:leakageEnergies} the corresponding eigenenergies are displayed. In lowest order in perturbation theory the interaction between qubit A and qubit B can be expressed as\cite{Doherty2013}
\begin{align}
H\st{int}=&\delta J_{0} + \delta J_{z}(\sigma_{z,A}+\sigma_{z,B})/2 +J_{zz}\sigma_{z,A}\sigma_{z,B}\nnb
&+J_{\perp}(\sigma_{x,A}\sigma_{x,B}+\sigma_{y,A}\sigma_{yB}),
\end{align}
where $\sigma_{i,Q}$ is the $i=x,y,z$ Pauli matrix acting on qubit $Q=A,B$.
Each of the coefficients $J_{0}$, $\delta J_{z}$, $J_{zz}$, and $J_{\perp}$ is proportional to the inter-qubit exchange interaction $J_{c}$. The parameters strongly depend on the chosen geometry, e.g., for a linear geometry (inter-qubit coupling between QD~3 of the first qubit and QD~1 of the second qubit) the parameters are as follows, $J_{z}/J_{c}=1/36$, $J_{zz}/J_{c}=1/36$, and $J_{\perp}/J_{c}=-1/24$. Very useful for the implementation of a \textsc{cphase} gate between the qubits is that for large inequality between the qubit splittings $|J_{z,A}-J_{z,B}|\gg J_{c}$ the degeneracy between the $\ket{01}$ and $\ket{10}$ two-qubit states is lifted, thus $J_{\perp}=0$. A \textsc{cphase}-gate can now be implemented in a single pulse for $\int_{0}^{\tau}dt^{\prime}J_{zz}(t^{\prime})=\pi/4$. For $J_{\perp}\neq 0$ single qubit operations are additionally needed to ``echo out'' the effects of the perpendicular interaction term\cite{Doherty2013} which is always possible\cite{Zhang2003}. Realistic values for the exchange interactions using the RX qubit encoding predict gate times $\tau\st{gate}=\unit[21]{ns}$ ($\tau\st{gate}=\unit[63]{ns}$) with a leakage error $L<1\%$ ($L<0.1\%$)\cite{Doherty2013}. Using realistic parameter setting for the AEON qubit the gates times are longer ($\tau\st{gate}>\unit[100]{ns}$)\cite{Shim2016} due to the weaker exchange splittings. A further improvement can be achieved by using different coupling geometries, especially the butterfly geometry (center QD of both qubits are connected) reduces the gate times significantly\cite{Doherty2013,Shim2016}. Even further improvement is obtained using different pulse shapes for the exchange pulse with the best having a sinusoidal shape, $J_{c}=J_{c,0}[1-\cos(2\pi\tau/\tau\st{gate})]$ allowing for single-pulse fidelities exceeding $0.9999\%$ for physically reasonable parameter settings\cite{Wardrop2016}. Leakage is increased by considering a realistic environment consisting of charge noise and Overhauser noise due to nuclear spins. Recent studies show that low-frequency charge noise has the strongest impact on the gate fidelity\cite{Wardrop2016}. 

The second approach uses a RF modulation of the exchange coupling $J_{c}(\tau)=J_{c,0}(\tau)+J_{c,\Delta}(\tau)\cos[(J_{z,A}-J_{c,B})\tau]$ between the qubits. Under a rotating wave approximation ($J_{c,0},J_{c,\Delta}\ll(J_{z,A}-J_{c,B})$) the two qubit interaction is given by\cite{Doherty2013}
\begin{align}
H\st{int}=\frac{J_{c,0}}{6}\sigma_{z,A}\sigma_{z,B}+\frac{J_{c,\Delta}}{24}(\sigma_{x,A}\sigma_{x,B}+\sigma_{y,A}\sigma_{yB}),
\end{align}
where we used the same expressions as in the paragraph above. The advantage of this approach is that both control parameters $J_{c,0}$ and $J_{c,\Delta}$ can be set individually allowing for more flexibility of controlling the two qubit gate. The only required condition is $|J_{c,\Delta}|<J_{c,0}$ due to the positive sign for the exchange interaction\cite{Doherty2013}.

\subsection{Long-ranged two-qubit gates}
\label{ssec:longTwo}

At the time of writing of this review the best available option for error correction techniques appears to be the surface codes which require a two-dimensional geometry of qubits\cite{Fowler2012,Lidar2013}. In realistic devices this is a challenge since each qubit must be accessed by multiple (gate) electrodes limiting the possibility to connect one qubit with more than two other qubits through exchange. This makes it more realistic to use a linear geometry. Since the exchange interaction is limited to adjacent QDs, other long-range interactions have to be considered to overcome this technical difficulty allowing for a two-dimensional array of qubits which are spatially separated\cite{Brecht2016}. There are several proposals for the achievement of such an interaction, e.g., tunneling mediated by a superconductor\cite{Leijnse2013,Hassler2015}, coupling though surface acoustic waves\cite{Stotz2005,Hermelin2011,McNeil2011,Schuetz2015,Benito2016,Bertrand2016}, ferromagnets\cite{Trifunovic2013}, superexchange mediated by an additional QD\cite{Braakman2013,Sanchez2014b,Mehl2014,Srinivasa2015b}, spatial adiabatic passage\cite{Kuo2014,Ferraro2015,Menchon2016}, photon assisted tunneling\cite{Braakman2014,Gallego2015,Stano2015}, and quantum Hall edge states\cite{Yang2016,Benito2016}. The most practical ideas (up to date) seem to be Coulomb-based dipole-dipole coupling\cite{Taylor2013,Pal2014,Pal2015} and cavity quantum electrodynamics (cQED) mediated coupling\cite{Amaha2012,Russ2015b,Beaudoin2016a,Beaudoin2016b,Kontos2016,Srinivasa2016,Puri2016,Schuetz2016,Viennot2016} which both use the electric dipole moment of the qubit, whereas in the second approach the interaction range is elongated by the use of a cavity as a mediator\cite{Childress2004,Burkard2006,Russ2015b,Srinivasa2016,Russ2016}. Three-spin qubits have (in certain parameter regimes) large electric dipole moments\cite{Shi2012,Taylor2013} which, combined with recent advances in superconducting microwave cavities, boost the vacuum coupling strength\cite{Guilherme2014,Samkharadze2016}, making the three-spin qubits a good candidate for the implementation of cQED. There are multiple ways to implement such two-qubit gates which we try to discuss in the following.

\paragraph{Qubit-cavity interaction}

Originally proposed for superconducting qubits\cite{Blais2004} due to their strong dipole coupling strength on the order of $g\approx \unit[200]{MHz}$\cite{Schoelkopf2008,Houck2012}, cQED can also be used for semiconductor spin qubits despite having a coupling strength at least one order of magnitude smaller\cite{Childress2004,Burkard2006}, i.e., $g\approx\unit[0.1-10]{MHz}$\cite{Taylor2013,Russ2015b,Russ2016,Srinivasa2016} for three-spin qubits, due to advances in the coherence times\cite{Zwanenburg2013} and cavity design\cite{Guilherme2014,Samkharadze2016}. It is crucial to achieve a coherent coupling between the qubit and the cavity, therefore, a coupling which is required to be stronger than the relaxation and dephasing mechanism in both the cavity and the qubit\cite{Kontos2016,Viennot2016,Mi2016}.

\begin{figure}
\begin{center}
\includegraphics[width=1.0\columnwidth]{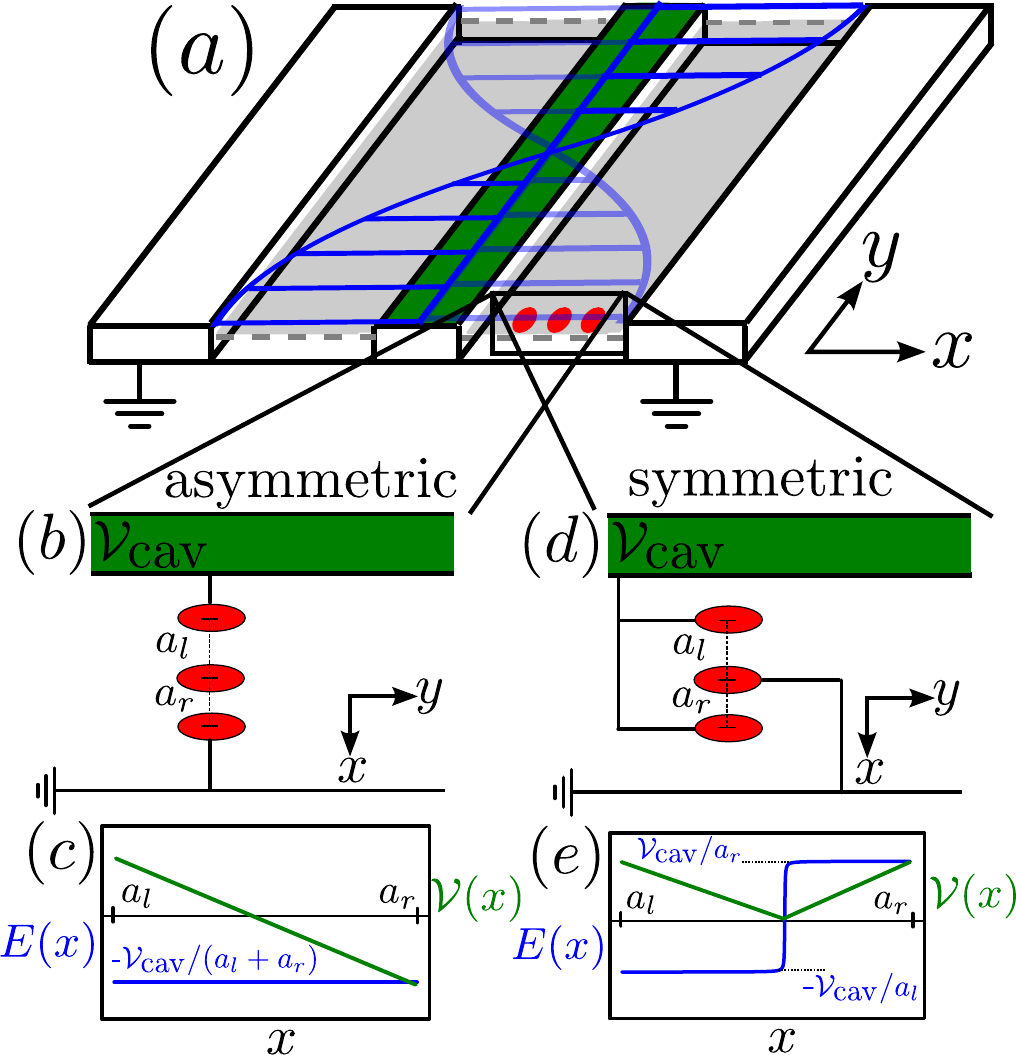}
\caption{$(a)$ Schematic illustration of a qubit implemented in a triple quantum dot coupled to the cavity and the architecture for a $(b)$ asymmetric and $(d)$ symmetric qubit-cavity coupling. The center conductor of the superconducting transmission line resonator is on the potential $\mathcal{V}\st{cav}$ while the outer conductors are connected to the ground to screen off surrounding fields. The corresponding potential (green) and electric field (blue) is shown for the asymmetric $(c)$ and symmetric $(e)$ arrangement as a function of the position $x$.  Figure taken from Ref.~\cite{Russ2016}.}
\label{fig:arch}
\end{center}
\end{figure}

For the purpose of its theoretical investigation, the cavity can be described as a resonator (see Fig.~\ref{fig:arch}~(a)) with only a single mode with frequency $\omega\st{ph}$ that lies nearby the resonant frequency of the qubit splitting $\omega$. Thus, the cavity is described without loss or decoherence effects by a quantum harmonic oscillator with this frequency $H\st{cav}=\hbar\omega\st{ph}(a^{\dagger}a+\frac{1}{2})$\cite{Cohen1989}, where $a^{\dagger}$ ($a$) creates (annihilates) a photon inside the cavity with the very same frequency. The corresponding energy of the cavity is $E\st{cav}=\hbar\omega\st{ph}(n\st{ph}+\frac{1}{2})$ which depends on the average number of photons $n\st{ph}=\braket{a^{\dagger}a}$. Many protocols for two-qubit gates\cite{Blais2004,Childress2004,Burkard2006} require the cavity to be in the ground state, therefore, depending on the resonance frequency to be cooled to very low temperatures, e.g., $T\ll \unit[50]{meV}$ for a $\unit[10]{GHz}$ cavity, while a few protocols also work with thermally populated cavities\cite{Sorensen2000,Schuetz2016}. 

In the approach of cQED the qubit-cavity interaction is described by the minimal coupling approach which replaces the momentum with the generalized momentum $\boldsymbol{p}\rightarrow\boldsymbol{p}-e\boldsymbol{A}$ that includes the electromagnetic vector potential $\boldsymbol{A}$ and the elementary charge $e$\cite{Cohen1989}. In the dipole interaction near the resonance the coupling is 
\begin{align}
H\st{dip}=-e\boldsymbol{E}\cdot \boldsymbol{d}
\end{align}
where $\boldsymbol{E}=\boldsymbol{\mathcal{E}}(a+a^{\dagger})$ denotes the electric field inside the cavity and $\boldsymbol{d}$ is the dipole operator of the qubit. Defining the qubit-cavity coupling strength as the transition amplitude between the qubit states $g\equiv -e\bra{0}\boldsymbol{\mathcal{E}}\cdot \boldsymbol{d}\ket{1}$ allows for a quantitative comparison\cite{Burkard2006}. In order to find the dipole operator $\boldsymbol{d}$ the microscopic wave functions of the three-spin qubit states are necessary which are in general rather difficult to obtain\cite{Hsieh2012}. Fortunately, there are a few approximations that help to overcome this difficulty. 

In a simplified picture, where the the spatial extension of the QD is much smaller than the wavelength of the resonator mode, the qubit-cavity interaction is derived from the oscillation of the electrostatic gate potentials\cite{Childress2004}. Depending on which gate electrode is connected to the cavity, thus, the architecture of the qubit-cavity system (see Figs.~\ref{fig:arch}~(b) and (d)), $\varepsilon$, $\varepsilon_{M}$ or both provide the coupling\cite{Russ2015b}. The corresponding dipole operator is $\boldsymbol{d}=d_{x} \boldsymbol{e}_{x}$ with $\boldsymbol{e}_{x}$ being the unit vector in $x$-direction and $d_{x}=\partial H(q) \nu(a^{\dagger}+a) /\partial q$, where the qubit Hamiltonian $H$ depends on the detuning $q=\alpha \varepsilon+\beta\varepsilon_{M}$ with $\alpha,\beta\in\mathbb{R}$ and $\alpha^{2}+\beta^{2}=1$. The phenomenological parameter $\nu$ describes the overall interaction strength and can be derived from the capacitances in the hybrid (qubit and cavity) system\cite{Childress2004}. For $\beta=0$ only $\varepsilon$ is relevant which leads to $d_{x}=\hbar g\sigma_{x}$ with the coupling strength\cite{Srinivasa2016}
\begin{align}
\frac{g}{g_{0}}=\frac{1}{2}\sqrt{\left(\frac{\partial J}{\partial \varepsilon}\right)^{2}+3\left(\frac{\partial j}{\partial \varepsilon}\right)^{2}}
\label{eq:gSimp}
\end{align}
with the exchange coupling $J=(J_{l}+J_{r})/2$ and $j=(J_{l}-J_{r})/2$ from Eq.~\eqref{eq:exchangeEO} and the vacuum coupling strength $g_{0}$\cite{Srinivasa2016}.

\begin{figure*}
\begin{center}
\includegraphics[width=0.9\textwidth]{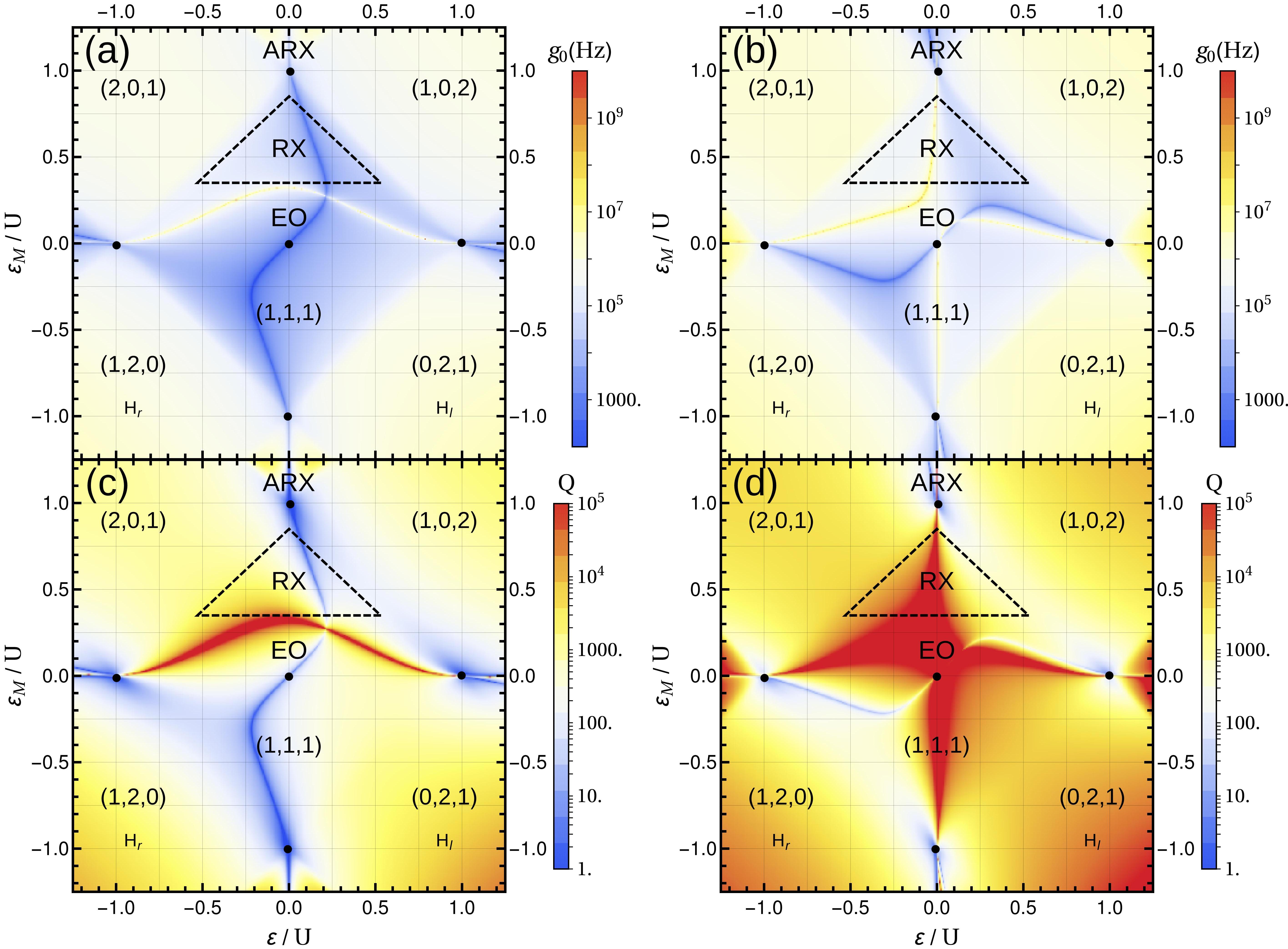}
\caption{(Top row) The minimal vacuum coupling $g_{0}$ needed to reach strong coupling between the qubit and the cavity under the assumption that qubit dephasing is the dominant loss mechanism. (Bottom row) Minimal Q-factor of the cavity needed for successful entanglement between two qubits in the same cavity using dispersive transversal coupling. Panels (a) and (c) show the results for the asymmetric architecture and with only noise in the asymmetric detuning parameter $\varepsilon$ while (b) and (d) show the results for the symmetric architecture and only noise in the symmetric detuning parameter $\varepsilon_{M}$. The parameters are chosen as follows; $\omega_{ph}=\unit[4.7]{GHz}$, $g_{0}=\unit[2\pi\times 10]{MHz}$, $t_{l}=0.022\tun$, $t_{r}=0.015\tun$, and $A_{q}=(10^{-3}\tun)^{2}$ where $q=\varepsilon$ in (a) and $q=\varepsilon_{M}$ in (b). The datasets for $T_{\varphi}$ are obtained from Ref.~\cite{Russ2016}. For the scale of $T_{\varphi}$ and $g$ an explicit value of $U=\unit[1]{meV}$ is used. Figure taken from Ref.~\cite{Russ2016}}\label{fig:plotQfac}
\end{center}
\end{figure*}

In a more realistic picture, the microscopic three-electron real-space wave-functions of the states $\ket{0}$, $\ket{1}$, $\ket{2}$, $\ket{3}$, $\ket{4}$, and $\ket{5}$ from Eqs.~\eqref{eq:basisstatesBeg}-\eqref{eq:basisstates} are constructed from the single-electron real-space wave-functions\cite{Burkard1999} $\ket{\psi_{i}}$ with $i=1,2,3$ needed for the dipole matrix elements\cite{Burkard2006,Russ2015b}. Using the formalism of orthonormalized Wannier orbitals\cite{Russ2015b,Russ2016}, the overlapping wave-functions $\ket{\psi_{i}}$ are transformed into a basis of orthonormalized maximally localized\cite{Marzari2012} wave-functions $\ket{\phi_{j}}$. Requirements for this transformation are a small overlap between the single-electron wave-functions, $|\braket{\psi_{i}|\psi_{j}}|=|S_{ij}|\ll1$ with $i,j=1,2,3$ \cite{Burkard1999,Burkard2006,Russ2015b}. The full expression of the dipole operator in the basis $\left\lbrace \ket{0},\ket{1},\ket{2},\ket{3},\ket{4},\ket{5}\right\rbrace$ can be found in Ref.~\cite{Russ2016} and depends solely on the transition dipole matrix elements $\boldsymbol{x}_{ij}=\bra{\phi_{i}}\boldsymbol{d}\ket{\phi_{j}}$ of the single-electron Wannier orbitals. In the next step the geometry of the qubit-cavity device is needed, since it enters the expression through the dependence of the electric field $\boldsymbol{E}$ from the position (see Figs.~\ref{fig:arch}~(c) and (e)). An analytical expression for the asymmetric case $\boldsymbol{E}=E(a^{\dagger}+a)\boldsymbol{e}_{x}$ with $\boldsymbol{e}_{x}$ being the unit vector in $x$-direction (see Fig.~\ref{fig:arch}~(c)) inside the (1,1,1) charge configuration regime is\cite{Russ2016}
\begin{align}
\begin{split}
	\frac{g_{A}}{g_{0}} =
	& -\sqrt{\frac{3}{2}}\left[ \frac{J_{l}}{t_{l}}\frac{\text{Re}(x_{12})}{2(a_{l}+a_{r})}-\frac{J_{r}}{t_{r}}\frac{\text{Re}(x_{23})}{2(a_{l}+a_{r})}\right]\\
		- \frac{\sqrt{3}}{4}\bigg[& \frac{\left(\varepsilon-\varepsilon_{M}\right)}{U}  \frac{J_{l}^{2}}{t_{l}^{2}}\frac{x_{11}}{a_{l}+a_{r}} +\frac{\left(\varepsilon+\varepsilon_{M}\right)}{U}  \frac{J_{r}^{2}}{t_{r}^{2}} \frac{x_{33}}{a_{l}+a_{r}}\bigg].
		\end{split}
		\label{eq:gSW}
\end{align}
Here, $g_{0}$ is again the vacuum coupling of the cavity, $a_{l}$ ($a_{r}$) is the inter-dot distance between QD~1 and QD~2  (QD~2 and QD~3) while $\text{Re}(\xi)$ denotes the real part of $\xi$. This result is consistent with the results in the simplified picture (see Eq.~\eqref{eq:gSimp}) under the assumptions $\text{Re}(x_{ij})=0$ for $i\neq j$, $x_{11}=-a_{l}$ and $x_{33}=a_{r}$ which corresponds to a vanishing overlap between the single-electron wave-functions. Obviously, this expression consists of two parts where each corresponds to the qubit-cavity coupling of a DQD\cite{Burkard2006}, thus, the combined effect of the coupling of two DQDs. 
For a symmetric architecture where the cavity is connected to the gate electrode of QD~2 (see Fig.~\ref{fig:arch}~(d)) the electric field is position dependent (see Fig.~\ref{fig:arch}~(e)), $\boldsymbol{E}=\frac{1}{\pi}\left\lbrace\tan^{-1}\left[\frac{\boldsymbol{d}\cdot\boldsymbol{e}_{x}}{T\left(a_{l}+a_{r}\right)}\right]+\frac{\pi(a_{l}-a_{r})}{2(a_{l}+a_{r})}\right\rbrace(a^{\dagger}+a)\boldsymbol{e}_{x}$ where $T$ is a dimensionless screening parameter. Approximate analytic expressions exist for large screening $T\gg 1$ 
\begin{align}
\frac{g_{S}}{g_{0}}=&  \frac{a_{l}-a_{r}}{2(a_{l}+a_{r})^{2}}\bra{g}\boldsymbol{d}\cdot\boldsymbol{e}_{x}\ket{e}\nnb
&+\frac{1}{\pi T(a_{l}+a_{r})^{2}}\bra{g}(\boldsymbol{d}\cdot\boldsymbol{e}_{x})^{2}\ket{e},
\label{eq:gsSW}
\end{align} 
where the full expression of $\bra{g}(\boldsymbol{d}\cdot\boldsymbol{e}_{x})^{2}\ket{e}$ is found in Ref.~\cite{Russ2016}. A comparison of the asymmetric and the symmetric coupling strength is seen in the top row of Fig.~\ref{fig:plotQfac} which shows the minimal vacuum coupling needed to reach strong coupling.

Instead of focusing solely on the transition dipole matrix elements typically used for (transversal) two-qubit entanglement protocols\cite{Childress2004,Burkard2006} one can also calculate the longitudinal\cite{Kerman2008,Kerman2013,Didier2015} dipole matrix element $g_{l}=(\bra{0}\boldsymbol{\mathcal{E}}\cdot \boldsymbol{d}\ket{0}-\bra{1}\boldsymbol{\mathcal{E}}\cdot \boldsymbol{d}\ket{1})/2$ used for longitudinal entanglement protocols\cite{Billangeon2015,Richer2016,Ruskov2016,Royer2016}. The crucial difference is that the first induces a transition between the qubit states $\ket{0}\leftrightarrow\ket{1}$ through the absorption/emission of a cavity photon, while the longitudinal dipole matrix element only changes the phase of the qubit state assisted by the cavity photon. Its strength can be estimated by the same procedure as for the transverse coupling.

Under realistic settings, both couplings, the transversal and the longitudinal, are permanently present, however, depending on the exact position in the $(\varepsilon,\varepsilon_{M})$-space their strength changes significantly, therefore, effectively turning off one kind of coupling\cite{Beaudoin2016b} which resembles a sweet spot (where first order effects vanish) for this type of coupling (see subsection~\ref{ssec:chargeDec}). Combining all the above elements, the qubit-cavity Hamiltonian in its eigenbasis, up to a constant shift in energy, is
\begin{align}
H=\hbar\omega\sigma_{z} + \hbar\omega\st{ph} a^{\dagger}a + g_{l}(a^{\dagger}+a)\sigma_{z} + g(a^{\dagger}\sigma_{-}+a\sigma_{+})
\label{eq:JC}
\end{align}
with the ladder operators $\sigma_{\pm}=(\sigma_{x}\pm\I \sigma_{y})/2$. This expression corresponds to the extended Jaynes-Cummings Hamiltonian\cite{Jaynes1963,Cummings1965,Shore1993} and is derived using a rotating wave approximation, $g\ll\omega\st{ph}$, whereby the counter-rotating terms $a^{\dagger}\sigma_{+}$ and $a\sigma_{-}$ are excluded since they oscillate with twice the cavity frequency and therefore average out\cite{Wu2007}.

\paragraph{Concepts for two-qubit gates}

In the conventional scheme for a long distant coupling the qubits are entangled using the photons as a carrier of information\cite{Imamoglu1999}. This concept is generally applicable for all two-level systems and only needs a sufficient strong qubit-cavity coupling outmatching the loss and dephasing effects\cite{Blais2004,Burkard2006,Blais2007}. The starting situation is as follows; two three-spin qubits in the same cavity that both are transversally coupled to the same cavity. Operating in the dispersive regime $g\ll|\Omega_{i}|\equiv|\omega\st{ph}-\omega_{1,2}|$, where $\hbar\omega_{i}$ is the qubit energy splitting of qubit $i=1,2$, the cavity mode can be eliminated by a Schrieffer-Wolff transformation\cite{Blais2004,Burkard2006,Blais2007} yielding the effective Hamiltonian\cite{Russ2015b,Srinivasa2016}
\begin{align}
H=\sum_{i=1,2}\left(\hbar\omega_{i}+\epsilon_{i}\right)\sigma_{z,i}+g\st{eff}(t)(\sigma_{+,1}\sigma_{-,2}+\sigma_{-,1}\sigma_{+,2}).
\end{align}
The ladder operators $\sigma_{\pm,i}$ act purely on qubit $i$ and the coupling strength is given by $g\st{eff}=g_{1}g_{2}(\Omega_{1}+\Omega_{2})/\Omega_{1}\Omega_{2}$ with the detuning $\Omega_{i}$ of QD~$i$. After the time $\tau_{g}=\hbar\pi/2g\st{eff}$ the two-qubit interaction yields the universal i\textsc{swap}-gate\cite{Schuch2003,Burkard2006,Russ2015b}. A sequence of two i\textsc{swap}-gates and two single-qubit rotations additionally form a \textsc{cnot}-gate\cite{Tanamoto2008}. In an earlier approach a \textsc{cnot}-gate gate is generated by the same Hamiltonian by using the two-qubit $\pi/4$-gate instead\cite{Imamoglu1999}. Figs.~\ref{fig:plotQfac}~(c) and (d) show qualitatively the required quality factor $\omega\st{ph}/\Omega_{1,2}$ for a successful entanglement considering dephasing due to charge noise through the respective detuning parameter, i.e., $\varepsilon$ for an asymmetric architecture and $\varepsilon_{M}$ for the symmetric architecture. Using realistic parameter settings the i\textsc{swap}-gate can be performed in $\tau_{g}=\unit[540]{ns}$ with a fidelity of $99\%$ while for faster gates the fidelity decreases\cite{Srinivasa2016}.

Instead of operating in the dispersive regime which is rather slow an alternative scheme that uses resonant driving provides faster two-qubit gates\cite{Srinivasa2016}. This scheme, based on the Cirac-Zoller gate for trapped ions\cite{Cirac1995,Childs2000}, uses sideband transitions\cite{Blais2007,Wallraff2007,Leek2009} that are generated when a external driving field $\nu$ is included in the qubit-cavity system. For resonant driving between the driving field and the qubit transition, $\nu=\omega$, the interaction Hamiltonian in a rotating frame is\cite{Srinivasa2016}
\begin{align}
H=\Delta_{0}a^{\dagger}a+g\left( \E^{\I\phi} a \sigma_{+}+\E^{-\I\phi} a^{\dagger} \sigma_{-} \right)+\Omega\sigma_{y},
\end{align}
where $\phi$ is the phase and $\epsilon$ the amplitude of the driving field, $\Delta_{0}$ is the detuning between the driving field and the cavity, and $\Omega=g\epsilon/\Delta_{0}$ is the Rabi frequency of the qubit. Switching into a second rotating frame of the Rabi frequency and carefully adjusting the detuning $\Delta_{0}=\pm2\Omega$ yields  ``red'' and ``blue'' sideband transition Hamiltonians\cite{Srinivasa2016}
\begin{align}
H_{\pm} =\frac{g}{2}\left(\E^{\mp\phi}a^{\dagger} \sigma_{\mp} +\E^{\pm\phi} a\sigma_{\pm}\right).
\end{align}
An entangling controlled-Z gate is constructed using pulses of ``red'' and ``blue'' sideband transition gates $S_{\pm}(\phi,\tau)=\exp(-\I H_{\pm}(g,\phi)\tau/\hbar)$ combined with single-qubit rotations. One of such a pulse sequence consists of seven pulses providing a controlled-Z gate time $\tau_{g}=\unit[270]{ns}$ with a fidelity of $99.6\%$ for realistic parameter settings\cite{Srinivasa2016}. 

The concept of cQED with longitudinal coupling was originally developed to read out the qubit states via a microwave cavity\cite{Kerman2008,Kerman2013,Didier2015,Beaudoin2016a}, however, can also be used to entangle multiple distant qubits\cite{Billangeon2015,Beaudoin2016a,Richer2016,Ruskov2016,Royer2016}. This concept is generally applicable for two-level systems, does not rely on perturbative arguments, and solely bases on the parametric modulation of the the longitudinal qubit-cavity coupling, therefore, does not produce any residual terms in the Hamiltonian\cite{Royer2016}. The starting situation is as follows; two three-spin qubits (two-level systems) are both longitudinally coupled to the same cavity (see Eq.~\eqref{eq:JC}) while the transversal coupling $g=0$. The longitudinal coupling leads to a small displacement of the oscillator field which can be significantly increased by resonant driving at the cavity frequency. Since the resonant driving leads also to a rapid dephasing, the modulation drive $\omega_{m}$ is to be chosen off-resonant that finally gives in the polaron frame rise to\cite{Royer2016}
\begin{align}
H=\sum_{i=1,2}\hbar\omega_{i}\sigma_{z,i}+\omega\st{ph}a^{\dagger}a+g\st{eff}(t)\sigma_{z,1}\sigma_{z,2},
\end{align}
where $\sigma_{z,i}$ acts on qubit $i$ whose states are split by $\hbar\omega_{i}$. The two-qubit interaction $\sigma_{z,1}\sigma_{z,2}$ yields, after time $\tau_{g}=\theta/4|g\st{eff}|$, the entangling controlled-phase gate $U=\text{diag}\left[1,1,1,\E^{\I\theta}\right]$ sufficient for universal quantum computing\cite{Royer2016}. An approximate expression for the coupling strength\cite{Royer2016} is $g\st{eff}=-g_{l,1}g_{l,2}\omega\st{ph}/(\omega\st{ph}-\omega_{m})(\omega\st{ph}+\omega_{m})$. While resembling similarities with transversal coupling\cite{Blais2007,Chow2013} the key difference is that for longitudinal coupling the results are exact and not only valid in a limited regime. Furthermore, under certain parameter settings the gate starts and ends in the vacuum state of the cavity, therefore, the gate can be performed non-adiabatically which yields, with realistic but optimistic parameter settings, $\tau_{g}=\unit[37]{ns}$ with a fidelity of $99.99\%$. Note, that there is no trade-off between gate time and fidelity for this scheme, thus, allowing for fast gates with arbitrarily small errors\cite{Royer2016}. Increased fidelity is achieved if squeezed photon states instead of coherent states are used\cite{Royer2016}. More specific investigations regarding three-spin qubits allow the operation of such a gate on such time-scales while operated on a charge noise sweet spot\cite{Ruskov2016}. However, all of the above requires a pure longitudinal coupling with no residual transversal coupling.

\section{Decoherence effects in three-spin qubits}
\label{sec:dec}

The main sources of decoherence in three-spin qubits are magnetic noise due to nuclear spin and charge noise originating from random fluctuations in the host material or transmitted via the electric gates\cite{Medford2013} and electron-phonon interaction. Electron-phonon interactions play a less important role in this review due to the choice of host material, typically GaAs and silicon, and design of the device, i.e., lateral quantum dots in a 2DEG\cite{Hanson2007,Zwanenburg2013} versus QDs in wires or nanotubes which usually have stronger spin-orbit interaction\cite{Sapmaz2006,Nadj2010,Laird2015}.

\subsection{Magnetic noise}
\label{ssec:magDec}

Since both qubit states of the three-spin qubits have identical spin quantum numbers $S$ and $S_{z}$ such three-spin qubits possess a natural protection against global magnetic fields\cite{Lidar2013}. Depending on the qubit implementation this degree of protection against global magnetic field fluctuations varies. The decoherence free subsystem (DFS) qubit is completely immune against general collective noise which includes all noise that couples identically to each spin in the system\cite{Zanardi1997,Lidar1998,Bacon2000,Viola2000,Nature2000,Kempe2001,Lidar2013}. The DFS Hamiltonian of a system coupled to a noise bath can then be expressed as $H=H\st{system}+H\st{bath}+H\st{int}$ and the interaction is $H\st{int}=\sum_{\alpha}S_{\alpha}\otimes B_\alpha$ where $S$ solely acts on the system and $B$ solely acts on the bath. The DFS qubit states both lie in the same subspace of such a $S_{\alpha}$, thus, both affected identically by the noise\cite{Lidar2013}. However, only global Overhauser (effective nuclear) fields\cite{Overhauser1953,Abragam1961,Slichter2010,Coish2009}, which would require a perfect polarization of the nuclear spins, are considered by general collective noise while static and fluctuating magnetic field gradients between the QDs are not considered, therefore, still inducing leakage\cite{Hung2014,Poulin2015} and dephasing\cite{Medford2013}. The general theory of DFS is already covered in a related review~\cite{Lidar2013} and we focus in this review on the effects due to Overhauser field gradients. 

\paragraph{Decoherence due to magnetic noise}

The main component of magnetic noise is induced by nuclear spins surrounding the nanostructures and coupling to the trapped electron spins in the QDs. These nuclear spins are present in almost all host materials with only a few having a nuclear spin free isotope, i.e., carbon ($\approx99\%$), silicon ($\approx 95\%$), and germanium ($\approx 91\%$), and unless one uses one of these materials interact with the trapped electrons through the hyperfine interaction\cite{Abragam1961,Slichter2010}. 
Due to their (almost) omnipresent nature the nuclear spins themselves and their effects on QDs are studied and reviewed very carefully in literature, e.g., by Coish and Baugh (see Ref.~\cite{Coish2009}), thus, we dispense with a repetition of the basics and focus on their effect on three-spin qubits. 

Considering only the contact hyperfine interaction of the ground-state orbital of the QD, which requires low temperature and large orbital level spacing\cite{Coish2009,Hung2013}, the interaction between the three-spin qubit and a bath of nuclear spins is effectively described by\cite{Hung2014} $H\st{HI}=\sum_{i=1}^{3}\boldsymbol{S}_{i}\cdot\boldsymbol{B}_{i}$ with $\boldsymbol{B}_{i}=\sum_{k\in i}A_{k,i}\boldsymbol{I}_{k}$. Here, $A_{k,i}$ is the hyperfine interaction constant of a nucleus with spin $\boldsymbol{I}_{k}$ interacting with the electron spin in QD$i$. In typical experiments the nuclear spins are randomly oriented $\braket{\boldsymbol{B_{i}}}=0$ with a finite standard deviation $\sqrt{\braket{\boldsymbol{B_{i}}^{2}}}\approx \mathcal{A}_{i}$ where $\mathcal{A}_{i}$ is the average hyperfine energy\cite{Hung2014} coupled to electron spin $i$. These Overhauser fields $\boldsymbol{B}_{i}$ can be correlated due to a finite overlap between the spin-wave-functions of the electron $i$ since a nucleus in an overlapping region affects both electron spin. In realizations using TQD devices these correlations can be small due to their small overlap. Depending on the material these coupling constants are rather strong $\mathcal{A}\approx\unit[85]{\mu eV}$ (for a full list see Ref.~\cite{Coish2009}) giving rise to effective Overhauser fields $|\boldsymbol{B}_{j}|\approx\unit[5]{T}$ in GaAs devices. These Overhauser fields have two main effects on the three-spin qubit. 

Differences in the Overhauser fields cause leakage into the non-computational space due to the spin non-preserving nature of $H\st{HI}$. Therefore, without an external magnetic field the Overhauser fields couples almost all spin states giving rise to leakage into nearly every state. Reverting the leakage requires complicated sequences of pulsed magnetic fields, thus, losing the benefit of the three-spin encoding\cite{Hung2014}. However, subject to a large magnetic field, the EO subspace qubit, $S=1/2$ and $S_{z}=+1/2$ leaks only into a single states, the $S=3/2$ and $S_{z}=1/2$ state
\begin{align}
\ket{L}=\frac{1}{\sqrt{3}}\left(\ket{\uparrow\uparrow\downarrow}+\ket{\uparrow\downarrow\uparrow}+\ket{\downarrow\uparrow\uparrow}\right),
\label{eq:leakState}
\end{align} 
giving rise to a three-level system. It is preferable to work in a regime where both Zeeman energy $E_{z}$ and exchange splitting $J$ is significant larger than the Overhauser fields, $J,E_{z}\ll \mathcal{A}_{i}$, since there leakage is suppressed $\propto \mathcal{A_{i}}/J$. Outside this regime the leakage dynamics occurs in timescales of nanoseconds\cite{Hung2014}.

Nuclear field gradients between the QDs make the qubit very vulnerable to inhomogeneous broadening which cause dephasing of the qubit states due to the acquisition of random local phases. Setting up a Ramsey free induction type measurement consisting of two $\pi/2$-pulses separated by the time $\tau$ allows for tracking of the dephasing. Considering Gaussian distributed Overhauser fields which is valid under typical experimental conditions\cite{Hanson2007} the resulting inhomogeneous broadening dephasing time is given by
\begin{align}
T^{\star}_{2}\propto\bigg(\sum_{i=1}^{3}\nu_{i}^{2}\mathcal{A}_{i}^{2}\bigg)^{-1/2}.
\end{align}
Here, $\mathcal{A}_{i}$ are the standard deviations of the Overhauser fields in QD~$j$ while their impact is given by the weighing factors $\nu_{1}=\nu_{3}=1$ and $\nu_{2}=2$. Therefore, the dephasing times of three-spin qubits are roughly on the same timescales as for spin-$\frac{1}{2}$ qubits\cite{Hung2014} assuming uncorrelated Overhauser fields in each dot. 

\paragraph{Suppressing magnetic noise}

Since nuclear noise has typically a very slow dynamics\cite{Delbecq2016} dynamical decoupling (DD)\cite{Falci2004,Faoro2004,Rebentrost2009,Lee2008,Du2009,Coish2009,Green2013} offers an efficient way to counteract the effects of noise. In simple words, they decouple the noise from the system, e.g., through gate sequences which swap the electrons in such a way that each electron ``feels'' the same noise, thus, inducing a symmetry to the noise bath. Under the assumption of static noise a simple example is the permutation sequence which cyclicly swaps the electrons after each time interval $\tau$. In this review we focus on recent advances in DD which only use the exchange interaction for the decoupling sequence in agreement with the concept of the EO qubit;  for the general case we refer to Ref.~\cite{Lidar2013}.

The starting situation is the following; a single three-spin qubit implemented in a linear TQD where the electron in each QD is coupled to a large number of nuclear spins. The three-spin qubit system in the (1,1,1) charge configuration is described by the Heisenberg exchange Hamiltonian plus a Zeeman term including the fluctuating Overhauser fields\cite{Rohling2016}
\begin{align}
H=\frac{J_{12}}{4}\bo{\sigma_1}\cdot\bo{\sigma_2}+\frac{J_{23}}{4}\bo{\sigma_2}\cdot\bo{\sigma_3} + \sum_{n=1,2,3}\boldsymbol{B}_{n}\cdot\boldsymbol{\sigma}_{n}.
\end{align}
Assuming a strong Zeeman splitting the dynamics of the qubit is described by the qubit states and single relevant leakage state $\ket{L}$ (see Eg.~\eqref{eq:leakState}).

Introduced by West and Fong\cite{West2012}, the DD sequence consists only of operations $\textsc{swap}_{ij}$ that interchange the spin of QD~$i$ and QD~$j$ and which are realized by the exchange interaction\cite{Loss1998}. During the \textsc{swap} operation leakage is suppressed by exchange\cite{Hung2014}, thus, dephasing is only possible in the remaining QD during the short time of an operation since exchange is completely turned in-between the pulses\cite{West2012}. Only sequences are considered which consists of the permutation $P=\textsc{swap}_{23}\textsc{swap}_{12}$ and its inverse $P^{-1}=\textsc{swap}_{12}\textsc{swap}_{23}$. Under the assumption of Gaussian distributed noise the variance of the gathered phase differences, which cause the dephasing, can be expressed in terms of switching functions $f_{j}(t)$ with $j=1,2,3$ which are defined according to the position of the spin states and their rescaled Fourier transforms $\gamma_{j}\equiv \frac{\omega}{\i} \int_{0}^{T} dt \E^{\I \omega t}f_{j}(t)$\cite{West2012,Rohling2016}. The resulting expression for the variance of the phase differences is
\begin{align}
\braket{\Delta\Phi(T)^{2}}=\frac{1}{\pi}\int_{0}^{\infty}d\omega \sum_{j=1}^{3}|\gamma_{j}(\omega T)|^{2}\frac{S(\omega)}{\omega^{2}},
\end{align}
where $S(\omega)$ is the power density noise spectrum of the noise which for simplicity is assumed to be identical in each QD whereas the noise in each dot is uncorrelated. The cross-correlation of the noise depends on the system and can be measured, e.g., using spatial separated QDs\cite{Szankowski2016}. For a simple permutation sequence and the West and Fong sequence of length $n$ the switching functions take the values $f_{i}=-1,0,1$ (for the exact definitions see Refs.~\cite{West2012,Rohling2016}) at the times $T\delta_{j}$ with $j=1,...,n$. 

\begin{figure}
\begin{center}
\includegraphics[width=1.\columnwidth]{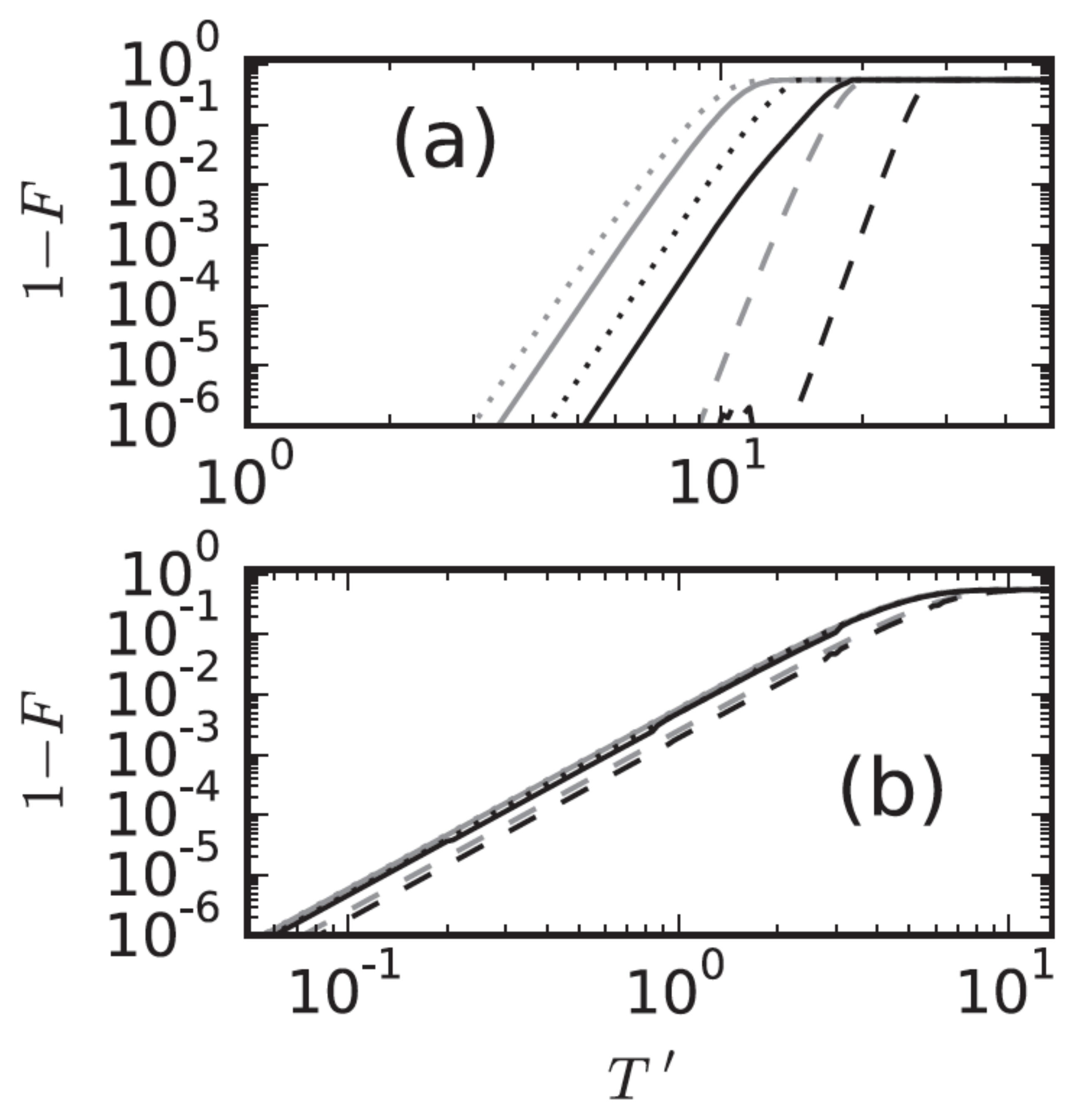}
\caption{Comparison of the infidelity $1-F$ as a function of the dimensionless storage time $T^{\prime}=T\omega_{1}$ for both noise decoupling strategies, Uhrig dynamical decoupling (gray) and optimized noise filtration dynamical decoupling (black), (a) assuming ohmic noise with $S(\omega)=\omega \theta(\omega-\omega_{1})/\omega_{1}$ and (b) Lorentzian noise $S(\omega)=1/(1+(\omega/\omega_{1})^{2})$ noise. Here, $\omega_{1}$ is a sharp cut-off frequency for ohmic noise while $\omega_{1}$ is the line-width for the Lorentzian noise. The dashed lines correspond to the single spin case, the solid lines correspond to the the simple permutation cycle, and the dotted lines correspond to the West and Fong sequence. Figure taken from Ref.~\cite{Rohling2016}.}
\label{fig:DD}
\end{center}
\end{figure}

There are many concepts for optimizing the waiting times $\delta_{j}$ the most popular being the CPMG sequence\cite{Carr1954,Meiboom1958} which uses equidistant time intervals $\Delta\delta_{j}=1/n$. However, CPMG is not optimized for the three-spin case, thus, does not lead to $\gamma_{i}=0$ in lowest order. Until now the best concepts are Uhrig dynamical decoupling (UDD)\cite{Uhrig2007,Uhrig2008,Uhrig2009} and optimized noise filtration dynamical decoupling (OFDD)\cite{Uys2009} depending on the given noise.
UDD requires that the Fourier transforms vanish up to an order $m$, thus, $\frac{\partial^{k}\gamma_{i}(q)}{\partial q^{k}}|_{q=0}=0$ for $k=0,...,m$ and $i=1,2,3$. In the single spin case $n=m$ pulses are required\cite{Uhrig2007} with the analytical expressions for the waiting times $\delta_{j}=[1-\cos(\pi j/(n+1))]/2$. For the three-spin case $n=2m$ pulses are required\cite{West2012} while the expression for the waiting times are obtained from the solution of the $2m$ polynomial functions up to order $m$. 
OFDD directly minimizes the integral $\int_{0}^{1}d\omega \sum_{j=1}^{3}|\gamma_{j}(\omega T)|^{2}$ finding a suitable set for the waiting times $\delta_{j}$. The integrals can be treated analytically, however, the values have to be determined using numerical optimization, e.g., one can use the values for CPMG and UDD as starting values. A comparison of both strategies is displayed in Fig.~\ref{fig:DD} which shows the better results of OFDD for both considered types of noise.

\subsection{Charge noise}
\label{ssec:chargeDec}

Charge noise or electrical noise, produced by fluctuations of charges or electric fields, is an omnipresent phenomenon if any form of electric control is used in experimental setups, directly, e.g., electric potentials to attract/deplete electrons or indirectly, e.g., background charge fluctuations. Thus, this also includes each device based on semiconductors or metals since electrons, the carriers of the charge, move around freely which unavoidably results in fluctuations due to the discreteness of the electric charge. These charge fluctuation can be correlated in time or space depending on the source of the noise. In semiconductors the background noise is typically dominated by low-frequency noise or colored noise which has a spectral density $S(\omega)=A\omega^{-\gamma}$, where $\gamma$ is the noise exponent\cite{Dutta1981,Weissman1988}. Noise which is induced into the qubit system through the gate electrodes necessary for confining and controlling the electrons is typically Nyquist-Johnson noise\cite{Johnson1928,Nyquist1928,Langsjoen2012,Poudel2013} due to the finite temperature and shot-noise\cite{Schottky1918} due to quantization (graininess) of electric charge. Like magnetic noise, charge noise also depends on the system\cite{Beaudoin2015,Ramon2015,Szankowski2016}, however, to a smaller degree than for magnetic noise from nuclear spins, e.g., charge noise can be enhanced in the presence of piezoelectric phonons and their coupling strength. Therefore, charge noise cannot be changed significantly if the host material is replaced, since freely-moving electrons exist in every semiconductor and every device is connected to wires.

In this subsection we investigate the effect of charge noise on three-spin qubits and look for approaches to avoid noise in the first order and consecutively, if this is not possible, to avoid the effects of such noise. The first mentioned approach is usually treated by working on ``sweet spots'', points in parameter space which are least susceptible to noise, while the latter is usually taken by using dynamical decoupling techniques\cite{Green2013,Cywinski2014,Ramon2015}. Since dynamical decoupling of charge noise is already presented in the previous subsection and in a related review~\cite{Lidar2012}, we mainly study the techniques for avoiding the noise in this review.

\begin{figure}[t]
\begin{center}
\includegraphics[width=1.0\columnwidth]{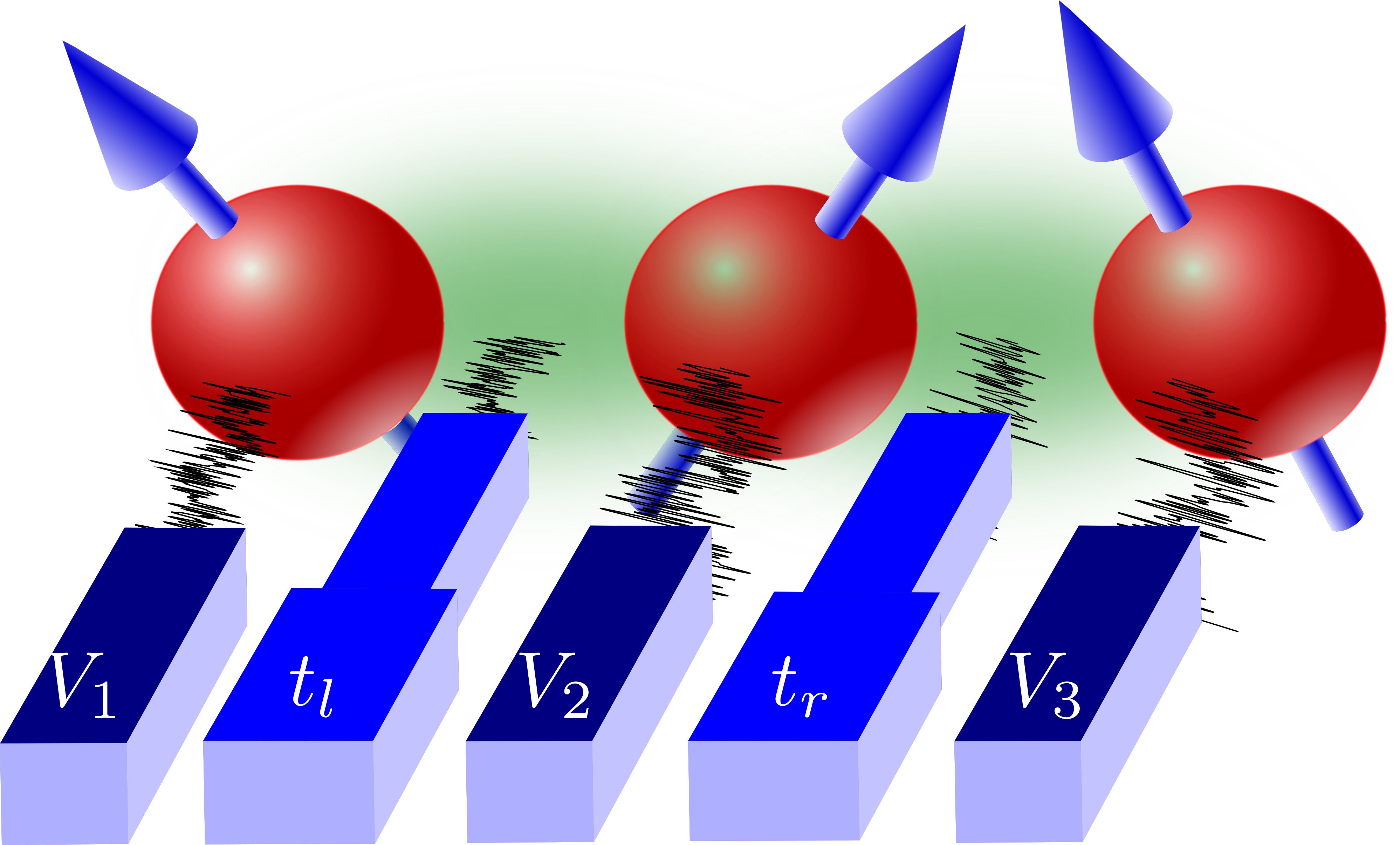}
\caption{Schematic illustration of a three-spin qubit coupled to charge noise. The environment mainly affects the electron spins directly through the gate voltages $V_{i}$ with $i\in\lbrace 1,2,3\rbrace$ of each quantum dot (QD) or the exchange coupling (green cloud) between the electron spins through the gate-controlled tunnel hopping ($t_{l}$ and $t_{r}$). Figure taken from Ref.~\cite{Russ2016}.}
\label{fig:modelCharge}
\end{center}
\end{figure}

\paragraph{Decoherence due to charge noise}

Electric noise affects the charge degrees of freedom in a quantum system, thus, allows for coupling through every electrically controlled parameter in the system. The dominant links which couple the charge noise to the TQD qubit are the detuning parameters, $\varepsilon$ (detuning between outer QDs) and $\varepsilon_{M}$ (detuning between center QD and mean of outer QDs) from Eqs.~\eqref{eq:defDet}, linked to the gate voltages underneath each QD\cite{Taylor2013,Fei2015,Russ2015,Friesen2016,Shim2016,Russ2016} (see Fig.~\ref{fig:modelCharge}). This is because these control parameters typically induce the qubit gates, thus, they have to be addressed fast and precisely over a large range which limits the amount of noise filtering\cite{Gaudreau2012,Medford2013b,Eng2015}. Furthermore, the tunneling parameters, $t_{l}$ and $t_{r}$, also provide a significant noise contribution if the qubit is operated symmetrically by controlling the tunnel barriers between the qubits\cite{Reed2016,Martins2016} (see Fig.~\ref{fig:modelCharge}). Additional parameters worth being considered are the charging and confinement energies\cite{Russ2016b} but these play a less important role since they are static, allowing for low-pass filtering. Formally, the noisy control parameters are described by considering noisy parameters $q(\delta q)=f(\delta q)$, where $f$ is some function which describes how the noisy parameter $q$ is affected by the corresponding fluctuations $\delta q$. Typically, one assumes that the strength of charge noise is unaffected by the strength of the noisy parameter, thus, $q(\delta q)=q+\delta q$\cite{Taylor2013,Fei2015,Russ2015,Shim2016,Russ2016,Friesen2016}. While by definition the average of these fluctuations vanishes, $\braket{\delta q}=0$, under realistic conditions there are no measurements yet in TQD devices of the standard deviations, $\sigma_{q}=\sqrt{\braket{\delta q^{2}}}$, and higher cumulants. For the detuning parameters, $\varepsilon$ and $\varepsilon_{M}$, measurements in single QDs and DQDs indicate values of $\sigma_{\varepsilon,\varepsilon_{M}}\simeq\unit[5]{\mu eV/\sqrt{Hz}}$ in a GaAs device\cite{Li1990,Dekker1991,Sakamoto1995,Kurdak1997,Hayashi2003,Jung2004,Buizert2008,Petersson2010} and $\sigma_{\varepsilon,\varepsilon_{M}}\simeq\unit[1]{\mu eV/\sqrt{Hz}}$ in SiO and Si/SiGe devices\cite{Takeda2013,Freeman2016}. An effective measurement of detuning charge noise in an isotopically purified Si/SiGe TQD indicates a higher value, $\sigma_{\varepsilon}\simeq\unit[15]{\mu eV/\sqrt{Hz}}$ for $1/f$-noise\cite{Eng2015}. While these values are likely to be accurate for $\varepsilon$, the noise coupling for the symmetric detuning $\varepsilon_{M}$ are claimed to be ten times smaller\cite{Friesen2016}. The fluctuations of the remaining parameters such as the tunnel couplings is still unknown, but the tools to measure these are already present\cite{Dial2013,Cywinski2014,Szankowski2016}. These charge fluctuations affect the qubit states but, unlike magnetic noise, do not cause any leakage out of the qubit space due to the spin conserving nature of charge noise. Leakage into other charge states with the same spin numbers such as (1,0,2) is still possible but ideally suppressed by detuning.

Since these longitudinal fluctuations are random the qubit states acquire local phases which are by definition unknown resulting in a dephasing of the qubit. A simple measurement to track these effects is provided by a Ramsey free decay sequence consisting of two $\pi/2$ pulses separated by the waiting time $\tau$\cite{Taylor2006}. Starting in the state $\ket{0}$, the first $\pi/2$-pulse produces a superposition of the qubit states, $\ket{+}=(\ket{0}+\ket{1})/\sqrt{2}$. After the second $\pi/2$-pulse the return probability into the $\ket{0}$ state is given by $P(\tau)=\frac{1}{2}\left(1+\text{Re}[f(\tau)\right)]$ with
\begin{align}
f(\tau)=\braket{\E^{\I \phi(\tau)}}=\E^{-\braket{\phi(\tau)}^{2}_{C}/2}
\end{align}
where  $\phi(\tau)$ is the average of the acquired phase difference between the qubit states and $\braket{\phi(\tau)}^{2}_{C}$ its second cumulant. For the second equality Gaussian distributed charge noise is assumed\cite{Russ2015,Russ2016}. The decay exponent itself strongly depends on the exact noise spectral density $S(\omega)$. Considering $1/f$-noise one finds $\braket{\phi(\tau)}^{2}_{C}=\tau^{2}/T_\varphi^{2}$ with the pure dephasing time $T_\varphi$.

The transversal effect of charge noise causes random transitions between the qubit states. However, since the time-scales of the transitions are rather long (milliseconds) the qubit-flip errors can often be neglected. 

\paragraph{Sweet spots and optimal working points}

The starting situation is: a single three-spin qubit implemented in a linear TQD affected by charge noise through various noisy parameters $q$. In general the noisy Hamiltonian is $H=H_{0}+H\st{noise}$ where $H_{0}=\hbar\omega\sigma_{z}/2$ is the unperturbed qubit Hamiltonian in its eigenbasis and
\begin{align}
H\st{noise}=\frac{\hbar}{2}\left[\delta\omega_{z}\sigma_{z} + \delta\omega_{x}\sigma_{x} + \delta\omega_{y}\sigma_{y}\right]
\label{eq:Hnoise}
\end{align}
is in the same basis and directly follows from perturbation theory with the single requirement that the fluctuations are small compared to the energy gap. The perturbation terms, one longitudinal and two transversal, are given by
\begin{align}
\delta\omega_{z}&=\frac{1}{2}\left(\bra{g} H_{1,q} \ket{g}-\bra{e} H_{1,q} \ket{e}\right),\nnb
			&\approx\sum_{q}\left( \omega_{q} \delta q+\frac{\omega_{q,q}}{2}\delta q^{2}\right)+\frac{1}{2}\sum_{p\neq q} \omega_{p,q}  \delta p\delta q,\label{eq:noiseGenz}\\
\delta\omega_{x}&=\frac{1}{2}\left(\bra{g} H_{1,q} \ket{e}+\bra{e} H_{1,q} \ket{g}\right),\\
\delta\omega_{y}&=\frac{1}{2\I}\left(\bra{g} H_{1,q} \ket{e}-\bra{e} H_{1,q} \ket{g}\right),
\end{align}
where $H_{1,q}=\frac{\partial}{\partial q}H\,\delta q$, $\omega_{q}\equiv\frac{\partial \omega}{\partial q}$, and $\left.\omega_{p,q}\equiv\frac{\partial^{2}\omega}{\partial p\partial q}\right.$. For the approximation in the second term the perturbation is expanded up to second order in the fluctuations $\delta q$. Therefore, the most devastating effect is contributed by the longitudinal charge noise $\delta\omega_{z}$ which becomes clear when expanding the eigenenergy gap
\begin{align}
\omega &= \sqrt{(\omega_0+\delta\omega_z)^2+\delta\omega_x^2+\delta\omega_y^2}\nonumber\\
&\simeq  \omega_0 + \delta\omega_z + \frac{\delta\omega_x^2}{2\omega_0} + \frac{\delta\omega_y^2}{2\omega_0}
+O(\delta\omega^3),
\end{align}
while the transversal charge noise, $\delta\omega_{x}$ and $\delta\omega_{y}$, contributes only second order and becomes smaller for large $\omega$\cite{Russ2015,Russ2016}. Using Eq.~\eqref{eq:EOmatrix8} and low-frequency noise $S(\omega)=A_{q}/\omega$, where $A_{q}=\sigma_{q}^{2}$ is the squared standard deviation of the noise, the Ramsey free decay pure dephasing time is\cite{Russ2016}
\begin{align}
	T_\varphi=& \hbar \bigg[\sum_{q}\frac{\omega_{q}^2}{2}\,A_p\log r+\frac{\omega_{q,q}^2}{4}A_q^2\log^2r\nnb
	&+\frac{1}{2}\sum_{p\neq q}\frac{ \omega_{p,q}^2}{2}A_{p}A_{q}\log^2r+ \frac{1}{8}\omega_{p,p} \omega_{q,q}A_p\,A_q\bigg]^{-\frac{1}{2}}.
\label{eq:dephasing}
\end{align}

\begin{figure*}
\begin{center}
\includegraphics[width=1.\textwidth]{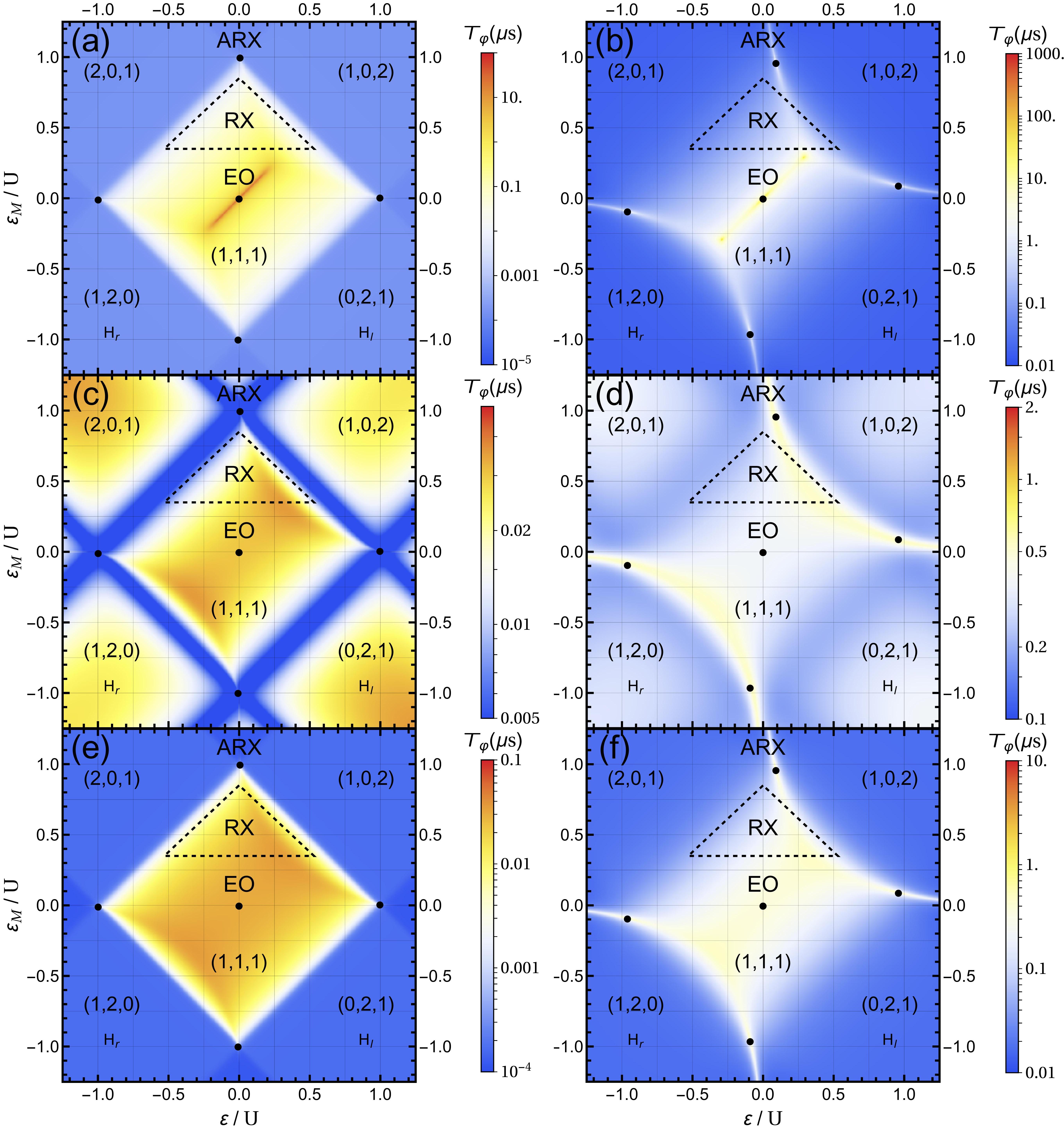}
\caption{Dephasing time $T_{\varphi}$ given by Eq.~\eqref{eq:dephasing} due to longitudinal noise as a function of the detuning parameters $\varepsilon$ and $\varepsilon_{M}$. In the top row ((a) and (b)) we plot $T_{\varphi}$ resulting from charge noise in the two detuning parameters $\varepsilon$ and $\varepsilon_{M}$, in the center row ((c) and (d)) we plot $T_{\varphi}$ resulting from charge noise in the two tunneling parameter $t_{l}$ and $t_{r}$, and in the bottom row ((e) and (f)) we plot $T_{\varphi}$ resulting from charge noise from all four parameters combined, where we choose the parameter settings identical in each column. The left column shows results for weak tunneling and strong noise while in the right column, results for strong tunneling and weak noise are plotted. Parameters are set as follows; $t_{l}=0.022\tun$, $t_{r}=0.015\tun$, $A_{q}=(10^{-3}\tun)^{2}$ where $q=\varepsilon,\varepsilon_{M}$ in (a) and (e), and $A_{q}= (10^{-4}\tun)^{2}$, where $q=t_{l},t_{r}$ in (c) and (e), for the left column and $t_{l}=0.22\tun$, $t_{r}=0.15\tun$, $A_{q}=(10^{-5}\tun)^{2}$ where $q=\varepsilon,\varepsilon_{M}$ in (b) and (f), and $A_{t_{l}}=A_{t_{r}}= (10^{-6}\tun)^{2}$, where $q=t_{l},t_{r}$ in (d) and (f), for the right column. To include a large frequency bandwidth we globally set the ratio of the lower and higher frequency cut-off $r=5\times10^{6}$. For the scale of $T_{\varphi}$ we used an explicit value of $U=\unit[1]{meV}$; note that $T_{\varphi}$ scales inversely proportional with $U$. The black dots indicate the DSS. Figure taken from Ref.~\cite{Russ2016}.}
\label{fig:Plotlong}
\end{center}
\end{figure*}

With this in mind, a formal definition for ``sweet spots''\cite{Taylor2013,Medford2013,Fei2015,Russ2015,Shim2016,Russ2016,Friesen2016,Wardrop2016} of order $n=1,2,3,\dots$ is 
\begin{align}
\sum_{q} \omega_{q} \delta q = 0,
\label{eq:SScond}
\end{align}
with the sum running over $n$~terms since then the dominating terms in the expressions above vanish (see Eq.~\eqref{eq:noiseGenz} and Eq.~\eqref{eq:dephasing})\cite{Russ2016}. A full sweet spot is only possible if each term in the sum vanishes simultaneously and sweet spot of order $n$ requires that $n$~terms are zero. This condition corresponds to an extremum of the qubit energy gap with respect to the noisy parameter $q$. In following we denote sweet spots of the order $n=1$ and $n=2$ as single sweet spots (SSSs) and double sweet spots (DSSs).

For the detuning parameters, $\varepsilon$ and $\varepsilon_{M}$, there are five known DSSs\cite{Russ2016}. One is located in the center of the (1,1,1) charge configuration regime\cite{Fei2015,Shim2016} (see Fig.~\ref{fig:Plotlong}~(a)), $\varepsilon=\varepsilon_{M}=0$ (for symmetric charging energies), and the motivation behind the AEON qubit (see subsection~\ref{ssec:AEON}). The reason for this is the high degree of symmetry this point possesses, being aligned symmetrically to the four charge configurations (1,0,2), (2,0,1), (1,2,0), and (0,2,1) and in the center of the (1,1,1) charge configuration regime yielding a real minimum of the energy gap. The other four DSS are located each at the charge transitions between two neighboring asymmetric [(1,0,2), (2,0,1), (1,2,0), or (0,2,1)] and the (1,1,1) charge configuration\cite{Russ2015,Russ2016}. They are approximately located at $(\varepsilon,\varepsilon_{M})=(0,U),\,(0,-U),\,(-U,0),\,(U,0)$ (see Fig.~\ref{fig:Plotlong}~(a)) whereby the real positions are slightly shifted due to the influence of the other charge configurations. Since these four DSSs possess less symmetry these positions are not minima of $\omega$ but correspond to saddle points\cite{Russ2016}. Taking into account also the second order effects, the center DSS is clearly favorable compared to the remaining four DSS (see Fig.~\ref{fig:Plotlong}~(a)). The explanation of such case, e.g., for the top DSS that corresponds to the $(2,0,1)\leftrightarrow(1,0,2)$ charge transition, would be the large electric dipole moment between the (2,0,1) and (1,0,2) states providing a channel through which charge noise couples to the qubit. On the other hand, the center DSS possesses only a vanishing electric dipole moment providing a better protection\cite{Shim2016,Russ2016}. Increasing the strength of the tunnel couplings, experimentally achieved by lowering of the tunnel barriers, smoothes out the curvature, giving rise to longer dephasing times and decreasing difference between the DSSs (see Fig.~\ref{fig:Plotlong}~(b)). The drawback of the center DSS is that its energy gap is minimal, making the gate operations slower. Since charge noise through $\varepsilon_{M}$ is claimed to be 10 times smaller\cite{Friesen2016} it can be sufficient to work on a SSS with respect to only $\varepsilon$ noise combined with a strong exchange splitting (RX regime) \cite{Taylor2013,Medford2013}. As experimentally demonstrated, when working at a SSS, the dephasing time is increased significantly, reaching $T_{2}=\unit[19]{\mu s}$ while measuring a larger spectral density exponent. This indicates that higher order effects play the dominating role, thus, a tell-tale sign of a sweet spot. However, this measurement includes also nuclear noise, making it difficualt to differentiate between these two\cite{Medford2013}.

For noisy tunneling parameters, $t_{l}$ and $t_{r}$, there exist no such DSSs where the impacts of both tunneling parameters are minimized simultaneously\cite{Russ2016} (see Fig.~\ref{fig:Plotlong}~(c) and Fig.~\ref{fig:Plotlong}~(d)). Therefore, only single SSSs exist which are located at the charge transitions associated with the tunneling parameter, e.g., $t_{l}$ is minimized nearby the $(2,0,1)\leftrightarrow(1,1,1)$ charge configuration\cite{Russ2016}. With their strong impact, fluctuations in the tunnel couplings altogether limit the dephasing times significantly and even more strongly than detuning noise if both fluctuations are comparable in strength. Since the effect of fluctuating tunnel barriers on the tunnel couplings is still under investigation, only qualitative conclusions are possible at this stage\cite{Russ2016}. Such a qualitative comparison is shown in Fig.~\ref{fig:Plotlong}~(e) and Fig.~\ref{fig:Plotlong}~(f).

\paragraph{Spin-phonon relaxation}
Another effect of charge noise are relaxation processes due to phonons interacting with the electrons\cite{Hanson2007,Hu2011,Mehl2013,Gamble2012,Huang2014} through the dipole moment of the qubit\cite{Taylor2013,Srinivasa2016}.
Theoretical estimations reveal that these relaxation rates are small in GaAs\cite{Taylor2013}, $\Gamma=\unit[1-100]{kHz}$ for typical parameter settings, and Si\cite{Srinivasa2016}, $\Gamma=\unit[1-100]{Hz}$ for similar parameter settings, but not completely negligible for GaAs devices while Si is better protected due to the absence of piezo-electric phonons in bulk silicon.

\section{Perspectives}
\label{sec:pers}

This review has described the recent experimental and theoretical progress and achievements in three-spin qubits that were initially proposed over a decade ago. Several realizations of the three-spin qubits have been discussed, such as the exchange-only qubit, the spin-charge qubit, the hybrid qubit, the resonant exchange qubit, and the always-on exchange only qubit, with special attention to their potential to fulfill the DiVincenzo criteria\cite{DiVincenzo2000}. We organize our summary according to these five criteria.
\begin{enumerate}
\item Having a scalable system with well-defined qubits. Electron spins fulfill definitely the requirement of well-defined qubits, thus, three-spin qubits inherit this property if encoded in a proper subspace which is the case. The requirement, however, can be violated by leakage since the three-spin qubit is encoded only in a subspace of the full three-spin space. But external global magnetic fields and the exchange interaction can be used to energetically separate the qubit space from the non-computational space, thus, reducing the leakage to a manageable quantity. Scalability follows from the geometry of the quantum dots and the gate electrodes.
\item Being able to initialize a proper state such as $\ket{00000\cdots}$. Spin-to-charge conversion allows to initialize a state with a probability close to 1. Two electrons filled in a single dot nearly always relax into a singlet ground-state after a sufficiently long time depending on the experimental setup which afterwards can be adiabatically transformed into a $\ket{0}$ state.
\item Having a long decoherence time, or more precisely a sufficient number of gate operations while the qubit is coherent. Three-spin qubits posses a natural robustness against some noise since they are encoded in a decoherence free subspace. Using isotopically purified host materials and operating on a charge noise sweet spot one can mitigate the two main sources of decoherence, allowing for large coherence times. The other possibility is to speed up the gate operations which is realized by utilizing the fast exchange interaction based operations.
\item Having a universal set of quantum gates. Three-spin qubits allow for a very fast and highly precise way for operating single-qubit gates. Utilizing the exchange interaction leads to gates in sub-nanosecond time scales $\tau_{g}\approx\unit[200]{ps}$, and fidelity exceeding $99\%$ or using resonant driving allows for gate times on the order of nanoseconds, $\tau_{g}\approx\unit[2.5]{ns}$. Both of these are significantly faster than qubit implementations using a single electron spin. However, fast exchange-based two-qubit gates either consist of complex sequences with more than 10 pulses or can leak into the non-computational space. Nonetheless, the large dipole moment of three-spin qubits makes up for this since it allows for fast long-ranged two-qubit interactions utilizing a cavity as a mediator. 
\item Having high fidelity and qubit-specific measurements. Spin-to-charge conversion combined with charge-sensitive detectors such as quantum point contacts or single electron transistors allow for precise and fast read-out schemes. Concepts using cavity quantum electrodynamics also allow for fast read-out.
\end{enumerate}

A yet rather unknown but nonetheless important factor is the six-fold (valley) degeneracy of the ground-state in bulk silicon originating from various minima in the conduction band. Applying strain raises four of the six valleys in energy such that there exists in silicon QDs an additional two-fold valley degeneracy which has the properties of a pseudo-spin\cite{Zwanenburg2013}. This two-fold valley degeneracy can be lifted by interfaces in the 2DEG, however, the exact orientation and splitting of the valley depends on atomistic steps of the interface\cite{Culcer2009,Culcer2010,Rohling2012,Rohling2014,Veldhorst2015b,Rancic2016,Boross2016,Zimmerman2016,Gamble2016}. This makes the valley splitting very unpredictable, and it is unwanted in qubit implementations in silicon as it boosts up the already large Hilbert space of three-spin qubits. In SiMOS devices the valley splitting is controllable to some degree by an electric field perpendicular to the interface\cite{Saraiva2011,Lim2011,Wu2012,Wu2012,Culcer2012,Yang2013,Kim2014,Hao2014,Zimmerman2016,Gamble2016}. There are theoretical concepts of utilizing the valley as an additional qubit\cite{Rohling2012,Rohling2014}, however, due to the unpredictable nature a large valley splitting is typically prefered in experiments involving silicon as the host material. 

\ack
We acknowledge funding from ARO through Grant No. W911NF-15-1-0149 and the DFG through SFB 767. We also thank M. Brooks, C. Peterfalvi, and M. Rancic for discussions contributing to this review.

\clearpage

\bibliographystyle{unsrturl}
\bibliography{lit5}

\begin{thebibliography}{100}

\bibitem{Bacon2000}
D.~Bacon, J.~Kempe, D.~A. Lidar, and K.~B. Whaley.
\newblock Universal fault-tolerant quantum computation on decoherence-free
  subspaces.
\newblock {\em Phys. Rev. Lett.}, 85:1758--1761, Aug 2000.
\newblock URL: \url{http://link.aps.org/doi/10.1103/PhysRevLett.85.1758}, \href
  {http://dx.doi.org/10.1103/PhysRevLett.85.1758}
  {\path{doi:10.1103/PhysRevLett.85.1758}}.

\bibitem{Nature2000}
D.~P. DiVincenzo, D.~Bacon, J.~Kempe, G.~Burkard, and K.~B. Whaley.
\newblock Universal quantum computation with the exchange interaction.
\newblock {\em Nature}, 408(6810):339, November 2000.
\newblock URL: \url{http://dx.doi.org/10.1038/35042541}.

\bibitem{Russ2016}
Maximilian Russ, Florian Ginzel, and Guido Burkard.
\newblock Coupling of three-spin qubits to their electric environment.
\newblock {\em Phys. Rev. B}, 94:165411, Oct 2016.
\newblock URL: \url{http://link.aps.org/doi/10.1103/PhysRevB.94.165411}, \href
  {http://dx.doi.org/10.1103/PhysRevB.94.165411}
  {\path{doi:10.1103/PhysRevB.94.165411}}.

\bibitem{Laird2010}
E.~A. Laird, J.~M. Taylor, D.~P. DiVincenzo, C.~M. Marcus, M.~P. Hanson, and
  A.~C. Gossard.
\newblock Coherent spin manipulation in an exchange-only qubit.
\newblock {\em Phys. Rev. B}, 82:075403, Aug 2010.
\newblock URL: \url{http://link.aps.org/doi/10.1103/PhysRevB.82.075403}, \href
  {http://dx.doi.org/10.1103/PhysRevB.82.075403}
  {\path{doi:10.1103/PhysRevB.82.075403}}.

\bibitem{Kyriakidis2007}
Jordan Kyriakidis and Guido Burkard.
\newblock Universal quantum computing with correlated spin-charge states.
\newblock {\em Phys. Rev. B}, 75:115324, Mar 2007.
\newblock URL: \url{http://link.aps.org/doi/10.1103/PhysRevB.75.115324}, \href
  {http://dx.doi.org/10.1103/PhysRevB.75.115324}
  {\path{doi:10.1103/PhysRevB.75.115324}}.

\bibitem{Shi2012}
Zhan Shi, C.~B. Simmons, J.~R. Prance, John~King Gamble, Teck~Seng Koh, Yun-Pil
  Shim, Xuedong Hu, D.~E. Savage, M.~G. Lagally, M.~A. Eriksson, Mark Friesen,
  and S.~N. Coppersmith.
\newblock Fast hybrid silicon double-quantum-dot qubit.
\newblock {\em Phys. Rev. Lett.}, 108:140503, Apr 2012.
\newblock URL: \url{http://link.aps.org/doi/10.1103/PhysRevLett.108.140503},
  \href {http://dx.doi.org/10.1103/PhysRevLett.108.140503}
  {\path{doi:10.1103/PhysRevLett.108.140503}}.

\bibitem{Koh2012}
Teck~Seng Koh, John~King Gamble, Mark Friesen, M.~A. Eriksson, and S.~N.
  Coppersmith.
\newblock Pulse-gated quantum-dot hybrid qubit.
\newblock {\em Phys. Rev. Lett.}, 109:250503, Dec 2012.
\newblock URL: \url{http://link.aps.org/doi/10.1103/PhysRevLett.109.250503},
  \href {http://dx.doi.org/10.1103/PhysRevLett.109.250503}
  {\path{doi:10.1103/PhysRevLett.109.250503}}.

\bibitem{Cao2016}
Gang Cao, Hai-Ou Li, Guo-Dong Yu, Bao-Chuan Wang, Bao-Bao Chen, Xiang-Xiang
  Song, Ming Xiao, Guang-Can Guo, Hong-Wen Jiang, Xuedong Hu, and Guo-Ping Guo.
\newblock Tunable hybrid qubit in a gaas double quantum dot.
\newblock {\em Phys. Rev. Lett.}, 116:086801, Feb 2016.
\newblock URL: \url{http://link.aps.org/doi/10.1103/PhysRevLett.116.086801},
  \href {http://dx.doi.org/10.1103/PhysRevLett.116.086801}
  {\path{doi:10.1103/PhysRevLett.116.086801}}.

\bibitem{Medford2013}
J.~Medford, J.~Beil, J.~M. Taylor, E.~I. Rashba, H.~Lu, A.~C. Gossard, and
  C.~M. Marcus.
\newblock Quantum-dot-based resonant exchange qubit.
\newblock {\em Phys. Rev. Lett.}, 111:050501, Jul 2013.
\newblock URL: \url{http://link.aps.org/doi/10.1103/PhysRevLett.111.050501},
  \href {http://dx.doi.org/10.1103/PhysRevLett.111.050501}
  {\path{doi:10.1103/PhysRevLett.111.050501}}.

\bibitem{Taylor2013}
J.~M. Taylor, V.~Srinivasa, and J.~Medford.
\newblock Electrically protected resonant exchange qubits in triple quantum
  dots.
\newblock {\em Phys. Rev. Lett.}, 111:050502, Jul 2013.
\newblock URL: \url{http://link.aps.org/doi/10.1103/PhysRevLett.111.050502},
  \href {http://dx.doi.org/10.1103/PhysRevLett.111.050502}
  {\path{doi:10.1103/PhysRevLett.111.050502}}.

\bibitem{Fong2011}
Bryan~H. Fong and Stephen~M. Wandzura.
\newblock Universal quantum computation and leakage reduction in the 3-qubit
  decoherence free subsystem.
\newblock {\em Quantum Info. Comput.}, 11(11-12):1003, November 2011.
\newblock URL: \url{http://dl.acm.org/citation.cfm?id=2230956.2230965}.

\bibitem{Setiawan2014}
F.~Setiawan, Hoi-Yin Hui, J.~P. Kestner, Xin Wang, and S.~Das Sarma.
\newblock Robust two-qubit gates for exchange-coupled qubits.
\newblock {\em Phys. Rev. B}, 89:085314, Feb 2014.
\newblock URL: \url{http://link.aps.org/doi/10.1103/PhysRevB.89.085314}, \href
  {http://dx.doi.org/10.1103/PhysRevB.89.085314}
  {\path{doi:10.1103/PhysRevB.89.085314}}.

\bibitem{Zeuch2014}
Daniel Zeuch, R.~Cipri, and N.~E. Bonesteel.
\newblock Analytic pulse-sequence construction for exchange-only quantum
  computation.
\newblock {\em Phys. Rev. B}, 90:045306, Jul 2014.
\newblock URL: \url{http://link.aps.org/doi/10.1103/PhysRevB.90.045306}, \href
  {http://dx.doi.org/10.1103/PhysRevB.90.045306}
  {\path{doi:10.1103/PhysRevB.90.045306}}.

\bibitem{Doherty2013}
Andrew~C. Doherty and Matthew~P. Wardrop.
\newblock Two-qubit gates for resonant exchange qubits.
\newblock {\em Phys. Rev. Lett.}, 111:050503, Jul 2013.
\newblock URL: \url{http://link.aps.org/doi/10.1103/PhysRevLett.111.050503},
  \href {http://dx.doi.org/10.1103/PhysRevLett.111.050503}
  {\path{doi:10.1103/PhysRevLett.111.050503}}.

\bibitem{Shim2016}
Yun-Pil Shim and Charles Tahan.
\newblock Charge-noise-insensitive gate operations for always-on, exchange-only
  qubits.
\newblock {\em Phys. Rev. B}, 93:121410, Mar 2016.
\newblock URL: \url{http://link.aps.org/doi/10.1103/PhysRevB.93.121410}, \href
  {http://dx.doi.org/10.1103/PhysRevB.93.121410}
  {\path{doi:10.1103/PhysRevB.93.121410}}.

\bibitem{Russ2015b}
Maximilian Russ and Guido Burkard.
\newblock Long distance coupling of resonant exchange qubits.
\newblock {\em Phys. Rev. B}, 92:205412, Nov 2015.
\newblock URL: \url{http://link.aps.org/doi/10.1103/PhysRevB.92.205412}, \href
  {http://dx.doi.org/10.1103/PhysRevB.92.205412}
  {\path{doi:10.1103/PhysRevB.92.205412}}.

\bibitem{Srinivasa2016}
V.~Srinivasa, J.~M. Taylor, and Charles Tahan.
\newblock Entangling distant resonant exchange qubits via circuit quantum
  electrodynamics.
\newblock {\em Phys. Rev. B}, 94:205421, Nov 2016.
\newblock URL: \url{http://link.aps.org/doi/10.1103/PhysRevB.94.205421}, \href
  {http://dx.doi.org/10.1103/PhysRevB.94.205421}
  {\path{doi:10.1103/PhysRevB.94.205421}}.

\bibitem{Kempe2001}
J.~Kempe, D.~Bacon, D.~A. Lidar, and K.~B. Whaley.
\newblock Theory of decoherence-free fault-tolerant universal quantum
  computation.
\newblock {\em Phys. Rev. A}, 63:042307, Mar 2001.
\newblock URL: \url{http://link.aps.org/doi/10.1103/PhysRevA.63.042307}, \href
  {http://dx.doi.org/10.1103/PhysRevA.63.042307}
  {\path{doi:10.1103/PhysRevA.63.042307}}.

\bibitem{Russ2015}
Maximilian Russ and Guido Burkard.
\newblock Asymmetric resonant exchange qubit under the influence of electrical
  noise.
\newblock {\em Phys. Rev. B}, 91:235411, Jun 2015.
\newblock URL: \url{http://link.aps.org/doi/10.1103/PhysRevB.91.235411}, \href
  {http://dx.doi.org/10.1103/PhysRevB.91.235411}
  {\path{doi:10.1103/PhysRevB.91.235411}}.

\bibitem{Ladd2012}
Thaddeus~D. Ladd.
\newblock Hyperfine-induced decay in triple quantum dots.
\newblock {\em Phys. Rev. B}, 86:125408, Sep 2012.
\newblock URL: \url{http://link.aps.org/doi/10.1103/PhysRevB.86.125408}, \href
  {http://dx.doi.org/10.1103/PhysRevB.86.125408}
  {\path{doi:10.1103/PhysRevB.86.125408}}.

\bibitem{Mehl2013}
S.~{Mehl} and D.~P. {DiVincenzo}.
\newblock Noise analysis of qubits implemented in triple quantum dot systems in
  a davies master equation approach.
\newblock {\em Phys. Rev. B}, 87(19):195309, May 2013.
\newblock \href {http://dx.doi.org/10.1103/PhysRevB.87.195309}
  {\path{doi:10.1103/PhysRevB.87.195309}}.

\bibitem{Hung2014}
Jo-Tzu Hung, Jianjia Fei, Mark Friesen, and Xuedong Hu.
\newblock Decoherence of an exchange qubit by hyperfine interaction.
\newblock {\em Phys. Rev. B}, 90:045308, Jul 2014.
\newblock URL: \url{http://link.aps.org/doi/10.1103/PhysRevB.90.045308}, \href
  {http://dx.doi.org/10.1103/PhysRevB.90.045308}
  {\path{doi:10.1103/PhysRevB.90.045308}}.

\bibitem{Fei2015}
Jianjia Fei, Jo-Tzu Hung, Teck~Seng Koh, Yun-Pil Shim, S.~N. Coppersmith,
  Xuedong Hu, and Mark Friesen.
\newblock Characterizing gate operations near the sweet spot of an
  exchange-only qubit.
\newblock {\em Phys. Rev. B}, 91:205434, May 2015.
\newblock URL: \url{http://link.aps.org/doi/10.1103/PhysRevB.91.205434}, \href
  {http://dx.doi.org/10.1103/PhysRevB.91.205434}
  {\path{doi:10.1103/PhysRevB.91.205434}}.

\bibitem{West2012}
J~R West and B~H Fong.
\newblock Exchange-only dynamical decoupling in the three-qubit decoherence
  free subsystem.
\newblock {\em New Journal of Physics}, 14(8):083002, 2012.
\newblock URL: \url{http://stacks.iop.org/1367-2630/14/i=8/a=083002}.

\bibitem{Rohling2016}
Niklas Rohling and Guido Burkard.
\newblock Optimizing electrically controlled echo sequences for the
  exchange-only qubit.
\newblock {\em Phys. Rev. B}, 93:205434, May 2016.
\newblock URL: \url{http://link.aps.org/doi/10.1103/PhysRevB.93.205434}, \href
  {http://dx.doi.org/10.1103/PhysRevB.93.205434}
  {\path{doi:10.1103/PhysRevB.93.205434}}.

\bibitem{nielsen2000}
M.A. Nielsen and I.L. Chuang.
\newblock {\em Quantum Computation and Quantum Information}.
\newblock Cambridge Series on Information and the Natural Sciences. Cambridge
  University Press, 2000.
\newblock URL: \url{http://books.google.de/books?id=65FqEKQOfP8C}.

\bibitem{Feynman1982}
Richard~P. Feynman.
\newblock Simulating physics with computers.
\newblock {\em International Journal of Theoretical Physics}, 21(6):467--488,
  1982.
\newblock URL: \url{http://dx.doi.org/10.1007/BF02650179}, \href
  {http://dx.doi.org/10.1007/BF02650179} {\path{doi:10.1007/BF02650179}}.

\bibitem{Lloyd1996}
Seth Lloyd.
\newblock Universal quantum simulators.
\newblock {\em Science}, 273(5278):1073--1078, 1996.
\newblock URL: \url{http://science.sciencemag.org/content/273/5278/1073}, \href
  {http://arxiv.org/abs/http://science.sciencemag.org/content/273/5278/1073.full.pdf}
  {\path{arXiv:http://science.sciencemag.org/content/273/5278/1073.full.pdf}},
  \href {http://dx.doi.org/10.1126/science.273.5278.1073}
  {\path{doi:10.1126/science.273.5278.1073}}.

\bibitem{Shor1997}
Peter~W. Shor.
\newblock Polynomial-time algorithms for prime factorization and discrete
  logarithms on a quantum computer.
\newblock {\em SIAM Journal on Computing}, 26(5):1484, 1997.
\newblock URL: \url{http://dx.doi.org/10.1137/S0097539795293172}, \href
  {http://dx.doi.org/10.1137/S0097539795293172}
  {\path{doi:10.1137/S0097539795293172}}.

\bibitem{Rivest1978}
R.~L. Rivest, A.~Shamir, and L.~Adleman.
\newblock A method for obtaining digital signatures and public-key
  cryptosystems.
\newblock {\em Commun. ACM}, 21(2):120, February 1978.
\newblock URL: \url{http://doi.acm.org/10.1145/359340.359342}, \href
  {http://dx.doi.org/10.1145/359340.359342} {\path{doi:10.1145/359340.359342}}.

\bibitem{Grover1996}
Lov~K. Grover.
\newblock A fast quantum mechanical algorithm for database search.
\newblock In {\em Proceedings of the Twenty-eighth Annual ACM Symposium on
  Theory of Computing}, STOC '96, page 212, New York, NY, USA, 1996. ACM.
\newblock URL: \url{http://doi.acm.org/10.1145/237814.237866}, \href
  {http://dx.doi.org/10.1145/237814.237866} {\path{doi:10.1145/237814.237866}}.

\bibitem{Grover1998}
Lov~K. Grover.
\newblock Quantum computers can search rapidly by using almost any
  transformation.
\newblock {\em Phys. Rev. Lett.}, 80:4329--4332, May 1998.
\newblock URL: \url{http://link.aps.org/doi/10.1103/PhysRevLett.80.4329}, \href
  {http://dx.doi.org/10.1103/PhysRevLett.80.4329}
  {\path{doi:10.1103/PhysRevLett.80.4329}}.

\bibitem{Lloyd1993}
Seth Lloyd.
\newblock A potentially realizable quantum computer.
\newblock {\em Science}, 261(5128):1569--1571, 1993.
\newblock URL: \url{http://science.sciencemag.org/content/261/5128/1569}, \href
  {http://arxiv.org/abs/http://science.sciencemag.org/content/261/5128/1569.full.pdf}
  {\path{arXiv:http://science.sciencemag.org/content/261/5128/1569.full.pdf}},
  \href {http://dx.doi.org/10.1126/science.261.5128.1569}
  {\path{doi:10.1126/science.261.5128.1569}}.

\bibitem{Kloeffel2013}
Christoph Kloeffel and Daniel Loss.
\newblock Prospects for spin-based quantum computing in quantum dots.
\newblock {\em Annual Review of Condensed Matter Physics}, 4(1):51--81, 2013.
\newblock URL:
  \url{http://dx.doi.org/10.1146/annurev-conmatphys-030212-184248}, \href
  {http://dx.doi.org/10.1146/annurev-conmatphys-030212-184248}
  {\path{doi:10.1146/annurev-conmatphys-030212-184248}}.

\bibitem{Vandersypen2005}
L.~M.~K. Vandersypen and I.~L. Chuang.
\newblock Nmr techniques for quantum control and computation.
\newblock {\em Rev. Mod. Phys.}, 76:1037--1069, Jan 2005.
\newblock URL: \url{http://link.aps.org/doi/10.1103/RevModPhys.76.1037}, \href
  {http://dx.doi.org/10.1103/RevModPhys.76.1037}
  {\path{doi:10.1103/RevModPhys.76.1037}}.

\bibitem{Miller2005}
R.~Miller, T.~E. Northup, K.~M. Birnbaum, A.~Boca, A.~D. Boozer, and H.~J.
  Kimble.
\newblock Trapped atoms in cavity qed: coupling quantized light and matter.
\newblock {\em Journal of Physics B: Atomic, Molecular and Optical Physics},
  38(9):S551, 2005.
\newblock URL: \url{http://stacks.iop.org/0953-4075/38/i=9/a=007}.

\bibitem{Cirac1995}
J.~I. Cirac and P.~Zoller.
\newblock Quantum computations with cold trapped ions.
\newblock {\em Phys. Rev. Lett.}, 74:4091--4094, May 1995.
\newblock URL: \url{http://link.aps.org/doi/10.1103/PhysRevLett.74.4091}, \href
  {http://dx.doi.org/10.1103/PhysRevLett.74.4091}
  {\path{doi:10.1103/PhysRevLett.74.4091}}.

\bibitem{Nakamura1999}
Y.~Nakamura, Yu.~A. Pashkin, and J.~S. Tsai.
\newblock Coherent control of macroscopic quantum states in a
  single-cooper-pair box.
\newblock {\em Nature}, 398(6730):786--788, 04 1999.
\newblock URL: \url{http://dx.doi.org/10.1038/19718}.

\bibitem{Makhlin2001}
Yuriy Makhlin, Gerd Sch\"on, and Alexander Shnirman.
\newblock Quantum-state engineering with josephson-junction devices.
\newblock {\em Rev. Mod. Phys.}, 73:357--400, May 2001.
\newblock URL: \url{http://link.aps.org/doi/10.1103/RevModPhys.73.357}, \href
  {http://dx.doi.org/10.1103/RevModPhys.73.357}
  {\path{doi:10.1103/RevModPhys.73.357}}.

\bibitem{Kane1998}
B.~E. Kane.
\newblock A silicon-based nuclear spin quantum computer.
\newblock {\em Nature}, 393(6681):133, May 1998.
\newblock URL: \url{http://dx.doi.org/10.1038/30156}.

\bibitem{Doherty2013b}
Marcus~W. Doherty, Neil~B. Manson, Paul Delaney, Fedor Jelezko, J{\"o}rg
  Wrachtrup, and Lloyd C.~L. Hollenberg.
\newblock The nitrogen-vacancy colour centre in diamond.
\newblock {\em Physics Reports}, 528(1):1--45, 7 2013.
\newblock URL:
  \url{http://www.sciencedirect.com/science/article/pii/S0370157313000562},
  \href {http://dx.doi.org/http://dx.doi.org/10.1016/j.physrep.2013.02.001}
  {\path{doi:http://dx.doi.org/10.1016/j.physrep.2013.02.001}}.

\bibitem{Dobrovitski2013}
V.~V. Dobrovitski, G.~D. Fuchs, A.~L. Falk, C.~Santori, and D.~D. Awschalom.
\newblock Quantum control over single spins in diamond.
\newblock {\em Annual Review of Condensed Matter Physics}, 4(1):23--50,
  2016/11/14 2013.
\newblock URL:
  \url{http://dx.doi.org/10.1146/annurev-conmatphys-030212-184238}, \href
  {http://dx.doi.org/10.1146/annurev-conmatphys-030212-184238}
  {\path{doi:10.1146/annurev-conmatphys-030212-184238}}.

\bibitem{Kok2007}
Pieter Kok, W.~J. Munro, Kae Nemoto, T.~C. Ralph, Jonathan~P. Dowling, and
  G.~J. Milburn.
\newblock Linear optical quantum computing with photonic qubits.
\newblock {\em Rev. Mod. Phys.}, 79:135--174, Jan 2007.
\newblock URL: \url{http://link.aps.org/doi/10.1103/RevModPhys.79.135}, \href
  {http://dx.doi.org/10.1103/RevModPhys.79.135}
  {\path{doi:10.1103/RevModPhys.79.135}}.

\bibitem{DiVincenzo2000}
D.~P. {DiVincenzo}.
\newblock {The Physical Implementation of Quantum Computation}.
\newblock {\em Fortschritte der Physik}, 48:771--783, 2000.
\newblock \href {http://arxiv.org/abs/quant-ph/0002077}
  {\path{arXiv:quant-ph/0002077}}, \href
  {http://dx.doi.org/10.1002/1521-3978(200009)48:9/11<771::AID-PROP771>3.0.CO;2-E}
  {\path{doi:10.1002/1521-3978(200009)48:9/11<771::AID-PROP771>3.0.CO;2-E}}.

\bibitem{Delgado2011}
Carlos~A. P\'erez-Delgado and Pieter Kok.
\newblock Quantum computers: Definition and implementations.
\newblock {\em Phys. Rev. A}, 83:012303, Jan 2011.
\newblock URL: \url{http://link.aps.org/doi/10.1103/PhysRevA.83.012303}, \href
  {http://dx.doi.org/10.1103/PhysRevA.83.012303}
  {\path{doi:10.1103/PhysRevA.83.012303}}.

\bibitem{Martinis2015}
John~M Martinis.
\newblock Qubit metrology for building a fault-tolerant quantum computer.
\newblock {\em Npj Quantum Information}, 1:15005 EP --, 10 2015.
\newblock URL: \url{http://dx.doi.org/10.1038/npjqi.2015.5}.

\bibitem{Bejanin2016}
J.~H. B\'ejanin, T.~G. McConkey, J.~R. Rinehart, C.~T. Earnest, C.~R.~H. McRae,
  D.~Shiri, J.~D. Bateman, Y.~Rohanizadegan, B.~Penava, P.~Breul, S.~Royak,
  M.~Zapatka, A.~G. Fowler, and M.~Mariantoni.
\newblock Three-dimensional wiring for extensible quantum computing: The
  quantum socket.
\newblock {\em Phys. Rev. Applied}, 6:044010, Oct 2016.
\newblock URL: \url{http://link.aps.org/doi/10.1103/PhysRevApplied.6.044010},
  \href {http://dx.doi.org/10.1103/PhysRevApplied.6.044010}
  {\path{doi:10.1103/PhysRevApplied.6.044010}}.

\bibitem{Petta2005}
J.~R. Petta, A.~C. Johnson, J.~M. Taylor, E.~A. Laird, A.~Yacoby, M.~D. Lukin,
  C.~M. Marcus, M.~P. Hanson, and A.~C. Gossard.
\newblock Coherent manipulation of coupled electron spins in semiconductor
  quantum dots.
\newblock {\em Science}, 309(5744):2180, 2005.
\newblock URL: \url{http://www.sciencemag.org/content/309/5744/2180.abstract},
  \href {http://dx.doi.org/10.1126/science.1116955}
  {\path{doi:10.1126/science.1116955}}.

\bibitem{Hanson2007}
R.~Hanson, L.~P. Kouwenhoven, J.~R. Petta, S.~Tarucha, and L.~M.~K.
  Vandersypen.
\newblock Spins in few-electron quantum dots.
\newblock {\em Rev. Mod. Phys.}, 79:1217, Oct 2007.
\newblock URL: \url{http://link.aps.org/doi/10.1103/RevModPhys.79.1217}, \href
  {http://dx.doi.org/10.1103/RevModPhys.79.1217}
  {\path{doi:10.1103/RevModPhys.79.1217}}.

\bibitem{Hubbard1963}
J.~Hubbard.
\newblock Electron correlations in narrow energy bands.
\newblock {\em Proceedings of the Royal Society of London. Series A.
  Mathematical and Physical Sciences}, 276(1365):238, 1963.
\newblock \href {http://dx.doi.org/10.1098/rspa.1963.0204}
  {\path{doi:10.1098/rspa.1963.0204}}.

\bibitem{Burkard1999}
Guido Burkard, Daniel Loss, and David~P. DiVincenzo.
\newblock Coupled quantum dots as quantum gates.
\newblock {\em Phys. Rev. B}, 59:2070, Jan 1999.
\newblock URL: \url{http://link.aps.org/doi/10.1103/PhysRevB.59.2070}, \href
  {http://dx.doi.org/10.1103/PhysRevB.59.2070}
  {\path{doi:10.1103/PhysRevB.59.2070}}.

\bibitem{Schrieffer1966}
J.~R. {Schrieffer} and P.~A. {Wolff}.
\newblock {Relation between the Anderson and Kondo Hamiltonians}.
\newblock {\em Physical Review}, 149:491--492, September 1966.
\newblock \href {http://dx.doi.org/10.1103/PhysRev.149.491}
  {\path{doi:10.1103/PhysRev.149.491}}.

\bibitem{Bravyi2011}
S.~{Bravyi}, D.~P. {Divincenzo}, and D.~{Loss}.
\newblock Schrieffer-wolff transformation for quantum many-body systems.
\newblock {\em Annals of Physics}, 326:2793, October 2011.
\newblock \href {http://arxiv.org/abs/1105.0675} {\path{arXiv:1105.0675}},
  \href {http://dx.doi.org/10.1016/j.aop.2011.06.004}
  {\path{doi:10.1016/j.aop.2011.06.004}}.

\bibitem{Loss1998}
Daniel Loss and David~P. DiVincenzo.
\newblock Quantum computation with quantum dots.
\newblock {\em Phys. Rev. A}, 57:120, Jan 1998.
\newblock URL: \url{http://link.aps.org/doi/10.1103/PhysRevA.57.120}, \href
  {http://dx.doi.org/10.1103/PhysRevA.57.120}
  {\path{doi:10.1103/PhysRevA.57.120}}.

\bibitem{Koppens2006}
F.~H.~L. Koppens, C.~Buizert, K.~J. Tielrooij, I.~T. Vink, K.~C. Nowack,
  T.~Meunier, L.~P. Kouwenhoven, and L.~M.~K. Vandersypen.
\newblock Driven coherent oscillations of a single electron spin in a quantum
  dot.
\newblock {\em Nature}, 442(7104):766, August 2006.
\newblock URL: \url{http://dx.doi.org/10.1038/nature05065}.

\bibitem{Golovach2006}
Vitaly~N. Golovach, Massoud Borhani, and Daniel Loss.
\newblock Electric-dipole-induced spin resonance in quantum dots.
\newblock {\em Phys. Rev. B}, 74:165319, Oct 2006.
\newblock URL: \url{http://link.aps.org/doi/10.1103/PhysRevB.74.165319}, \href
  {http://dx.doi.org/10.1103/PhysRevB.74.165319}
  {\path{doi:10.1103/PhysRevB.74.165319}}.

\bibitem{Brunner2011}
R.~Brunner, Y.-S. Shin, T.~Obata, M.~Pioro-Ladri\`ere, T.~Kubo, K.~Yoshida,
  T.~Taniyama, Y.~Tokura, and S.~Tarucha.
\newblock Two-qubit gate of combined single-spin rotation and interdot spin
  exchange in a double quantum dot.
\newblock {\em Phys. Rev. Lett.}, 107:146801, Sep 2011.
\newblock URL: \url{http://link.aps.org/doi/10.1103/PhysRevLett.107.146801},
  \href {http://dx.doi.org/10.1103/PhysRevLett.107.146801}
  {\path{doi:10.1103/PhysRevLett.107.146801}}.

\bibitem{Obata2010}
Toshiaki Obata, Michel Pioro-Ladri\`ere, Yasuhiro Tokura, Yun-Sok Shin,
  Toshihiro Kubo, Katsuharu Yoshida, Tomoyasu Taniyama, and Seigo Tarucha.
\newblock Coherent manipulation of individual electron spin in a double quantum
  dot integrated with a micromagnet.
\newblock {\em Phys. Rev. B}, 81:085317, Feb 2010.
\newblock URL: \url{http://link.aps.org/doi/10.1103/PhysRevB.81.085317}, \href
  {http://dx.doi.org/10.1103/PhysRevB.81.085317}
  {\path{doi:10.1103/PhysRevB.81.085317}}.

\bibitem{Kawakami2014}
Kawakami E., Scarlino P., Ward~D. R., Braakman~F. R., Savage~D. E., Lagally~M.
  G., Mark Friesen, Coppersmith~S. N., Eriksson~M. A., and Vandersypen L.~M. K.
\newblock Electrical control of a long-lived spin qubit in a si/sige quantum
  dot.
\newblock {\em Nat Nano}, 9(9):666--670, 09 2014.
\newblock URL: \url{http://dx.doi.org/10.1038/nnano.2014.153}.

\bibitem{Khomitsky2012}
D.~V. Khomitsky, L.~V. Gulyaev, and E.~Ya. Sherman.
\newblock Spin dynamics in a strongly driven system: Very slow rabi
  oscillations.
\newblock {\em Phys. Rev. B}, 85:125312, Mar 2012.
\newblock URL: \url{http://link.aps.org/doi/10.1103/PhysRevB.85.125312}, \href
  {http://dx.doi.org/10.1103/PhysRevB.85.125312}
  {\path{doi:10.1103/PhysRevB.85.125312}}.

\bibitem{Li2013}
Rui Li, J.~Q. You, C.~P. Sun, and Franco Nori.
\newblock Controlling a nanowire spin-orbit qubit via electric-dipole spin
  resonance.
\newblock {\em Phys. Rev. Lett.}, 111:086805, Aug 2013.
\newblock URL: \url{http://link.aps.org/doi/10.1103/PhysRevLett.111.086805},
  \href {http://dx.doi.org/10.1103/PhysRevLett.111.086805}
  {\path{doi:10.1103/PhysRevLett.111.086805}}.

\bibitem{Nadj2010}
S.~Nadj-Perge, S.~M. Frolov, E.~P. A.~M. Bakkers, and L.~P. Kouwenhoven.
\newblock Spin-orbit qubit in a semiconductor nanowire.
\newblock {\em Nature}, 468(7327):1084--1087, 12 2010.
\newblock URL: \url{http://dx.doi.org/10.1038/nature09682}.

\bibitem{Tokura2006}
Yasuhiro Tokura, Wilfred~G. van~der Wiel, Toshiaki Obata, and Seigo Tarucha.
\newblock Coherent single electron spin control in a slanting zeeman field.
\newblock {\em Phys. Rev. Lett.}, 96:047202, Jan 2006.
\newblock URL: \url{http://link.aps.org/doi/10.1103/PhysRevLett.96.047202},
  \href {http://dx.doi.org/10.1103/PhysRevLett.96.047202}
  {\path{doi:10.1103/PhysRevLett.96.047202}}.

\bibitem{Nowack2007}
K.~C. Nowack, F.~H.~L. Koppens, Yu.~V. Nazarov, and L.~M.~K. Vandersypen.
\newblock Coherent control of a single electron spin with electric fields.
\newblock {\em Science}, 318(5855):1430, 2007.
\newblock URL: \url{http://www.sciencemag.org/content/318/5855/1430.abstract},
  \href {http://dx.doi.org/10.1126/science.1148092}
  {\path{doi:10.1126/science.1148092}}.

\bibitem{Laird2007}
E.~A. Laird, C.~Barthel, E.~I. Rashba, C.~M. Marcus, M.~P. Hanson, and A.~C.
  Gossard.
\newblock Hyperfine-mediated gate-driven electron spin resonance.
\newblock {\em Phys. Rev. Lett.}, 99:246601, Dec 2007.
\newblock URL: \url{http://link.aps.org/doi/10.1103/PhysRevLett.99.246601},
  \href {http://dx.doi.org/10.1103/PhysRevLett.99.246601}
  {\path{doi:10.1103/PhysRevLett.99.246601}}.

\bibitem{Pioro2008}
M.~Pioro-Ladriere, T.~Obata, Y.~Tokura, Y.~S. Shin, T.~Kubo, K.~Yoshida,
  T.~Taniyama, and S.~Tarucha.
\newblock Electrically driven single-electron spin resonance in a slanting
  zeeman field.
\newblock {\em Nat Phys}, 4(10):776--779, 10 2008.
\newblock URL: \url{http://dx.doi.org/10.1038/nphys1053}.

\bibitem{Foletti2009}
Sandra Foletti, Hendrik Bluhm, Diana Mahalu, Vladimir Umansky, and Amir Yacoby.
\newblock Universal quantum control of two-electron spin quantum bits using
  dynamic nuclear polarization.
\newblock {\em Nat Phys}, 5(12):903, December 2009.
\newblock URL: \url{http://dx.doi.org/10.1038/nphys1424}.

\bibitem{Culcer2009b}
Dimitrie Culcer, Xuedong Hu, and S.~Das~Sarma.
\newblock Dephasing of si spin qubits due to charge noise.
\newblock {\em Applied Physics Letters}, 95(7), 2009.
\newblock URL:
  \url{http://scitation.aip.org/content/aip/journal/apl/95/7/10.1063/1.3194778},
  \href {http://dx.doi.org/http://dx.doi.org/10.1063/1.3194778}
  {\path{doi:http://dx.doi.org/10.1063/1.3194778}}.

\bibitem{Vion2002}
D.~Vion, A.~Aassime, A.~Cottet, P.~Joyez, H.~Pothier, C.~Urbina, D.~Esteve, and
  M.~H. Devoret.
\newblock Manipulating the quantum state of an electrical circuit.
\newblock {\em Science}, 296(5569):886, 2002.
\newblock URL: \url{http://www.sciencemag.org/content/296/5569/886.abstract},
  \href {http://dx.doi.org/10.1126/science.1069372}
  {\path{doi:10.1126/science.1069372}}.

\bibitem{Paladino2014}
E.~Paladino, Y.~M. Galperin, G.~Falci, and B.~L. Altshuler.
\newblock $1/f$ noise: Implications for solid-state quantum information.
\newblock {\em Rev. Mod. Phys.}, 86:361--418, Apr 2014.
\newblock URL: \url{http://link.aps.org/doi/10.1103/RevModPhys.86.361}, \href
  {http://dx.doi.org/10.1103/RevModPhys.86.361}
  {\path{doi:10.1103/RevModPhys.86.361}}.

\bibitem{Gong2016}
Bo~Gong, Li~Wang, Tao Tu, Chuan-Feng Li, and Guang-Can Guo.
\newblock Robust universal gates for quantum-dot spin qubits using tunable
  adiabatic passages.
\newblock {\em Phys. Rev. A}, 94:032311, Sep 2016.
\newblock URL: \url{http://link.aps.org/doi/10.1103/PhysRevA.94.032311}, \href
  {http://dx.doi.org/10.1103/PhysRevA.94.032311}
  {\path{doi:10.1103/PhysRevA.94.032311}}.

\bibitem{Chirolli2008}
Luca Chirolli and Guido Burkard.
\newblock Decoherence in solid-state qubits.
\newblock {\em Advances in Physics}, 57(3):225--285, 2008.
\newblock URL: \url{http://dx.doi.org/10.1080/00018730802218067}, \href
  {http://arxiv.org/abs/http://dx.doi.org/10.1080/00018730802218067}
  {\path{arXiv:http://dx.doi.org/10.1080/00018730802218067}}, \href
  {http://dx.doi.org/10.1080/00018730802218067}
  {\path{doi:10.1080/00018730802218067}}.

\bibitem{Merkulov2002}
I.~A. Merkulov, Al.~L. Efros, and M.~Rosen.
\newblock Electron spin relaxation by nuclei in semiconductor quantum dots.
\newblock {\em Phys. Rev. B}, 65:205309, Apr 2002.
\newblock URL: \url{http://link.aps.org/doi/10.1103/PhysRevB.65.205309}, \href
  {http://dx.doi.org/10.1103/PhysRevB.65.205309}
  {\path{doi:10.1103/PhysRevB.65.205309}}.

\bibitem{Coish2004}
W.~A. Coish and Daniel Loss.
\newblock Hyperfine interaction in a quantum dot: Non-markovian electron spin
  dynamics.
\newblock {\em Phys. Rev. B}, 70:195340, Nov 2004.
\newblock URL: \url{http://link.aps.org/doi/10.1103/PhysRevB.70.195340}, \href
  {http://dx.doi.org/10.1103/PhysRevB.70.195340}
  {\path{doi:10.1103/PhysRevB.70.195340}}.

\bibitem{Fischer2009}
Jan Fischer, Mircea Trif, W.A. Coish, and Daniel Loss.
\newblock Spin interactions, relaxation and decoherence in quantum dots.
\newblock {\em Solid State Communications}, 149(35--36):1443 -- 1450, 2009.
\newblock Fundamental Phenomena and Applications of Quantum Dots.
\newblock URL:
  \url{http://www.sciencedirect.com/science/article/pii/S0038109809002610},
  \href {http://dx.doi.org/http://dx.doi.org/10.1016/j.ssc.2009.04.033}
  {\path{doi:http://dx.doi.org/10.1016/j.ssc.2009.04.033}}.

\bibitem{Coish2009}
W.~A. Coish and J.~Baugh.
\newblock Nuclear spins in nanostructures.
\newblock {\em physica status solidi (b)}, 246(10):2203--2215, 2009.
\newblock URL: \url{http://dx.doi.org/10.1002/pssb.200945229}, \href
  {http://dx.doi.org/10.1002/pssb.200945229}
  {\path{doi:10.1002/pssb.200945229}}.

\bibitem{Khaetskii2001}
Alexander~V. Khaetskii and Yuli~V. Nazarov.
\newblock Spin-flip transitions between zeeman sublevels in semiconductor
  quantum dots.
\newblock {\em Phys. Rev. B}, 64:125316, Sep 2001.
\newblock URL: \url{http://link.aps.org/doi/10.1103/PhysRevB.64.125316}, \href
  {http://dx.doi.org/10.1103/PhysRevB.64.125316}
  {\path{doi:10.1103/PhysRevB.64.125316}}.

\bibitem{Golovach2004}
Vitaly~N. Golovach, Alexander Khaetskii, and Daniel Loss.
\newblock Phonon-induced decay of the electron spin in quantum dots.
\newblock {\em Phys. Rev. Lett.}, 93:016601, Jun 2004.
\newblock URL: \url{http://link.aps.org/doi/10.1103/PhysRevLett.93.016601},
  \href {http://dx.doi.org/10.1103/PhysRevLett.93.016601}
  {\path{doi:10.1103/PhysRevLett.93.016601}}.

\bibitem{Bulaev2005}
Denis~V. Bulaev and Daniel Loss.
\newblock Spin relaxation and anticrossing in quantum dots: Rashba versus
  dresselhaus spin-orbit coupling.
\newblock {\em Phys. Rev. B}, 71:205324, May 2005.
\newblock URL: \url{http://link.aps.org/doi/10.1103/PhysRevB.71.205324}, \href
  {http://dx.doi.org/10.1103/PhysRevB.71.205324}
  {\path{doi:10.1103/PhysRevB.71.205324}}.

\bibitem{Falko2005}
Vladimir~I. Fal'ko, B.~L. Altshuler, and O.~Tsyplyatyev.
\newblock Anisotropy of spin splitting and spin relaxation in lateral quantum
  dots.
\newblock {\em Phys. Rev. Lett.}, 95:076603, Aug 2005.
\newblock URL: \url{http://link.aps.org/doi/10.1103/PhysRevLett.95.076603},
  \href {http://dx.doi.org/10.1103/PhysRevLett.95.076603}
  {\path{doi:10.1103/PhysRevLett.95.076603}}.

\bibitem{Prada2016}
M.~{Prada} and D.~{Pfannkuche}.
\newblock {Anisotropy of spin coherence in high mobility quantum wells with
  arbitrary magnetic fields}.
\newblock {\em ArXiv e-prints}, May 2016.
\newblock \href {http://arxiv.org/abs/1605.03399} {\path{arXiv:1605.03399}}.

\bibitem{Amasha2008}
S.~Amasha, K.~MacLean, Iuliana~P. Radu, D.~M. Zumb\"uhl, M.~A. Kastner, M.~P.
  Hanson, and A.~C. Gossard.
\newblock Electrical control of spin relaxation in a quantum dot.
\newblock {\em Phys. Rev. Lett.}, 100:046803, Jan 2008.
\newblock URL: \url{http://link.aps.org/doi/10.1103/PhysRevLett.100.046803},
  \href {http://dx.doi.org/10.1103/PhysRevLett.100.046803}
  {\path{doi:10.1103/PhysRevLett.100.046803}}.

\bibitem{Zwanenburg2013}
Floris~A. Zwanenburg, Andrew~S. Dzurak, Andrea Morello, Michelle~Y. Simmons,
  Lloyd C.~L. Hollenberg, Gerhard Klimeck, Sven Rogge, Susan~N. Coppersmith,
  and Mark~A. Eriksson.
\newblock Silicon quantum electronics.
\newblock {\em Rev. Mod. Phys.}, 85:961, Jul 2013.
\newblock URL: \url{http://link.aps.org/doi/10.1103/RevModPhys.85.961}, \href
  {http://dx.doi.org/10.1103/RevModPhys.85.961}
  {\path{doi:10.1103/RevModPhys.85.961}}.

\bibitem{Levy2002}
Jeremy Levy.
\newblock Universal quantum computation with spin-$1/2$ pairs and heisenberg
  exchange.
\newblock {\em Phys. Rev. Lett.}, 89:147902, Sep 2002.
\newblock URL: \url{http://link.aps.org/doi/10.1103/PhysRevLett.89.147902},
  \href {http://dx.doi.org/10.1103/PhysRevLett.89.147902}
  {\path{doi:10.1103/PhysRevLett.89.147902}}.

\bibitem{Barenco1995}
Adriano Barenco, David Deutsch, Artur Ekert, and Richard Jozsa.
\newblock Conditional quantum dynamics and logic gates.
\newblock {\em Phys. Rev. Lett.}, 74:4083, May 1995.
\newblock URL: \url{http://link.aps.org/doi/10.1103/PhysRevLett.74.4083}, \href
  {http://dx.doi.org/10.1103/PhysRevLett.74.4083}
  {\path{doi:10.1103/PhysRevLett.74.4083}}.

\bibitem{Klinovaja2012}
Jelena Klinovaja, Dimitrije Stepanenko, Bertrand~I. Halperin, and Daniel Loss.
\newblock Exchange-based cnot gates for singlet-triplet qubits with spin-orbit
  interaction.
\newblock {\em Phys. Rev. B}, 86(8):085423, August 2012.
\newblock URL: \url{http://link.aps.org/doi/10.1103/PhysRevB.86.085423}.

\bibitem{Wardrop2014}
Matthew~P. Wardrop and Andrew~C. Doherty.
\newblock Exchange-based two-qubit gate for singlet-triplet qubits.
\newblock {\em Phys. Rev. B}, 90:045418, Jul 2014.
\newblock URL: \url{http://link.aps.org/doi/10.1103/PhysRevB.90.045418}, \href
  {http://dx.doi.org/10.1103/PhysRevB.90.045418}
  {\path{doi:10.1103/PhysRevB.90.045418}}.

\bibitem{Taylor2005}
J.~M. Taylor, H.-A. Engel, W.~Dur, A.~Yacoby, C.~M. Marcus, P.~Zoller, and
  M.~D. Lukin.
\newblock Fault-tolerant architecture for quantum computation using
  electrically controlled semiconductor spins.
\newblock {\em Nat Phys}, 1(3):177, December 2005.
\newblock URL: \url{http://dx.doi.org/10.1038/nphys174}.

\bibitem{Coish2005}
W.~A. Coish and Daniel Loss.
\newblock Singlet-triplet decoherence due to nuclear spins in a double quantum
  dot.
\newblock {\em Phys. Rev. B}, 72:125337, Sep 2005.
\newblock URL: \url{http://link.aps.org/doi/10.1103/PhysRevB.72.125337}, \href
  {http://dx.doi.org/10.1103/PhysRevB.72.125337}
  {\path{doi:10.1103/PhysRevB.72.125337}}.

\bibitem{Stepanenko2007}
Dimitrije Stepanenko and Guido Burkard.
\newblock Quantum gates between capacitively coupled double quantum dot
  two-spin qubits.
\newblock {\em Phys. Rev. B}, 75:085324, Feb 2007.
\newblock URL: \url{http://link.aps.org/doi/10.1103/PhysRevB.75.085324}, \href
  {http://dx.doi.org/10.1103/PhysRevB.75.085324}
  {\path{doi:10.1103/PhysRevB.75.085324}}.

\bibitem{Calderon2015}
F.~A. Calderon-Vargas and J.~P. Kestner.
\newblock Directly accessible entangling gates for capacitively coupled
  singlet-triplet qubits.
\newblock {\em Phys. Rev. B}, 91:035301, Jan 2015.
\newblock URL: \url{http://link.aps.org/doi/10.1103/PhysRevB.91.035301}, \href
  {http://dx.doi.org/10.1103/PhysRevB.91.035301}
  {\path{doi:10.1103/PhysRevB.91.035301}}.

\bibitem{Mehl2014}
Sebastian Mehl, Hendrik Bluhm, and David~P. DiVincenzo.
\newblock Two-qubit couplings of singlet-triplet qubits mediated by one quantum
  state.
\newblock {\em Phys. Rev. B}, 90:045404, Jul 2014.
\newblock URL: \url{http://link.aps.org/doi/10.1103/PhysRevB.90.045404}, \href
  {http://dx.doi.org/10.1103/PhysRevB.90.045404}
  {\path{doi:10.1103/PhysRevB.90.045404}}.

\bibitem{Taylor2007}
J.~M. Taylor, J.~R. Petta, A.~C. Johnson, A.~Yacoby, C.~M. Marcus, and M.~D.
  Lukin.
\newblock Relaxation, dephasing, and quantum control of electron spins in
  double quantum dots.
\newblock {\em Phys. Rev. B}, 76:035315, Jul 2007.
\newblock URL: \url{http://link.aps.org/doi/10.1103/PhysRevB.76.035315}, \href
  {http://dx.doi.org/10.1103/PhysRevB.76.035315}
  {\path{doi:10.1103/PhysRevB.76.035315}}.

\bibitem{Shulman2012}
M.~D. Shulman, O.~E. Dial, S.~P. Harvey, H.~Bluhm, V.~Umansky, and A.~Yacoby.
\newblock Demonstration of entanglement of electrostatically coupled
  singlet-triplet qubits.
\newblock {\em Science}, 336(6078):202, 2012.
\newblock URL: \url{http://www.sciencemag.org/content/336/6078/202.abstract},
  \href {http://dx.doi.org/10.1126/science.1217692}
  {\path{doi:10.1126/science.1217692}}.

\bibitem{Srinivasa2015}
V.~Srinivasa and J.~M. Taylor.
\newblock Capacitively coupled singlet-triplet qubits in the double charge
  resonant regime.
\newblock {\em Phys. Rev. B}, 92:235301, Dec 2015.
\newblock URL: \url{http://link.aps.org/doi/10.1103/PhysRevB.92.235301}, \href
  {http://dx.doi.org/10.1103/PhysRevB.92.235301}
  {\path{doi:10.1103/PhysRevB.92.235301}}.

\bibitem{Imamoglu1999}
A.~Imamoglu, D.~D. Awschalom, G.~Burkard, D.~P. DiVincenzo, D.~Loss,
  M.~Sherwin, and A.~Small.
\newblock Quantum information processing using quantum dot spins and cavity
  qed.
\newblock {\em Phys. Rev. Lett.}, 83:4204, Nov 1999.
\newblock URL: \url{http://link.aps.org/doi/10.1103/PhysRevLett.83.4204}, \href
  {http://dx.doi.org/10.1103/PhysRevLett.83.4204}
  {\path{doi:10.1103/PhysRevLett.83.4204}}.

\bibitem{Childress2004}
L.~Childress, A.~S. S\o{}rensen, and M.~D. Lukin.
\newblock Mesoscopic cavity quantum electrodynamics with quantum dots.
\newblock {\em Phys. Rev. A}, 69:042302, Apr 2004.
\newblock URL: \url{http://link.aps.org/doi/10.1103/PhysRevA.69.042302}, \href
  {http://dx.doi.org/10.1103/PhysRevA.69.042302}
  {\path{doi:10.1103/PhysRevA.69.042302}}.

\bibitem{Burkard2006}
Guido Burkard and Atac Imamoglu.
\newblock Ultra-long-distance interaction between spin qubits.
\newblock {\em Phys. Rev. B}, 74:041307, Jul 2006.
\newblock URL: \url{http://link.aps.org/doi/10.1103/PhysRevB.74.041307}, \href
  {http://dx.doi.org/10.1103/PhysRevB.74.041307}
  {\path{doi:10.1103/PhysRevB.74.041307}}.

\bibitem{Taylor2006}
J.M. Taylor and M.D. Lukin.
\newblock Dephasing of quantum bits by a quasi-static mesoscopic environment.
\newblock {\em Quantum Information Processing}, 5(6):503, 2006.
\newblock URL: \url{http://dx.doi.org/10.1007/s11128-006-0036-z}, \href
  {http://dx.doi.org/10.1007/s11128-006-0036-z}
  {\path{doi:10.1007/s11128-006-0036-z}}.

\bibitem{Hu2012}
Xuedong Hu, Yu-xi Liu, and Franco Nori.
\newblock Strong coupling of a spin qubit to a superconducting stripline
  cavity.
\newblock {\em Phys. Rev. B}, 86:035314, Jul 2012.
\newblock URL: \url{http://link.aps.org/doi/10.1103/PhysRevB.86.035314}, \href
  {http://dx.doi.org/10.1103/PhysRevB.86.035314}
  {\path{doi:10.1103/PhysRevB.86.035314}}.

\bibitem{Petersson2012}
K.~D. Petersson, L.~W. McFaul, M.~D. Schroer, M.~Jung, J.~M. Taylor, A.~A.
  Houck, and J.~R. Petta.
\newblock Circuit quantum electrodynamics with a spin qubit.
\newblock {\em Nature}, 490(7420):380, Oct 2012.
\newblock URL: \url{http://dx.doi.org/10.1038/nature11559}, \href
  {http://dx.doi.org/10.1038/nature11559} {\path{doi:10.1038/nature11559}}.

\bibitem{Viennot2015}
J.~J. Viennot, M.~C. Dartiailh, A.~Cottet, and T.~Kontos.
\newblock Coherent coupling of a single spin to microwave cavity photons.
\newblock {\em Science}, 349(6246):408--411, 2015.
\newblock URL: \url{http://science.sciencemag.org/content/349/6246/408}, \href
  {http://dx.doi.org/10.1126/science.aaa3786}
  {\path{doi:10.1126/science.aaa3786}}.

\bibitem{Guilherme2014}
Guilherme Tosi, Fahd~A. Mohiyaddin, Hans Huebl, and Andrea Morello.
\newblock Circuit-quantum electrodynamics with direct magnetic coupling to
  single-atom spin qubits in isotopically enriched 28si.
\newblock {\em AIP Advances}, 4(8), 2014.
\newblock URL:
  \url{http://scitation.aip.org/content/aip/journal/adva/4/8/10.1063/1.4893242},
  \href {http://dx.doi.org/10.1063/1.4893242} {\path{doi:10.1063/1.4893242}}.

\bibitem{Samkharadze2016}
N.~Samkharadze, A.~Bruno, P.~Scarlino, G.~Zheng, D.~P. DiVincenzo, L.~DiCarlo,
  and L.~M.~K. Vandersypen.
\newblock High-kinetic-inductance superconducting nanowire resonators for
  circuit qed in a magnetic field.
\newblock {\em Phys. Rev. Applied}, 5:044004, Apr 2016.
\newblock URL: \url{http://link.aps.org/doi/10.1103/PhysRevApplied.5.044004},
  \href {http://dx.doi.org/10.1103/PhysRevApplied.5.044004}
  {\path{doi:10.1103/PhysRevApplied.5.044004}}.

\bibitem{Kontos2016}
T.~Kontos.
\newblock Cavity quantum electrodynamics with carbon nanotubes: From
  atomic-like systems to condensed matter (invited).
\newblock {\em Invited talk, ICPS 2016, Beijing}, 2016.

\bibitem{Viennot2016}
Jeremie~J. Viennot, Matthieu~R. Delbecq, Laure~E. Bruhat, Matthieu~C.
  Dartiailh, Matthieu~M. Desjardins, Matthieu Baillergeau, Audrey Cottet, and
  Takis Kontos.
\newblock Towards hybrid circuit quantum electrodynamics with quantum dots.
\newblock {\em Comptes Rendus Physique}, 17(7):705 -- 717, 2016.
\newblock Quantum microwaves / Micro-ondes quantiques.
\newblock URL:
  \url{http://www.sciencedirect.com/science/article/pii/S1631070516300597},
  \href {http://dx.doi.org/http://dx.doi.org/10.1016/j.crhy.2016.07.008}
  {\path{doi:http://dx.doi.org/10.1016/j.crhy.2016.07.008}}.

\bibitem{Mi2016}
X.~{Mi}, J.~V. {Cady}, D.~M. {Zajac}, J.~{Stehlik}, L.~F. {Edge}, and J.~R.
  {Petta}.
\newblock {Circuit Quantum Electrodynamics Architecture for Gate-Defined
  Quantum Dots in Silicon}.
\newblock {\em ArXiv e-prints}, October 2016.
\newblock \href {http://arxiv.org/abs/1610.05571} {\path{arXiv:1610.05571}}.

\bibitem{Jung2004}
S.~W. Jung, T.~Fujisawa, Y.~Hirayama, and Y.~H. Jeong.
\newblock Background charge fluctuation in a gaas quantum dot device.
\newblock {\em Applied Physics Letters}, 85(5):768, 2004.
\newblock URL:
  \url{http://scitation.aip.org/content/aip/journal/apl/85/5/10.1063/1.1777802},
  \href {http://dx.doi.org/10.1063/1.1777802} {\path{doi:10.1063/1.1777802}}.

\bibitem{Mueller2006}
Jens M\"uller, Stephan von Moln\'ar, Yuzo Ohno, and Hideo Ohno.
\newblock Decomposition of $1/f$ noise in
  ${\mathrm{al}}_{x}{\mathrm{ga}}_{1-x}\mathrm{As}/\mathrm{GaAs}$ hall devices.
\newblock {\em Phys. Rev. Lett.}, 96:186601, May 2006.
\newblock URL: \url{http://link.aps.org/doi/10.1103/PhysRevLett.96.186601},
  \href {http://dx.doi.org/10.1103/PhysRevLett.96.186601}
  {\path{doi:10.1103/PhysRevLett.96.186601}}.

\bibitem{Dutta1981}
P.~Dutta and P.~M. Horn.
\newblock Low-frequency fluctuations in solids: $\frac{1}{f}$ noise.
\newblock {\em Rev. Mod. Phys.}, 53:497--516, Jul 1981.
\newblock URL: \url{http://link.aps.org/doi/10.1103/RevModPhys.53.497}, \href
  {http://dx.doi.org/10.1103/RevModPhys.53.497}
  {\path{doi:10.1103/RevModPhys.53.497}}.

\bibitem{Weissman1988}
M.~B. Weissman.
\newblock $\frac{1}{f}$ noise and other slow, nonexponential kinetics in
  condensed matter.
\newblock {\em Rev. Mod. Phys.}, 60:537--571, Apr 1988.
\newblock URL: \url{http://link.aps.org/doi/10.1103/RevModPhys.60.537}, \href
  {http://dx.doi.org/10.1103/RevModPhys.60.537}
  {\path{doi:10.1103/RevModPhys.60.537}}.

\bibitem{Beaudoin2015}
F\'elix Beaudoin and W.~A. Coish.
\newblock Microscopic models for charge-noise-induced dephasing of solid-state
  qubits.
\newblock {\em Phys. Rev. B}, 91:165432, Apr 2015.
\newblock URL: \url{http://link.aps.org/doi/10.1103/PhysRevB.91.165432}, \href
  {http://dx.doi.org/10.1103/PhysRevB.91.165432}
  {\path{doi:10.1103/PhysRevB.91.165432}}.

\bibitem{Ramon2010}
Guy Ramon and Xuedong Hu.
\newblock Decoherence of spin qubits due to a nearby charge fluctuator in
  gate-defined double dots.
\newblock {\em Phys. Rev. B}, 81:045304, Jan 2010.
\newblock URL: \url{http://link.aps.org/doi/10.1103/PhysRevB.81.045304}, \href
  {http://dx.doi.org/10.1103/PhysRevB.81.045304}
  {\path{doi:10.1103/PhysRevB.81.045304}}.

\bibitem{Ramon2011}
Guy Ramon.
\newblock Electrically controlled quantum gates for two-spin qubits in two
  double quantum dots.
\newblock {\em Phys. Rev. B}, 84:155329, Oct 2011.
\newblock URL: \url{http://link.aps.org/doi/10.1103/PhysRevB.84.155329}, \href
  {http://dx.doi.org/10.1103/PhysRevB.84.155329}
  {\path{doi:10.1103/PhysRevB.84.155329}}.

\bibitem{Hiltunen2015}
Tuukka Hiltunen, Hendrik Bluhm, Sebastian Mehl, and Ari Harju.
\newblock Charge-noise tolerant exchange gates of singlet-triplet qubits in
  asymmetric double quantum dots.
\newblock {\em Phys. Rev. B}, 91:075301, Feb 2015.
\newblock URL: \url{http://link.aps.org/doi/10.1103/PhysRevB.91.075301}, \href
  {http://dx.doi.org/10.1103/PhysRevB.91.075301}
  {\path{doi:10.1103/PhysRevB.91.075301}}.

\bibitem{Reed2016}
M.~D. Reed, B.~M. Maune, R.~W. Andrews, M.~G. Borselli, K.~Eng, M.~P. Jura,
  A.~A. Kiselev, T.~D. Ladd, S.~T. Merkel, I.~Milosavljevic, E.~J. Pritchett,
  M.~T. Rakher, R.~S. Ross, A.~E. Schmitz, A.~Smith, J.~A. Wright, M.~F. Gyure,
  and A.~T. Hunter.
\newblock Reduced sensitivity to charge noise in semiconductor spin qubits via
  symmetric operation.
\newblock {\em Phys. Rev. Lett.}, 116:110402, Mar 2016.
\newblock URL: \url{http://link.aps.org/doi/10.1103/PhysRevLett.116.110402},
  \href {http://dx.doi.org/10.1103/PhysRevLett.116.110402}
  {\path{doi:10.1103/PhysRevLett.116.110402}}.

\bibitem{Martins2016}
Frederico Martins, Filip~K. Malinowski, Peter~D. Nissen, Edwin Barnes, Saeed
  Fallahi, Geoffrey~C. Gardner, Michael~J. Manfra, Charles~M. Marcus, and
  Ferdinand Kuemmeth.
\newblock Noise suppression using symmetric exchange gates in spin qubits.
\newblock {\em Phys. Rev. Lett.}, 116:116801, Mar 2016.
\newblock URL: \url{http://link.aps.org/doi/10.1103/PhysRevLett.116.116801},
  \href {http://dx.doi.org/10.1103/PhysRevLett.116.116801}
  {\path{doi:10.1103/PhysRevLett.116.116801}}.

\bibitem{Lidar2012}
Daniel~A. Lidar.
\newblock {\em Review of Decoherence-Free Subspaces, Noiseless Subsystems, and
  Dynamical Decoupling}, pages 295--354.
\newblock John Wiley \& Sons, Inc., 2014.
\newblock URL: \url{http://dx.doi.org/10.1002/9781118742631.ch11}, \href
  {http://dx.doi.org/10.1002/9781118742631.ch11}
  {\path{doi:10.1002/9781118742631.ch11}}.

\bibitem{Buchachenko2002}
Anatoly~L. Buchachenko and Vitaly~L. Berdinsky.
\newblock Electron spin catalysis.
\newblock {\em Chemical Reviews}, 102(3):603, 2002.
\newblock URL: \url{http://dx.doi.org/10.1021/cr010370l}, \href
  {http://dx.doi.org/10.1021/cr010370l} {\path{doi:10.1021/cr010370l}}.

\bibitem{Waugh1995}
F.~R. Waugh, M.~J. Berry, D.~J. Mar, R.~M. Westervelt, K.~L. Campman, and A.~C.
  Gossard.
\newblock Single-electron charging in double and triple quantum dots with
  tunable coupling.
\newblock {\em Phys. Rev. Lett.}, 75:705--708, Jul 1995.
\newblock URL: \url{http://link.aps.org/doi/10.1103/PhysRevLett.75.705}, \href
  {http://dx.doi.org/10.1103/PhysRevLett.75.705}
  {\path{doi:10.1103/PhysRevLett.75.705}}.

\bibitem{Usukura2005}
Junko Usukura, Yasuhiro Saiga, and Dai~S. Hirashima.
\newblock Three-electron systems in quantum nanostructures: Ground state
  transition from a spin-doublet state to a spin-quartet state.
\newblock {\em Journal of the Physical Society of Japan}, 74(4):1231--1239,
  2005.
\newblock URL: \url{http://dx.doi.org/10.1143/JPSJ.74.1231}, \href
  {http://arxiv.org/abs/http://dx.doi.org/10.1143/JPSJ.74.1231}
  {\path{arXiv:http://dx.doi.org/10.1143/JPSJ.74.1231}}, \href
  {http://dx.doi.org/10.1143/JPSJ.74.1231} {\path{doi:10.1143/JPSJ.74.1231}}.

\bibitem{Korkusinski2007}
Marek Korkusinski, Irene~Puerto Gimenez, Pawel Hawrylak, Louis Gaudreau,
  Sergei~A. Studenikin, and Andrew~S. Sachrajda.
\newblock Topological hunds rules and the electronic properties of a triple
  lateral quantum dot molecule.
\newblock {\em Phys. Rev. B}, 75:115301, Mar 2007.
\newblock URL: \url{http://link.aps.org/doi/10.1103/PhysRevB.75.115301}, \href
  {http://dx.doi.org/10.1103/PhysRevB.75.115301}
  {\path{doi:10.1103/PhysRevB.75.115301}}.

\bibitem{Delgado2007}
F.~Delgado, Y.-P. Shim, M.~Korkusinski, and P.~Hawrylak.
\newblock Theory of spin, electronic, and transport properties of the lateral
  triple quantum dot molecule in a magnetic field.
\newblock {\em Phys. Rev. B}, 76:115332, Sep 2007.
\newblock URL: \url{http://link.aps.org/doi/10.1103/PhysRevB.76.115332}, \href
  {http://dx.doi.org/10.1103/PhysRevB.76.115332}
  {\path{doi:10.1103/PhysRevB.76.115332}}.

\bibitem{Li2007}
Yuesong Li, Constantine Yannouleas, and Uzi Landman.
\newblock Three-electron anisotropic quantum dots in variable magnetic fields:
  Exact results for excitation spectra, spin structures, and entanglement.
\newblock {\em Phys. Rev. B}, 76:245310, Dec 2007.
\newblock URL: \url{http://link.aps.org/doi/10.1103/PhysRevB.76.245310}, \href
  {http://dx.doi.org/10.1103/PhysRevB.76.245310}
  {\path{doi:10.1103/PhysRevB.76.245310}}.

\bibitem{Hsieh2010}
Chang-Yu Hsieh and Pawel Hawrylak.
\newblock Quantum circuits based on coded qubits encoded in chirality of
  electron spin complexes in triple quantum dots.
\newblock {\em Phys. Rev. B}, 82:205311, Nov 2010.
\newblock URL: \url{http://link.aps.org/doi/10.1103/PhysRevB.82.205311}, \href
  {http://dx.doi.org/10.1103/PhysRevB.82.205311}
  {\path{doi:10.1103/PhysRevB.82.205311}}.

\bibitem{Hsieh2012}
Chang-Yu Hsieh, Yun-Pil Shim, Marek Korkusinski, and Pawel Hawrylak.
\newblock Physics of lateral triple quantum-dot molecules with controlled
  electron numbers.
\newblock {\em Reports on Progress in Physics}, 75(11):114501, 2012.
\newblock URL: \url{http://stacks.iop.org/0034-4885/75/i=11/a=114501}.

\bibitem{Gaudreau2006}
L.~Gaudreau, S.~A. Studenikin, A.~S. Sachrajda, P.~Zawadzki, A.~Kam,
  J.~Lapointe, M.~Korkusinski, and P.~Hawrylak.
\newblock Stability diagram of a few-electron triple dot.
\newblock {\em Phys. Rev. Lett.}, 97:036807, Jul 2006.
\newblock URL: \url{http://link.aps.org/doi/10.1103/PhysRevLett.97.036807},
  \href {http://dx.doi.org/10.1103/PhysRevLett.97.036807}
  {\path{doi:10.1103/PhysRevLett.97.036807}}.

\bibitem{Schroeer2007}
D.~Schr\"oer, A.~D. Greentree, L.~Gaudreau, K.~Eberl, L.~C.~L. Hollenberg,
  J.~P. Kotthaus, and S.~Ludwig.
\newblock Electrostatically defined serial triple quantum dot charged with few
  electrons.
\newblock {\em Phys. Rev. B}, 76:075306, Aug 2007.
\newblock URL: \url{http://link.aps.org/doi/10.1103/PhysRevB.76.075306}, \href
  {http://dx.doi.org/10.1103/PhysRevB.76.075306}
  {\path{doi:10.1103/PhysRevB.76.075306}}.

\bibitem{Ihn2007}
Thomas Ihn, Martin Sigrist, Klaus Ensslin, Werner Wegscheider, and Matthias
  Reinwald.
\newblock Interference in a quantum dot molecule embedded in a ring
  interferometer.
\newblock {\em New Journal of Physics}, 9(5):111, 2007.
\newblock URL: \url{http://stacks.iop.org/1367-2630/9/i=5/a=111}.

\bibitem{Rogge2008}
M.~C. Rogge and R.~J. Haug.
\newblock Noninvasive detection of molecular bonds in quantum dots.
\newblock {\em Phys. Rev. B}, 78:153310, Oct 2008.
\newblock URL: \url{http://link.aps.org/doi/10.1103/PhysRevB.78.153310}, \href
  {http://dx.doi.org/10.1103/PhysRevB.78.153310}
  {\path{doi:10.1103/PhysRevB.78.153310}}.

\bibitem{Gaudreau2009}
L.~Gaudreau, A.~Kam, G.~Granger, S.~A. Studenikin, P.~Zawadzki, and A.~S.
  Sachrajda.
\newblock A tunable few electron triple quantum dot.
\newblock {\em Applied Physics Letters}, 95(19), 2009.
\newblock URL:
  \url{http://scitation.aip.org/content/aip/journal/apl/95/19/10.1063/1.3258663},
  \href {http://dx.doi.org/http://dx.doi.org/10.1063/1.3258663}
  {\path{doi:http://dx.doi.org/10.1063/1.3258663}}.

\bibitem{Yamahata2009}
Gento Yamahata, Yoshishige Tsuchiya, Hiroshi Mizuta, Ken Uchida, and Shunri
  Oda.
\newblock Electron transport through silicon serial triple quantum dots.
\newblock {\em Solid-State Electronics}, 53(7):779 -- 785, 2009.
\newblock Papers Selected from the 38th European Solid-State Device Research
  Conference -- ESSDERC'08.
\newblock URL:
  \url{http://www.sciencedirect.com/science/article/pii/S0038110109000835},
  \href {http://dx.doi.org/http://dx.doi.org/10.1016/j.sse.2009.03.009}
  {\path{doi:http://dx.doi.org/10.1016/j.sse.2009.03.009}}.

\bibitem{Pierre2009}
M.~Pierre, R.~Wacquez, B.~Roche, X.~Jehl, M.~Sanquer, M.~Vinet, E.~Prati,
  M.~Belli, and M.~Fanciulli.
\newblock Compact silicon double and triple dots realized with only two gates.
\newblock {\em Applied Physics Letters}, 95(24), 2009.
\newblock URL:
  \url{http://scitation.aip.org/content/aip/journal/apl/95/24/10.1063/1.3273857},
  \href {http://dx.doi.org/http://dx.doi.org/10.1063/1.3273857}
  {\path{doi:http://dx.doi.org/10.1063/1.3273857}}.

\bibitem{Takakura2010}
T.~Takakura, M.~Pioro-Ladri{\`e}re, T.~Obata, Y.-S. Shin, R.~Brunner,
  K.~Yoshida, T.~Taniyama, and S.~Tarucha.
\newblock Triple quantum dot device designed for three spin qubits.
\newblock {\em Applied Physics Letters}, 97(21), 2010.
\newblock URL:
  \url{http://scitation.aip.org/content/aip/journal/apl/97/21/10.1063/1.3518919}.

\bibitem{Granger2010}
G.~Granger, L.~Gaudreau, A.~Kam, M.~Pioro-Ladri\`ere, S.~A. Studenikin, Z.~R.
  Wasilewski, P.~Zawadzki, and A.~S. Sachrajda.
\newblock Three-dimensional transport diagram of a triple quantum dot.
\newblock {\em Phys. Rev. B}, 82:075304, Aug 2010.
\newblock URL: \url{http://link.aps.org/doi/10.1103/PhysRevB.82.075304}, \href
  {http://dx.doi.org/10.1103/PhysRevB.82.075304}
  {\path{doi:10.1103/PhysRevB.82.075304}}.

\bibitem{Gaudreau2012}
L.~Gaudreau, G.~Granger, A.~Kam, G.~C. Aers, S.~A. Studenikin, P.~Zawadzki,
  M.~Pioro-Ladriere, Z.~R. Wasilewski, and A.~S. Sachrajda.
\newblock Coherent control of three-spin states in a triple quantum dot.
\newblock {\em Nat Phys}, 8(1):54, January 2012.
\newblock URL: \url{http://dx.doi.org/10.1038/nphys2149}.

\bibitem{Amaha2012}
S.~Amaha, T.~Hatano, H.~Tamura, S.~Teraoka, T.~Kubo, Y.~Tokura, D.~G. Austing,
  and S.~Tarucha.
\newblock Resonance-hybrid states in a triple quantum dot.
\newblock {\em Phys. Rev. B}, 85:081301, Feb 2012.
\newblock URL: \url{http://link.aps.org/doi/10.1103/PhysRevB.85.081301}, \href
  {http://dx.doi.org/10.1103/PhysRevB.85.081301}
  {\path{doi:10.1103/PhysRevB.85.081301}}.

\bibitem{Pan2012}
H.~Pan, M.~G. House, X.~Hao, and H.~W. Jiang.
\newblock Fabrication and characterization of a silicon
  metal-oxide-semiconductor based triple quantum dot.
\newblock {\em Applied Physics Letters}, 100(26), 2012.
\newblock URL:
  \url{http://scitation.aip.org/content/aip/journal/apl/100/26/10.1063/1.4731275},
  \href {http://dx.doi.org/http://dx.doi.org/10.1063/1.4731275}
  {\path{doi:http://dx.doi.org/10.1063/1.4731275}}.

\bibitem{Busl2013}
Busl M., Granger G., Gaudreau L., Sanchez R., Kam A., Pioro-Ladriere M.,
  Studenikin~S. A., Zawadzki P., Wasilewski~Z. R., Sachrajda~A. S., and Platero
  G.
\newblock Bipolar spin blockade and coherent state superpositions in a triple
  quantum dot.
\newblock {\em Nat Nano}, 8(4):261--265, 04 2013.
\newblock URL: \url{http://dx.doi.org/10.1038/nnano.2013.7}.

\bibitem{Kim2014}
Dohun Kim, Zhan Shi, C.~B. Simmons, D.~R. Ward, J.~R. Prance, Teck~Seng Koh,
  John~King Gamble, D.~E. Savage, M.~G. Lagally, Mark Friesen, S.~N.
  Coppersmith, and Mark~A. Eriksson.
\newblock Quantum control and process tomography of a semiconductor quantum dot
  hybrid qubit.
\newblock {\em Nature}, 511(7507):70--74, 07 2014.
\newblock URL: \url{http://dx.doi.org/10.1038/nature13407}.

\bibitem{Shi2014}
Zhan Shi, C.~B. Simmons, Daniel~R. Ward, J.~R. Prance, Xian Wu, Teck~Seng Koh,
  John~King Gamble, D.~E. Savage, M.~G. Lagally, Mark Friesen, S.~N.
  Coppersmith, and M.~A. Eriksson.
\newblock Fast coherent manipulation of three-electron states in a double
  quantum dot.
\newblock {\em Nat Commun}, 5, 01 2014.
\newblock URL: \url{http://dx.doi.org/10.1038/ncomms4020}.

\bibitem{Kim2015b}
Dohun Kim, Daniel~R Ward, Christie~B Simmons, Don~E Savage, Max~G Lagally, Mark
  Friesen, Susan~N Coppersmith, and Mark~A Eriksson.
\newblock High-fidelity resonant gating of a silicon-based quantum dot hybrid
  qubit.
\newblock {\em Npj Quantum Information}, 1:15004 EP --, 10 2015.
\newblock URL: \url{http://dx.doi.org/10.1038/npjqi.2015.4}.

\bibitem{Aers2012}
G.~C. Aers, S.~A. Studenikin, G.~Granger, A.~Kam, P.~Zawadzki, Z.~R.
  Wasilewski, and A.~S. Sachrajda.
\newblock Coherent exchange and double beam splitter oscillations in a triple
  quantum dot.
\newblock {\em Phys. Rev. B}, 86:045316, Jul 2012.
\newblock URL: \url{http://link.aps.org/doi/10.1103/PhysRevB.86.045316}, \href
  {http://dx.doi.org/10.1103/PhysRevB.86.045316}
  {\path{doi:10.1103/PhysRevB.86.045316}}.

\bibitem{Medford2013b}
J.~Medford, J.~Beil, J.~M. Taylor, S.~D. Bartlett, A.~C. Doherty, E.~I. Rashba,
  D.~P. DiVincenzo, H.~Lu, A.~C. Gossard, and C.~M. Marcus.
\newblock Self-consistent measurement and state tomography of an exchange-only
  spin qubit.
\newblock {\em Nat Nano}, 8(9):654, Sep 2013.
\newblock URL: \url{http://dx.doi.org/10.1038/nnano.2013.168}.

\bibitem{Sanchez2014}
R.~Sanchez, G.~Granger, L.~Gaudreau, A.~Kam, M.~Pioro-Ladriere, S.~A.
  Studenikin, P.~Zawadzki, A.~S. Sachrajda, and G.~Platero.
\newblock Long-range spin transfer in triple quantum dots.
\newblock {\em Phys. Rev. Lett.}, 112:176803, May 2014.
\newblock URL: \url{http://link.aps.org/doi/10.1103/PhysRevLett.112.176803},
  \href {http://dx.doi.org/10.1103/PhysRevLett.112.176803}
  {\path{doi:10.1103/PhysRevLett.112.176803}}.

\bibitem{Eng2015}
Kevin Eng, Thaddeus~D. Ladd, Aaron Smith, Matthew~G. Borselli, Andrey~A.
  Kiselev, Bryan~H. Fong, Kevin~S. Holabird, Thomas~M. Hazard, Biqin Huang,
  Peter~W. Deelman, Ivan Milosavljevic, Adele~E. Schmitz, Richard~S. Ross,
  Mark~F. Gyure, and Andrew~T. Hunter.
\newblock Isotopically enhanced triple-quantum-dot qubit.
\newblock {\em Science Advances}, 1(4), 2015.
\newblock URL: \url{http://advances.sciencemag.org/content/1/4/e1500214}, \href
  {http://dx.doi.org/10.1126/sciadv.1500214}
  {\path{doi:10.1126/sciadv.1500214}}.

\bibitem{Poulin2015}
G.~Poulin-Lamarre, J.~Thorgrimson, S.~A. Studenikin, G.~C. Aers, A.~Kam,
  P.~Zawadzki, Z.~R. Wasilewski, and A.~S. Sachrajda.
\newblock Three-spin coherent oscillations and interference.
\newblock {\em Phys. Rev. B}, 91:125417, Mar 2015.
\newblock URL: \url{http://link.aps.org/doi/10.1103/PhysRevB.91.125417}, \href
  {http://dx.doi.org/10.1103/PhysRevB.91.125417}
  {\path{doi:10.1103/PhysRevB.91.125417}}.

\bibitem{Scarola2004}
V.~W. Scarola, K.~Park, and S.~Das Sarma.
\newblock Chirality in quantum computation with spin cluster qubits.
\newblock {\em Phys. Rev. Lett.}, 93:120503, Sep 2004.
\newblock URL: \url{http://link.aps.org/doi/10.1103/PhysRevLett.93.120503},
  \href {http://dx.doi.org/10.1103/PhysRevLett.93.120503}
  {\path{doi:10.1103/PhysRevLett.93.120503}}.

\bibitem{Scarola2005}
V.~W. Scarola and S.~Das~Sarma.
\newblock Exchange gate in solid-state spin-quantum computation: The
  applicability of the heisenberg model.
\newblock {\em Phys. Rev. A}, 71:032340, Mar 2005.
\newblock URL: \url{http://link.aps.org/doi/10.1103/PhysRevA.71.032340}, \href
  {http://dx.doi.org/10.1103/PhysRevA.71.032340}
  {\path{doi:10.1103/PhysRevA.71.032340}}.

\bibitem{Hsieh2012b}
Chang-Yu Hsieh, Alexandre Rene, and Pawel Hawrylak.
\newblock Herzberg circuit and berry's phase in chirality-based coded qubit in
  a triangular triple quantum dot.
\newblock {\em Phys. Rev. B}, 86:115312, Sep 2012.
\newblock URL: \url{http://link.aps.org/doi/10.1103/PhysRevB.86.115312}, \href
  {http://dx.doi.org/10.1103/PhysRevB.86.115312}
  {\path{doi:10.1103/PhysRevB.86.115312}}.

\bibitem{Luczak2012}
Jakub {\L}uczak and Bogdan~R Bu{\l}ka.
\newblock Entanglement in a three spin system controlled by electric and
  magnetic fields.
\newblock {\em Journal of Physics: Condensed Matter}, 24(37):375303, 2012.
\newblock URL: \url{http://stacks.iop.org/0953-8984/24/i=37/a=375303}.

\bibitem{Urbaniak2013}
M.~Urbaniak, S.~B. Tooski, A.~Ram{\v{s}}ak, and B.~R. Bu{\l}ka.
\newblock Thermal entanglement in a triple quantum dot system.
\newblock {\em The European Physical Journal B}, 86(12):1--8, 2013.
\newblock URL: \url{http://dx.doi.org/10.1140/epjb/e2013-40761-3}, \href
  {http://dx.doi.org/10.1140/epjb/e2013-40761-3}
  {\path{doi:10.1140/epjb/e2013-40761-3}}.

\bibitem{Tooski2014}
S.~B. Tooski, Bogdan~R. Bu{\l}ka, Rok {\v{Z}}itko, and Anton Ram{\v{s}}ak.
\newblock Entanglement switching via the kondo effect in triple quantum dots.
\newblock {\em The European Physical Journal B}, 87(6):1--8, 2014.
\newblock URL: \url{http://dx.doi.org/10.1140/epjb/e2014-41025-6}, \href
  {http://dx.doi.org/10.1140/epjb/e2014-41025-6}
  {\path{doi:10.1140/epjb/e2014-41025-6}}.

\bibitem{Luczak2014}
Jakub \L{}uczak and Bogdan~R. Bu\l{}ka.
\newblock Readout and dynamics of a qubit built on three quantum dots.
\newblock {\em Phys. Rev. B}, 90:165427, Oct 2014.
\newblock URL: \url{http://link.aps.org/doi/10.1103/PhysRevB.90.165427}, \href
  {http://dx.doi.org/10.1103/PhysRevB.90.165427}
  {\path{doi:10.1103/PhysRevB.90.165427}}.

\bibitem{Recher2009}
Patrik Recher, Johan Nilsson, Guido Burkard, and Bj√∂rn Trauzettel.
\newblock Bound states and magnetic field induced valley splitting in
  gate-tunable graphene quantum dots.
\newblock {\em Phys. Rev. B}, 79:085407, Feb 2009.
\newblock URL: \url{http://link.aps.org/doi/10.1103/PhysRevB.79.085407}, \href
  {http://dx.doi.org/10.1103/PhysRevB.79.085407}
  {\path{doi:10.1103/PhysRevB.79.085407}}.

\bibitem{Churchill2009}
H.~O.~H. Churchill, F.~Kuemmeth, J.~W. Harlow, A.~J. Bestwick, E.~I. Rashba,
  K.~Flensberg, C.~H. Stwertka, T.~Taychatanapat, S.~K. Watson, and C.~M.
  Marcus.
\newblock Relaxation and dephasing in a two-electron 13c nanotube double
  quantum dot.
\newblock {\em Phys. Rev. Lett.}, 102(16):166802{\"A}{\`\i}, April 2009.
\newblock URL: \url{http://link.aps.org/doi/10.1103/PhysRevLett.102.166802}.

\bibitem{Reynoso2011}
Andres~A. Reynoso and Karsten Flensberg.
\newblock Dephasing and hyperfine interaction in carbon nanotube double quantum
  dots: The clean limit.
\newblock {\em Phys. Rev. B}, 84(20):205449{\"A}{\`\i}, November 2011.
\newblock URL: \url{http://link.aps.org/doi/10.1103/PhysRevB.84.205449}.

\bibitem{Reynoso2012}
Andres~A. Reynoso and Karsten Flensberg.
\newblock Dephasing and hyperfine interaction in carbon nanotubes double
  quantum dots: Disordered case.
\newblock {\em Phys. Rev. B}, 85(19):195441{\"A}{\`\i}, May 2012.
\newblock URL: \url{http://link.aps.org/doi/10.1103/PhysRevB.85.195441}.

\bibitem{Song2015}
Xiang-Xiang Song, Di~Liu, Vahid Mosallanejad, Jie You, Tian-Yi Han, Dian-Teng
  Chen, Hai-Ou Li, Gang Cao, Ming Xiao, Guang-Can Guo, and Guo-Ping Guo.
\newblock A gate defined quantum dot on the two-dimensional transition metal
  dichalcogenide semiconductor wse2.
\newblock {\em Nanoscale}, 7:16867--16873, 2015.
\newblock URL: \url{http://dx.doi.org/10.1039/C5NR04961J}, \href
  {http://dx.doi.org/10.1039/C5NR04961J} {\path{doi:10.1039/C5NR04961J}}.

\bibitem{Song2015b}
Xiang-Xiang Song, Zhuo-Zhi Zhang, Jie You, Di~Liu, Hai-Ou Li, Gang Cao, Ming
  Xiao, and Guo-Ping Guo.
\newblock Temperature dependence of coulomb oscillations in a few-layer
  two-dimensional ws2 quantum dot.
\newblock {\em Scientific Reports}, 5:16113 EP --, 11 2015.
\newblock URL: \url{http://dx.doi.org/10.1038/srep16113}.

\bibitem{Laird2015}
Edward~A. Laird, Ferdinand Kuemmeth, Gary~A. Steele, Kasper Grove-Rasmussen,
  Jesper Nyg\aa{}rd, Karsten Flensberg, and Leo~P. Kouwenhoven.
\newblock Quantum transport in carbon nanotubes.
\newblock {\em Rev. Mod. Phys.}, 87:703--764, Jul 2015.
\newblock URL: \url{http://link.aps.org/doi/10.1103/RevModPhys.87.703}, \href
  {http://dx.doi.org/10.1103/RevModPhys.87.703}
  {\path{doi:10.1103/RevModPhys.87.703}}.

\bibitem{Wang2016}
K.~{Wang}, T.~{Taniguchi}, K.~{Watanabe}, and P.~{Kim}.
\newblock {Engineering Quantum Confinement in Semiconducting van der Waals
  Heterostructure}.
\newblock {\em ArXiv e-prints}, October 2016.
\newblock \href {http://arxiv.org/abs/1610.02929} {\path{arXiv:1610.02929}}.

\bibitem{Angus2007}
Susan~J. Angus, Andrew~J. Ferguson, Andrew~S. Dzurak, and Robert~G. Clark.
\newblock Gate-defined quantum dots in intrinsic silicon.
\newblock {\em Nano Letters}, 7(7):2051--2055, 2007.
\newblock PMID: 17567176.
\newblock URL: \url{http://dx.doi.org/10.1021/nl070949k}, \href
  {http://arxiv.org/abs/http://dx.doi.org/10.1021/nl070949k}
  {\path{arXiv:http://dx.doi.org/10.1021/nl070949k}}, \href
  {http://dx.doi.org/10.1021/nl070949k} {\path{doi:10.1021/nl070949k}}.

\bibitem{Jones2016}
C.~{Jones}, M.~F. {Gyure}, T.~D. {Ladd}, M.~A. {Fogarty}, A.~{Morello}, and
  A.~S. {Dzurak}.
\newblock {A logical qubit in a linear array of semiconductor quantum dots}.
\newblock {\em ArXiv e-prints}, August 2016.
\newblock \href {http://arxiv.org/abs/1608.06335} {\path{arXiv:1608.06335}}.

\bibitem{Yang2013}
C.~H. Yang, A.~Rossi, R.~Ruskov, N.~S. Lai, F.~A. Mohiyaddin, S.~Lee, C.~Tahan,
  G.~Klimeck, A.~Morello, and A.~S. Dzurak.
\newblock Spin-valley lifetimes in a silicon quantum dot with tunable valley
  splitting.
\newblock {\em Nat Commun}, 4, 06 2013.
\newblock URL: \url{http://dx.doi.org/10.1038/ncomms3069}.

\bibitem{Veldhorst2014}
Veldhorst M., Hwang J.~C. C., Yang~C. H., Leenstra~A. W., de~Ronde~B.,
  Dehollain~J. P., Muhonen~J. T., Hudson~F. E., Itoh~K. M., Morello A., and
  Dzurak~A. S.
\newblock An addressable quantum dot qubit with fault-tolerant
  control-fidelity.
\newblock {\em Nat Nano}, 9(12):981--985, 12 2014.
\newblock URL: \url{http://dx.doi.org/10.1038/nnano.2014.216}.

\bibitem{Muhonen2014}
Juha~T. Muhonen, Juan~P. Dehollain, Arne Laucht, Fay~E. Hudson, Rachpon Kalra,
  Takeharu Sekiguchi, Kohei~M. Itoh, David~N. Jamieson, Jeffrey~C. McCallum,
  Andrew~S. Dzurak, and Andrea Morello.
\newblock Storing quantum information for 30 seconds in a nanoelectronic
  device.
\newblock {\em Nat Nano}, 9(12):986--991, 12 2014.
\newblock URL: \url{http://dx.doi.org/10.1038/nnano.2014.211}.

\bibitem{Morello2015}
Andrea Morello.
\newblock Silicon quantum dots: fine-tuning to maturity.
\newblock {\em Nanotechnology}, 26(50):502501, 2015.
\newblock URL: \url{http://stacks.iop.org/0957-4484/26/i=50/a=502501}.

\bibitem{Borselli2015}
M~G Borselli, K~Eng, R~S Ross, T~M Hazard, K~S Holabird, B~Huang, A~A Kiselev,
  P~W Deelman, L~D Warren, I~Milosavljevic, A~E Schmitz, M~Sokolich, M~F Gyure,
  and A~T Hunter.
\newblock Undoped accumulation-mode si/sige quantum dots.
\newblock {\em Nanotechnology}, 26(37):375202, 2015.
\newblock URL: \url{http://stacks.iop.org/0957-4484/26/i=37/a=375202}.

\bibitem{Zajac2015}
D.~M. Zajac, T.~M. Hazard, X.~Mi, K.~Wang, and J.~R. Petta.
\newblock A reconfigurable gate architecture for si/sige quantum dots.
\newblock {\em Applied Physics Letters}, 106(22), 2015.
\newblock URL:
  \url{http://scitation.aip.org/content/aip/journal/apl/106/22/10.1063/1.4922249},
  \href {http://dx.doi.org/http://dx.doi.org/10.1063/1.4922249}
  {\path{doi:http://dx.doi.org/10.1063/1.4922249}}.

\bibitem{Ward2016}
Daniel~R. Ward, Dohun Kim, Donald~E. Savage, Max~G. Lagally, Ryan~H. Foote,
  Mark Friesen, Susan~N. Coppersmith, and Mark~A. Eriksson.
\newblock State-conditional coherent charge qubit oscillations in a si/sige
  quadruple quantum dot.
\newblock {\em Npj Quantum Information}, 2:16032 EP --, 10 2016.
\newblock URL: \url{http://dx.doi.org/10.1038/npjqi.2016.32}.

\bibitem{Knapp2016}
T~J Knapp, R~T Mohr, Yize~Stephanie Li, Brandur Thorgrimsson, Ryan~H Foote,
  Xian Wu, Daniel~R Ward, D~E Savage, M~G Lagally, Mark Friesen, S~N
  Coppersmith, and M~A Eriksson.
\newblock Characterization of a gate-defined double quantum dot in a si/sige
  nanomembrane.
\newblock {\em Nanotechnology}, 27(15):154002, 2016.
\newblock URL: \url{http://stacks.iop.org/0957-4484/27/i=15/a=154002}.

\bibitem{Zajac2016}
D.~M. {Zajac}, T.~M. {Hazard}, X.~{Mi}, E.~{Nielsen}, and J.~R. {Petta}.
\newblock {Scalable gate architecture for densely packed semiconductor spin
  qubits}.
\newblock {\em ArXiv e-prints}, July 2016.
\newblock \href {http://arxiv.org/abs/1607.07025} {\path{arXiv:1607.07025}}.

\bibitem{Itoh2014}
Kohei~M. Itoh and Hideyuki Watanabe.
\newblock Isotope engineering of silicon and diamond for quantum computing and
  sensing applications.
\newblock {\em MRS Communications}, 4(4):143--157, 11 2014.
\newblock URL:
  \url{https://www.cambridge.org/core/journals/mrs-communications/article/isotope-engineering-of-silicon-and-diamond-for-quantum-computing-and-sensing-applications/3E3632E92A5002BFBEE1FF65741C4B54},
  \href {http://dx.doi.org/10.1557/mrc.2014.32}
  {\path{doi:10.1557/mrc.2014.32}}.

\bibitem{Field1993}
M.~Field, C.~G. Smith, M.~Pepper, D.~A. Ritchie, J.~E.~F. Frost, G.~A.~C.
  Jones, and D.~G. Hasko.
\newblock Measurements of coulomb blockade with a noninvasive voltage probe.
\newblock {\em Phys. Rev. Lett.}, 70:1311--1314, Mar 1993.
\newblock URL: \url{http://link.aps.org/doi/10.1103/PhysRevLett.70.1311}, \href
  {http://dx.doi.org/10.1103/PhysRevLett.70.1311}
  {\path{doi:10.1103/PhysRevLett.70.1311}}.

\bibitem{Barthel2009}
C.~Barthel, D.~J. Reilly, C.~M. Marcus, M.~P. Hanson, and A.~C. Gossard.
\newblock Rapid single-shot measurement of a singlet-triplet qubit.
\newblock {\em Phys. Rev. Lett.}, 103:160503, Oct 2009.
\newblock URL: \url{http://link.aps.org/doi/10.1103/PhysRevLett.103.160503},
  \href {http://dx.doi.org/10.1103/PhysRevLett.103.160503}
  {\path{doi:10.1103/PhysRevLett.103.160503}}.

\bibitem{Devoret2000}
Michel~H. Devoret and Robert~J. Schoelkopf.
\newblock Amplifying quantum signals with the single-electron transistor.
\newblock {\em Nature}, 406(6799):1039--1046, 08 2000.
\newblock URL: \url{http://dx.doi.org/10.1038/35023253}.

\bibitem{Barthel2010}
C.~Barthel, M.~Kj\ae{}rgaard, J.~Medford, M.~Stopa, C.~M. Marcus, M.~P. Hanson,
  and A.~C. Gossard.
\newblock Fast sensing of double-dot charge arrangement and spin state with a
  radio-frequency sensor quantum dot.
\newblock {\em Phys. Rev. B}, 81:161308, Apr 2010.
\newblock URL: \url{http://link.aps.org/doi/10.1103/PhysRevB.81.161308}, \href
  {http://dx.doi.org/10.1103/PhysRevB.81.161308}
  {\path{doi:10.1103/PhysRevB.81.161308}}.

\bibitem{House2016}
M.~G. House, I.~Bartlett, P.~Pakkiam, M.~Koch, E.~Peretz, J.~van~der Heijden,
  T.~Kobayashi, S.~Rogge, and M.~Y. Simmons.
\newblock High-sensitivity charge detection with a single-lead quantum dot for
  scalable quantum computation.
\newblock {\em Phys. Rev. Applied}, 6:044016, Oct 2016.
\newblock URL: \url{http://link.aps.org/doi/10.1103/PhysRevApplied.6.044016},
  \href {http://dx.doi.org/10.1103/PhysRevApplied.6.044016}
  {\path{doi:10.1103/PhysRevApplied.6.044016}}.

\bibitem{House2014}
M.~G. House, E.~Peretz, J.~G. Keizer, S.~J. Hile, and M.~Y. Simmons.
\newblock Single-charge detection by an atomic precision tunnel junction.
\newblock {\em Applied Physics Letters}, 104(11), 2014.
\newblock URL:
  \url{http://scitation.aip.org/content/aip/journal/apl/104/11/10.1063/1.4869032},
  \href {http://dx.doi.org/http://dx.doi.org/10.1063/1.4869032}
  {\path{doi:http://dx.doi.org/10.1063/1.4869032}}.

\bibitem{Kerman2008}
Andrew~J. Kerman and William~D. Oliver.
\newblock High-fidelity quantum operations on superconducting qubits in the
  presence of noise.
\newblock {\em Phys. Rev. Lett.}, 101:070501, Aug 2008.
\newblock URL: \url{http://link.aps.org/doi/10.1103/PhysRevLett.101.070501},
  \href {http://dx.doi.org/10.1103/PhysRevLett.101.070501}
  {\path{doi:10.1103/PhysRevLett.101.070501}}.

\bibitem{Delbecq2011}
M.~R. Delbecq, V.~Schmitt, F.~D. Parmentier, N.~Roch, J.~J. Viennot, G.~F\`eve,
  B.~Huard, C.~Mora, A.~Cottet, and T.~Kontos.
\newblock Coupling a quantum dot, fermionic leads, and a microwave cavity on a
  chip.
\newblock {\em Phys. Rev. Lett.}, 107:256804, Dec 2011.
\newblock URL: \url{http://link.aps.org/doi/10.1103/PhysRevLett.107.256804},
  \href {http://dx.doi.org/10.1103/PhysRevLett.107.256804}
  {\path{doi:10.1103/PhysRevLett.107.256804}}.

\bibitem{Kerman2013}
Andrew~J Kerman.
\newblock Quantum information processing using quasiclassical electromagnetic
  interactions between qubits and electrical resonators.
\newblock {\em New Journal of Physics}, 15(12):123011, 2013.
\newblock URL: \url{http://stacks.iop.org/1367-2630/15/i=12/a=123011}.

\bibitem{Gonzalez2014}
M.~F. Gonzalez-Zalba, S.~Barraud, A.~J. Ferguson, and A.~C. Betz.
\newblock Probing the limits of gate-based charge sensing.
\newblock {\em Nature Communications}, 6:6084 EP --, 01 2015.
\newblock URL: \url{http://dx.doi.org/10.1038/ncomms7084}.

\bibitem{Didier2015}
Nicolas Didier, J\'er\^ome Bourassa, and Alexandre Blais.
\newblock Fast quantum nondemolition readout by parametric modulation of
  longitudinal qubit-oscillator interaction.
\newblock {\em Phys. Rev. Lett.}, 115:203601, Nov 2015.
\newblock URL: \url{http://link.aps.org/doi/10.1103/PhysRevLett.115.203601},
  \href {http://dx.doi.org/10.1103/PhysRevLett.115.203601}
  {\path{doi:10.1103/PhysRevLett.115.203601}}.

\bibitem{Burkard2016}
Guido Burkard and J.~R. Petta.
\newblock Dispersive readout of valley splittings in cavity-coupled silicon
  quantum dots.
\newblock {\em Phys. Rev. B}, 94:195305, Nov 2016.
\newblock URL: \url{http://link.aps.org/doi/10.1103/PhysRevB.94.195305}, \href
  {http://dx.doi.org/10.1103/PhysRevB.94.195305}
  {\path{doi:10.1103/PhysRevB.94.195305}}.

\bibitem{Beaudoin2016a}
F.~{Beaudoin}, A.~{Blais}, and W.~A. {Coish}.
\newblock {Hamiltonian engineering for robust quantum state transfer and qubit
  readout in cavity QED}.
\newblock {\em ArXiv e-prints}, February 2016.
\newblock \href {http://arxiv.org/abs/1602.05090} {\path{arXiv:1602.05090}}.

\bibitem{Wiel2002}
W.~G. van~der Wiel, S.~De~Franceschi, J.~M. Elzerman, T.~Fujisawa, S.~Tarucha,
  and L.~P. Kouwenhoven.
\newblock Electron transport through double quantum dots.
\newblock {\em Rev. Mod. Phys.}, 75:1--22, Dec 2002.
\newblock URL: \url{http://link.aps.org/doi/10.1103/RevModPhys.75.1}, \href
  {http://dx.doi.org/10.1103/RevModPhys.75.1}
  {\path{doi:10.1103/RevModPhys.75.1}}.

\bibitem{Vidan2005}
A.~Vidan, R.~M. Westervelt, M.~Stopa, M.~Hanson, and A.~C. Gossard.
\newblock Charging and spin effects in triple dot artificial molecules.
\newblock {\em Journal of Superconductivity}, 18(2):223--227, 2005.
\newblock URL: \url{http://dx.doi.org/10.1007/s10948-005-3373-8}, \href
  {http://dx.doi.org/10.1007/s10948-005-3373-8}
  {\path{doi:10.1007/s10948-005-3373-8}}.

\bibitem{Rogge2009}
M.~C. Rogge and R.~J. Haug.
\newblock The three dimensionality of triple quantum dot stability diagrams.
\newblock {\em New Journal of Physics}, 11(11):113037, 2009.
\newblock URL: \url{http://stacks.iop.org/1367-2630/11/i=11/a=113037}.

\bibitem{Seo2013}
M.~Seo, H.~K. Choi, S.-Y. Lee, N.~Kim, Y.~Chung, H.-S. Sim, V.~Umansky, and
  D.~Mahalu.
\newblock Charge frustration in a triangular triple quantum dot.
\newblock {\em Phys. Rev. Lett.}, 110:046803, Jan 2013.
\newblock URL: \url{http://link.aps.org/doi/10.1103/PhysRevLett.110.046803},
  \href {http://dx.doi.org/10.1103/PhysRevLett.110.046803}
  {\path{doi:10.1103/PhysRevLett.110.046803}}.

\bibitem{Broome2016}
M.~A. Broome, S.~K. Gorman, J.~G. Keizer, T.~F. Watson, S.~J. Hile, W.~J.
  Baker, and M.~Y. Simmons.
\newblock Mapping the chemical potential landscape of a triple quantum dot.
\newblock {\em Phys. Rev. B}, 94:054314, Aug 2016.
\newblock URL: \url{http://link.aps.org/doi/10.1103/PhysRevB.94.054314}, \href
  {http://dx.doi.org/10.1103/PhysRevB.94.054314}
  {\path{doi:10.1103/PhysRevB.94.054314}}.

\bibitem{Khoi2013}
Khoi~T. Nguyen, Michael~P. Lilly, Erik Nielsen, Nathan Bishop, Rajib Rahman,
  Ralph Young, Joel Wendt, Jason Dominguez, Tammy Pluym, Jeffery Stevens,
  Tzu-Ming Lu, Richard Muller, and Malcolm~S. Carroll.
\newblock Charge sensed pauli blockade in a metal--oxide--semiconductor lateral
  double quantum dot.
\newblock {\em Nano Letters}, 13(12):5785--5790, 2013.
\newblock URL: \url{http://dx.doi.org/10.1021/nl4020759}, \href
  {http://arxiv.org/abs/http://dx.doi.org/10.1021/nl4020759}
  {\path{arXiv:http://dx.doi.org/10.1021/nl4020759}}, \href
  {http://dx.doi.org/10.1021/nl4020759} {\path{doi:10.1021/nl4020759}}.

\bibitem{Powell2014}
Ian~E. Powell.
\newblock Simulating charge stability diagrams for double and triple quantum
  dot systems.
\newblock {\em (unpublished)}, September 2014.
\newblock
  http://www.pa.ucla.edu/\-sites/\-default/\-files/\-files/\-REU/\-Papers\%202014/\-powell.pdf.
\newblock URL:
  \url{http://www.pa.ucla.edu/sites/default/files/files/REU/Papers\%202014/powell.pdf}.

\bibitem{Russ2016b}
Maximilian Russ and Guido Burkard.
\newblock Fluctuating charging energies in three-spin qubits.
\newblock {\em unpublished}, 2016.

\bibitem{Mizel2004}
Ari Mizel and Daniel~A. Lidar.
\newblock Three- and four-body interactions in spin-based quantum computers.
\newblock {\em Phys. Rev. Lett.}, 92:077903, Feb 2004.
\newblock URL: \url{http://link.aps.org/doi/10.1103/PhysRevLett.92.077903},
  \href {http://dx.doi.org/10.1103/PhysRevLett.92.077903}
  {\path{doi:10.1103/PhysRevLett.92.077903}}.

\bibitem{Hawrylak2005}
Pawel Hawrylak and Marek Korkusinski.
\newblock Voltage-controlled coded qubit based on electron spin.
\newblock {\em Solid State Communications}, 136(9--10):508 -- 512, 2005.
\newblock URL:
  \url{http://www.sciencedirect.com/science/article/pii/S0038109805007994},
  \href {http://dx.doi.org/http://dx.doi.org/10.1016/j.ssc.2005.09.026}
  {\path{doi:http://dx.doi.org/10.1016/j.ssc.2005.09.026}}.

\bibitem{Ren2014}
Y.F. Ren, L.~Wang, Z.~Liu, and M.W. Wu.
\newblock Energy spectra of three electrons in sige/si/sige laterally coupled
  triple quantum dots.
\newblock {\em Physica E: Low-dimensional Systems and Nanostructures}, 63:329
  -- 336, 2014.
\newblock URL:
  \url{http://www.sciencedirect.com/science/article/pii/S1386947714002288},
  \href {http://dx.doi.org/http://dx.doi.org/10.1016/j.physe.2014.06.012}
  {\path{doi:http://dx.doi.org/10.1016/j.physe.2014.06.012}}.

\bibitem{Kawano2006}
Y.~Kawano and M.~Ozawa.
\newblock Quantum gates generated by rotationally invariant operators in a
  decoherence-free subsystem.
\newblock {\em Phys. Rev. A}, 73:012339, Jan 2006.
\newblock URL: \url{http://link.aps.org/doi/10.1103/PhysRevA.73.012339}, \href
  {http://dx.doi.org/10.1103/PhysRevA.73.012339}
  {\path{doi:10.1103/PhysRevA.73.012339}}.

\bibitem{Lee2008}
B.~Lee, W.~M. Witzel, and S.~Das~Sarma.
\newblock Universal pulse sequence to minimize spin dephasing in the central
  spin decoherence problem.
\newblock {\em Phys. Rev. Lett.}, 100:160505, Apr 2008.
\newblock URL: \url{http://link.aps.org/doi/10.1103/PhysRevLett.100.160505},
  \href {http://dx.doi.org/10.1103/PhysRevLett.100.160505}
  {\path{doi:10.1103/PhysRevLett.100.160505}}.

\bibitem{Hickman2013}
G.~T. Hickman, Xin Wang, J.~P. Kestner, and S.~Das~Sarma.
\newblock Dynamically corrected gates for an exchange-only qubit.
\newblock {\em Phys. Rev. B}, 88:161303, Oct 2013.
\newblock URL: \url{http://link.aps.org/doi/10.1103/PhysRevB.88.161303}, \href
  {http://dx.doi.org/10.1103/PhysRevB.88.161303}
  {\path{doi:10.1103/PhysRevB.88.161303}}.

\bibitem{Takakura2014}
T.~Takakura, A.~Noiri, T.~Obata, T.~Otsuka, J.~Yoneda, K.~Yoshida, and
  S.~Tarucha.
\newblock Single to quadruple quantum dots with tunable tunnel couplings.
\newblock {\em Applied Physics Letters}, 104(11), 2014.
\newblock URL:
  \url{http://scitation.aip.org/content/aip/journal/apl/104/11/10.1063/1.4869108},
  \href {http://dx.doi.org/http://dx.doi.org/10.1063/1.4869108}
  {\path{doi:http://dx.doi.org/10.1063/1.4869108}}.

\bibitem{Noiri2016}
A.~Noiri, J.~Yoneda, T.~Nakajima, T.~Otsuka, M.~R. Delbecq, K.~Takeda,
  S.~Amaha, G.~Allison, A.~Ludwig, A.~D. Wieck, and S.~Tarucha.
\newblock Coherent electron-spin-resonance manipulation of three individual
  spins in a triple quantum dot.
\newblock {\em Applied Physics Letters}, 108(15), 2016.
\newblock URL:
  \url{http://scitation.aip.org/content/aip/journal/apl/108/15/10.1063/1.4945592},
  \href {http://dx.doi.org/http://dx.doi.org/10.1063/1.4945592}
  {\path{doi:http://dx.doi.org/10.1063/1.4945592}}.

\bibitem{Anderson1964}
S.~Alexander and P.~W. Anderson.
\newblock Interaction between localized states in metals.
\newblock {\em Phys. Rev.}, 133:A1594--A1603, Mar 1964.
\newblock URL: \url{http://link.aps.org/doi/10.1103/PhysRev.133.A1594}, \href
  {http://dx.doi.org/10.1103/PhysRev.133.A1594}
  {\path{doi:10.1103/PhysRev.133.A1594}}.

\bibitem{Stoehr2006}
Joachim St{\"o}hr and Hans~Christoph Siegmann.
\newblock Magnetism.
\newblock {\em Solid-State Sciences. Springer, Berlin, Heidelberg}, 5, 2006.

\bibitem{Kostyrko2011}
Tomasz Kostyrko and Bogdan~R. Bu\l{}ka.
\newblock Canonical perturbation theory for inhomogeneous systems of
  interacting fermions.
\newblock {\em Phys. Rev. B}, 84:035123, Jul 2011.
\newblock URL: \url{http://link.aps.org/doi/10.1103/PhysRevB.84.035123}, \href
  {http://dx.doi.org/10.1103/PhysRevB.84.035123}
  {\path{doi:10.1103/PhysRevB.84.035123}}.

\bibitem{Trif2008}
Mircea Trif, Filippo Troiani, Dimitrije Stepanenko, and Daniel Loss.
\newblock Spin-electric coupling in molecular magnets.
\newblock {\em Phys. Rev. Lett.}, 101:217201, Nov 2008.
\newblock URL: \url{http://link.aps.org/doi/10.1103/PhysRevLett.101.217201},
  \href {http://dx.doi.org/10.1103/PhysRevLett.101.217201}
  {\path{doi:10.1103/PhysRevLett.101.217201}}.

\bibitem{Kyriakidis2005}
Jordan Kyriakidis and Stephen~J. Penney.
\newblock Coherent rotations of a single spin-based qubit in a single quantum
  dot at fixed zeeman energy.
\newblock {\em Phys. Rev. B}, 71:125332, Mar 2005.
\newblock URL: \url{http://link.aps.org/doi/10.1103/PhysRevB.71.125332}, \href
  {http://dx.doi.org/10.1103/PhysRevB.71.125332}
  {\path{doi:10.1103/PhysRevB.71.125332}}.

\bibitem{Elzerman2004}
J.~M. Elzerman, R.~Hanson, L.~H. {Willems van Beveren}, B.~Witkamp, L.~M.~K.
  Vandersypen, and L.~P. Kouwenhoven.
\newblock Single-shot read-out of an individual electron spin in a quantum dot.
\newblock {\em Nature}, 430(6998):431, Jul 2004.
\newblock URL: \url{http://dx.doi.org/10.1038/nature02693}, \href
  {http://dx.doi.org/10.1038/nature02693} {\path{doi:10.1038/nature02693}}.

\bibitem{Royer2016}
B.~{Royer}, A.~L. {Grimsmo}, N.~{Didier}, and A.~{Blais}.
\newblock {Fast and High-Fidelity Entangling Gate through Parametrically
  Modulated Longitudinal Coupling}.
\newblock {\em ArXiv e-prints}, March 2016.
\newblock \href {http://arxiv.org/abs/1603.04424} {\path{arXiv:1603.04424}}.

\bibitem{Beaudoin2016b}
F{\'e}lix Beaudoin, Dany Lachance-Quirion, W~A Coish, and Michel
  Pioro-Ladri{\`e}re.
\newblock Coupling a single electron spin to a microwave resonator: controlling
  transverse and longitudinal couplings.
\newblock {\em Nanotechnology}, 27(46):464003, 2016.
\newblock URL: \url{http://stacks.iop.org/0957-4484/27/i=46/a=464003}.

\bibitem{Borselli2011}
M.~G. Borselli, R.~S. Ross, A.~A. Kiselev, E.~T. Croke, K.~S. Holabird, P.~W.
  Deelman, L.~D. Warren, I.~Alvarado-Rodriguez, I.~Milosavljevic, F.~C. Ku,
  W.~S. Wong, A.~E. Schmitz, M.~Sokolich, M.~F. Gyure, and A.~T. Hunter.
\newblock Measurement of valley splitting in high-symmetry si/sige quantum
  dots.
\newblock {\em Applied Physics Letters}, 98(12), 2011.
\newblock URL:
  \url{http://scitation.aip.org/content/aip/journal/apl/98/12/10.1063/1.3569717},
  \href {http://dx.doi.org/http://dx.doi.org/10.1063/1.3569717}
  {\path{doi:http://dx.doi.org/10.1063/1.3569717}}.

\bibitem{Liu2012}
Z.~Liu, L.~Wang, and K.~Shen.
\newblock Energy spectra of three electrons in si/sige single and vertically
  coupled double quantum dots.
\newblock {\em Phys. Rev. B}, 85:045311, Jan 2012.
\newblock URL: \url{http://link.aps.org/doi/10.1103/PhysRevB.85.045311}, \href
  {http://dx.doi.org/10.1103/PhysRevB.85.045311}
  {\path{doi:10.1103/PhysRevB.85.045311}}.

\bibitem{Tahan2014}
Charles Tahan and Robert Joynt.
\newblock Relaxation of excited spin, orbital, and valley qubit states in ideal
  silicon quantum dots.
\newblock {\em Phys. Rev. B}, 89:075302, Feb 2014.
\newblock URL: \url{http://link.aps.org/doi/10.1103/PhysRevB.89.075302}, \href
  {http://dx.doi.org/10.1103/PhysRevB.89.075302}
  {\path{doi:10.1103/PhysRevB.89.075302}}.

\bibitem{Ferraro2014}
E.~Ferraro, M.~De~Michielis, G.~Mazzeo, M.~Fanciulli, and E.~Prati.
\newblock Effective hamiltonian for the hybrid double quantum dot qubit.
\newblock {\em Quantum Information Processing}, 13(5):1155--1173, 2014.
\newblock URL: \url{http://dx.doi.org/10.1007/s11128-013-0718-2}, \href
  {http://dx.doi.org/10.1007/s11128-013-0718-2}
  {\path{doi:10.1007/s11128-013-0718-2}}.

\bibitem{Shi2011}
Zhan Shi, C.~B. Simmons, J.~R. Prance, John King~Gamble, Mark Friesen, D.~E.
  Savage, M.~G. Lagally, S.~N. Coppersmith, and M.~A. Eriksson.
\newblock Tunable singlet-triplet splitting in a few-electron si/sige quantum
  dot.
\newblock {\em Applied Physics Letters}, 99(23), 2011.
\newblock URL:
  \url{http://scitation.aip.org/content/aip/journal/apl/99/23/10.1063/1.3666232},
  \href {http://dx.doi.org/http://dx.doi.org/10.1063/1.3666232}
  {\path{doi:http://dx.doi.org/10.1063/1.3666232}}.

\bibitem{Boykin2004}
Timothy~B. Boykin, Gerhard Klimeck, Mark Friesen, S.~N. Coppersmith, Paul von
  Allmen, Fabiano Oyafuso, and Seungwon Lee.
\newblock Valley splitting in low-density quantum-confined heterostructures
  studied using tight-binding models.
\newblock {\em Phys. Rev. B}, 70:165325, Oct 2004.
\newblock URL: \url{http://link.aps.org/doi/10.1103/PhysRevB.70.165325}, \href
  {http://dx.doi.org/10.1103/PhysRevB.70.165325}
  {\path{doi:10.1103/PhysRevB.70.165325}}.

\bibitem{Saraiva2009}
A.~L. Saraiva, M.~J. Calder\'on, Xuedong Hu, S.~Das~Sarma, and Belita Koiller.
\newblock Physical mechanisms of interface-mediated intervalley coupling in si.
\newblock {\em Phys. Rev. B}, 80:081305, Aug 2009.
\newblock URL: \url{http://link.aps.org/doi/10.1103/PhysRevB.80.081305}, \href
  {http://dx.doi.org/10.1103/PhysRevB.80.081305}
  {\path{doi:10.1103/PhysRevB.80.081305}}.

\bibitem{Lei2014}
Yan Lei, Yin Wen, and Wang Fang-Wei.
\newblock Hybrid double-dot qubit measurement with a quantum point contact.
\newblock {\em Chinese Physics B}, 23(10):100303, 2014.
\newblock URL: \url{http://stacks.iop.org/1674-1056/23/i=10/a=100303}.

\bibitem{Hanson2005}
R.~Hanson, L.~H.~Willems van Beveren, I.~T. Vink, J.~M. Elzerman, W.~J.~M.
  Naber, F.~H.~L. Koppens, L.~P. Kouwenhoven, and L.~M.~K. Vandersypen.
\newblock Single-shot readout of electron spin states in a quantum dot using
  spin-dependent tunnel rates.
\newblock {\em Phys. Rev. Lett.}, 94:196802, May 2005.
\newblock URL: \url{http://link.aps.org/doi/10.1103/PhysRevLett.94.196802},
  \href {http://dx.doi.org/10.1103/PhysRevLett.94.196802}
  {\path{doi:10.1103/PhysRevLett.94.196802}}.

\bibitem{Simmons2011}
C.~B. Simmons, J.~R. Prance, B.~J. Van~Bael, Teck~Seng Koh, Zhan Shi, D.~E.
  Savage, M.~G. Lagally, R.~Joynt, Mark Friesen, S.~N. Coppersmith, and M.~A.
  Eriksson.
\newblock Tunable spin loading and ${T}_{1}$ of a silicon spin qubit measured
  by single-shot readout.
\newblock {\em Phys. Rev. Lett.}, 106:156804, Apr 2011.
\newblock URL: \url{http://link.aps.org/doi/10.1103/PhysRevLett.106.156804},
  \href {http://dx.doi.org/10.1103/PhysRevLett.106.156804}
  {\path{doi:10.1103/PhysRevLett.106.156804}}.

\bibitem{Ferraro2015}
E.~Ferraro, M.~De~Michielis, M.~Fanciulli, and E.~Prati.
\newblock Coherent tunneling by adiabatic passage of an exchange-only spin
  qubit in a double quantum dot chain.
\newblock {\em Phys. Rev. B}, 91:075435, Feb 2015.
\newblock URL: \url{http://link.aps.org/doi/10.1103/PhysRevB.91.075435}, \href
  {http://dx.doi.org/10.1103/PhysRevB.91.075435}
  {\path{doi:10.1103/PhysRevB.91.075435}}.

\bibitem{Mehl2015}
Sebastian Mehl.
\newblock Two-qubit pulse gate for the three-electron double quantum dot qubit.
\newblock {\em Phys. Rev. B}, 91:035430, Jan 2015.
\newblock URL: \url{http://link.aps.org/doi/10.1103/PhysRevB.91.035430}, \href
  {http://dx.doi.org/10.1103/PhysRevB.91.035430}
  {\path{doi:10.1103/PhysRevB.91.035430}}.

\bibitem{Serina2016}
M.~{Serina}, L.~{Trifunovic}, C.~{Kloeffel}, and D.~{Loss}.
\newblock {Long-Range Interaction between Charge and Spin Qubits in Quantum
  Dots}.
\newblock {\em ArXiv e-prints}, January 2016.
\newblock \href {http://arxiv.org/abs/1601.03564} {\path{arXiv:1601.03564}}.

\bibitem{Rashba2008}
Emmanuel~I. Rashba.
\newblock Theory of electric dipole spin resonance in quantum dots: Mean field
  theory with gaussian fluctuations and beyond.
\newblock {\em Phys. Rev. B}, 78:195302, Nov 2008.
\newblock URL: \url{http://link.aps.org/doi/10.1103/PhysRevB.78.195302}, \href
  {http://dx.doi.org/10.1103/PhysRevB.78.195302}
  {\path{doi:10.1103/PhysRevB.78.195302}}.

\bibitem{Gordon1958}
J.~P. Gordon and K.~D. Bowers.
\newblock Microwave spin echoes from donor electrons in silicon.
\newblock {\em Phys. Rev. Lett.}, 1:368--370, Nov 1958.
\newblock URL: \url{http://link.aps.org/doi/10.1103/PhysRevLett.1.368}, \href
  {http://dx.doi.org/10.1103/PhysRevLett.1.368}
  {\path{doi:10.1103/PhysRevLett.1.368}}.

\bibitem{Chiba1972}
Meiro Chiba and Akira Hirai.
\newblock Electron spin echo decay behaviours of phosphorus doped silicon.
\newblock {\em Journal of the Physical Society of Japan}, 33(3):730--738, 1972.
\newblock URL: \url{http://dx.doi.org/10.1143/JPSJ.33.730}, \href
  {http://arxiv.org/abs/http://dx.doi.org/10.1143/JPSJ.33.730}
  {\path{arXiv:http://dx.doi.org/10.1143/JPSJ.33.730}}, \href
  {http://dx.doi.org/10.1143/JPSJ.33.730} {\path{doi:10.1143/JPSJ.33.730}}.

\bibitem{Tezuka2010}
H.~Tezuka, A.~R. Stegner, A.~M. Tyryshkin, S.~Shankar, M.~L.~W. Thewalt, S.~A.
  Lyon, K.~M. Itoh, and M.~S. Brandt.
\newblock Electron paramagnetic resonance of boron acceptors in isotopically
  purified silicon.
\newblock {\em Phys. Rev. B}, 81:161203, Apr 2010.
\newblock URL: \url{http://link.aps.org/doi/10.1103/PhysRevB.81.161203}, \href
  {http://dx.doi.org/10.1103/PhysRevB.81.161203}
  {\path{doi:10.1103/PhysRevB.81.161203}}.

\bibitem{DiVincenzo1995}
David~P. DiVincenzo.
\newblock Two-bit gates are universal for quantum computation.
\newblock {\em Phys. Rev. A}, 51:1015, Feb 1995.
\newblock URL: \url{http://link.aps.org/doi/10.1103/PhysRevA.51.1015}, \href
  {http://dx.doi.org/10.1103/PhysRevA.51.1015}
  {\path{doi:10.1103/PhysRevA.51.1015}}.

\bibitem{Lloyd1995}
Seth Lloyd.
\newblock Almost any quantum logic gate is universal.
\newblock {\em Phys. Rev. Lett.}, 75:346--349, Jul 1995.
\newblock URL: \url{http://link.aps.org/doi/10.1103/PhysRevLett.75.346}, \href
  {http://dx.doi.org/10.1103/PhysRevLett.75.346}
  {\path{doi:10.1103/PhysRevLett.75.346}}.

\bibitem{Zeuch2016}
Daniel Zeuch and N.~E. Bonesteel.
\newblock Simple derivation of the fong-wandzura pulse sequence.
\newblock {\em Phys. Rev. A}, 93:010303, Jan 2016.
\newblock URL: \url{http://link.aps.org/doi/10.1103/PhysRevA.93.010303}, \href
  {http://dx.doi.org/10.1103/PhysRevA.93.010303}
  {\path{doi:10.1103/PhysRevA.93.010303}}.

\bibitem{Wardrop2016}
Matthew~P. Wardrop and Andrew~C. Doherty.
\newblock Characterization of an exchange-based two-qubit gate for resonant
  exchange qubits.
\newblock {\em Phys. Rev. B}, 93:075436, Feb 2016.
\newblock URL: \url{http://link.aps.org/doi/10.1103/PhysRevB.93.075436}, \href
  {http://dx.doi.org/10.1103/PhysRevB.93.075436}
  {\path{doi:10.1103/PhysRevB.93.075436}}.

\bibitem{Zhang2003}
Jun Zhang, Jiri Vala, Shankar Sastry, and K.~Birgitta Whaley.
\newblock Geometric theory of nonlocal two-qubit operations.
\newblock {\em Phys. Rev. A}, 67:042313, Apr 2003.
\newblock URL: \url{http://link.aps.org/doi/10.1103/PhysRevA.67.042313}, \href
  {http://dx.doi.org/10.1103/PhysRevA.67.042313}
  {\path{doi:10.1103/PhysRevA.67.042313}}.

\bibitem{Fowler2012}
Austin~G. Fowler, Matteo Mariantoni, John~M. Martinis, and Andrew~N. Cleland.
\newblock Surface codes: Towards practical large-scale quantum computation.
\newblock {\em Phys. Rev. A}, 86:032324, Sep 2012.
\newblock URL: \url{http://link.aps.org/doi/10.1103/PhysRevA.86.032324}, \href
  {http://dx.doi.org/10.1103/PhysRevA.86.032324}
  {\path{doi:10.1103/PhysRevA.86.032324}}.

\bibitem{Lidar2013}
D.A. Lidar and T.A. Brun.
\newblock {\em Quantum Error Correction}.
\newblock Cambridge University Press, 2013.
\newblock URL: \url{https://books.google.de/books?id=XV9sAAAAQBAJ}.

\bibitem{Brecht2016}
Teresa Brecht, Wolfgang Pfaff, Chen Wang, Yiwen Chu, Luigi Frunzio, Michel~H
  Devoret, and Robert~J Schoelkopf.
\newblock Multilayer microwave integrated quantum circuits for scalable quantum
  computing.
\newblock {\em Npj Quantum Information}, 2:16002 EP --, 02 2016.
\newblock URL: \url{http://dx.doi.org/10.1038/npjqi.2016.2}.

\bibitem{Leijnse2013}
Martin Leijnse and Karsten Flensberg.
\newblock Coupling spin qubits via superconductors.
\newblock {\em Phys. Rev. Lett.}, 111:060501, Aug 2013.
\newblock URL: \url{http://link.aps.org/doi/10.1103/PhysRevLett.111.060501},
  \href {http://dx.doi.org/10.1103/PhysRevLett.111.060501}
  {\path{doi:10.1103/PhysRevLett.111.060501}}.

\bibitem{Hassler2015}
Fabian Hassler, Gianluigi Catelani, and Hendrik Bluhm.
\newblock Exchange interaction of two spin qubits mediated by a superconductor.
\newblock {\em Phys. Rev. B}, 92:235401, Dec 2015.
\newblock URL: \url{http://link.aps.org/doi/10.1103/PhysRevB.92.235401}, \href
  {http://dx.doi.org/10.1103/PhysRevB.92.235401}
  {\path{doi:10.1103/PhysRevB.92.235401}}.

\bibitem{Stotz2005}
James A.~H. Stotz, Rudolf Hey, Paulo~V. Santos, and Klaus~H. Ploog.
\newblock Coherent spin transport through dynamic quantum dots.
\newblock {\em Nat Mater}, 4(8):585--588, 08 2005.
\newblock URL: \url{http://dx.doi.org/10.1038/nmat1430}.

\bibitem{Hermelin2011}
Sylvain Hermelin, Shintaro Takada, Michihisa Yamamoto, Seigo Tarucha,
  Andreas~D. Wieck, Laurent Saminadayar, Christopher Bauerle, and Tristan
  Meunier.
\newblock Electrons surfing on a sound wave as a platform for quantum optics
  with flying electrons.
\newblock {\em Nature}, 477(7365):435--438, 09 2011.
\newblock URL: \url{http://dx.doi.org/10.1038/nature10416}.

\bibitem{McNeil2011}
R.~P.~G. McNeil, M.~Kataoka, C.~J.~B. Ford, C.~H.~W. Barnes, D.~Anderson,
  G.~A.~C. Jones, I.~Farrer, and D.~A. Ritchie.
\newblock On-demand single-electron transfer between distant quantum dots.
\newblock {\em Nature}, 477(7365):439--442, 09 2011.
\newblock URL: \url{http://dx.doi.org/10.1038/nature10444}.

\bibitem{Schuetz2015}
M.~J.~A. Schuetz, E.~M. Kessler, G.~Giedke, L.~M.~K. Vandersypen, M.~D. Lukin,
  and J.~I. Cirac.
\newblock Universal quantum transducers based on surface acoustic waves.
\newblock {\em Phys. Rev. X}, 5:031031, Sep 2015.
\newblock URL: \url{http://link.aps.org/doi/10.1103/PhysRevX.5.031031}, \href
  {http://dx.doi.org/10.1103/PhysRevX.5.031031}
  {\path{doi:10.1103/PhysRevX.5.031031}}.

\bibitem{Benito2016}
M.~Benito, M.~J.~A. Schuetz, J.~I. Cirac, G.~Platero, and G.~Giedke.
\newblock Dissipative long-range entanglement generation between electronic
  spins.
\newblock {\em Phys. Rev. B}, 94:115404, Sep 2016.
\newblock URL: \url{http://link.aps.org/doi/10.1103/PhysRevB.94.115404}, \href
  {http://dx.doi.org/10.1103/PhysRevB.94.115404}
  {\path{doi:10.1103/PhysRevB.94.115404}}.

\bibitem{Bertrand2016}
Bertrand B., Hermelin S., Takada S., Yamamoto M., Tarucha S., Ludwig A.,
  Wieck~A. D., B{\"a}uerle C., and Meunier T.
\newblock Fast spin information transfer between distant quantum dots using
  individual electrons.
\newblock {\em Nat Nano}, 11(8):672--676, 08 2016.
\newblock URL: \url{http://dx.doi.org/10.1038/nnano.2016.82}.

\bibitem{Trifunovic2013}
Luka Trifunovic, Fabio~L. Pedrocchi, and Daniel Loss.
\newblock Long-distance entanglement of spin qubits via ferromagnet.
\newblock {\em Phys. Rev. X}, 3:041023, Dec 2013.
\newblock URL: \url{http://link.aps.org/doi/10.1103/PhysRevX.3.041023}, \href
  {http://dx.doi.org/10.1103/PhysRevX.3.041023}
  {\path{doi:10.1103/PhysRevX.3.041023}}.

\bibitem{Braakman2013}
Braakman~F. R., Barthelemy P., Reichl C., Wegscheider W., and Vandersypen L.~M.
  K.
\newblock Long-distance coherent coupling in a quantum dot array.
\newblock {\em Nat Nano}, 8(6):432--437, 06 2013.
\newblock URL: \url{http://dx.doi.org/10.1038/nnano.2013.67}.

\bibitem{Sanchez2014b}
Rafael S\'anchez, Fernando Gallego-Marcos, and Gloria Platero.
\newblock Superexchange blockade in triple quantum dots.
\newblock {\em Phys. Rev. B}, 89:161402, Apr 2014.
\newblock URL: \url{http://link.aps.org/doi/10.1103/PhysRevB.89.161402}, \href
  {http://dx.doi.org/10.1103/PhysRevB.89.161402}
  {\path{doi:10.1103/PhysRevB.89.161402}}.

\bibitem{Srinivasa2015b}
V.~Srinivasa, H.~Xu, and J.~M. Taylor.
\newblock Tunable spin-qubit coupling mediated by a multielectron quantum dot.
\newblock {\em Phys. Rev. Lett.}, 114:226803, Jun 2015.
\newblock URL: \url{http://link.aps.org/doi/10.1103/PhysRevLett.114.226803},
  \href {http://dx.doi.org/10.1103/PhysRevLett.114.226803}
  {\path{doi:10.1103/PhysRevLett.114.226803}}.

\bibitem{Kuo2014}
David M.~T. Kuo and Yia-chung Chang.
\newblock Long-distance coherent tunneling effect on the charge and heat
  currents in serially coupled triple quantum dots.
\newblock {\em Phys. Rev. B}, 89:115416, Mar 2014.
\newblock URL: \url{http://link.aps.org/doi/10.1103/PhysRevB.89.115416}, \href
  {http://dx.doi.org/10.1103/PhysRevB.89.115416}
  {\path{doi:10.1103/PhysRevB.89.115416}}.

\bibitem{Menchon2016}
R~Menchon-Enrich, A~Benseny, V~Ahufinger, A~D Greentree, Th~Busch, and
  J~Mompart.
\newblock Spatial adiabatic passage: a review of recent progress.
\newblock {\em Reports on Progress in Physics}, 79(7):074401, 2016.
\newblock URL: \url{http://stacks.iop.org/0034-4885/79/i=7/a=074401}.

\bibitem{Braakman2014}
F.~R. Braakman, J.~Danon, L.~R. Schreiber, W.~Wegscheider, and L.~M.~K.
  Vandersypen.
\newblock Dynamics of spin-flip photon-assisted tunneling.
\newblock {\em Phys. Rev. B}, 89:075417, Feb 2014.
\newblock URL: \url{http://link.aps.org/doi/10.1103/PhysRevB.89.075417}, \href
  {http://dx.doi.org/10.1103/PhysRevB.89.075417}
  {\path{doi:10.1103/PhysRevB.89.075417}}.

\bibitem{Gallego2015}
Fernando Gallego-Marcos, Rafael SÃ¡nchez, and Gloria Platero.
\newblock Photon assisted long-range tunneling.
\newblock {\em Journal of Applied Physics}, 117(11), 2015.
\newblock URL:
  \url{http://scitation.aip.org/content/aip/journal/jap/117/11/10.1063/1.4913834},
  \href {http://dx.doi.org/http://dx.doi.org/10.1063/1.4913834}
  {\path{doi:http://dx.doi.org/10.1063/1.4913834}}.

\bibitem{Stano2015}
Peter Stano, Jelena Klinovaja, Floris~R. Braakman, Lieven M.~K. Vandersypen,
  and Daniel Loss.
\newblock Fast long-distance control of spin qubits by photon-assisted
  cotunneling.
\newblock {\em Phys. Rev. B}, 92:075302, Aug 2015.
\newblock URL: \url{http://link.aps.org/doi/10.1103/PhysRevB.92.075302}, \href
  {http://dx.doi.org/10.1103/PhysRevB.92.075302}
  {\path{doi:10.1103/PhysRevB.92.075302}}.

\bibitem{Yang2016}
Guang Yang, Chen-Hsuan Hsu, Peter Stano, Jelena Klinovaja, and Daniel Loss.
\newblock Long-distance entanglement of spin qubits via quantum hall edge
  states.
\newblock {\em Phys. Rev. B}, 93:075301, Feb 2016.
\newblock URL: \url{http://link.aps.org/doi/10.1103/PhysRevB.93.075301}, \href
  {http://dx.doi.org/10.1103/PhysRevB.93.075301}
  {\path{doi:10.1103/PhysRevB.93.075301}}.

\bibitem{Pal2014}
Arijeet Pal, Emmanuel~I. Rashba, and Bertrand~I. Halperin.
\newblock Driven nonlinear dynamics of two coupled exchange-only qubits.
\newblock {\em Phys. Rev. X}, 4:011012, Jan 2014.
\newblock URL: \url{http://link.aps.org/doi/10.1103/PhysRevX.4.011012}, \href
  {http://dx.doi.org/10.1103/PhysRevX.4.011012}
  {\path{doi:10.1103/PhysRevX.4.011012}}.

\bibitem{Pal2015}
Arijeet Pal, Emmanuel~I. Rashba, and Bertrand~I. Halperin.
\newblock Exact cnot gates with a single nonlocal rotation for quantum-dot
  qubits.
\newblock {\em Phys. Rev. B}, 92:125409, Sep 2015.
\newblock URL: \url{http://link.aps.org/doi/10.1103/PhysRevB.92.125409}, \href
  {http://dx.doi.org/10.1103/PhysRevB.92.125409}
  {\path{doi:10.1103/PhysRevB.92.125409}}.

\bibitem{Puri2016}
Shruti Puri and Alexandre Blais.
\newblock High-fidelity resonator-induced phase gate with single-mode
  squeezing.
\newblock {\em Phys. Rev. Lett.}, 116:180501, May 2016.
\newblock URL: \url{http://link.aps.org/doi/10.1103/PhysRevLett.116.180501},
  \href {http://dx.doi.org/10.1103/PhysRevLett.116.180501}
  {\path{doi:10.1103/PhysRevLett.116.180501}}.

\bibitem{Schuetz2016}
M.~J.~A. {Schuetz}, G.~{Giedke}, L.~M.~K. {Vandersypen}, and J.~I. {Cirac}.
\newblock {High-Fidelity Hot Gates for Generic Spin-Resonator Systems}.
\newblock {\em ArXiv e-prints}, July 2016.
\newblock \href {http://arxiv.org/abs/1607.01614} {\path{arXiv:1607.01614}}.

\bibitem{Blais2004}
Alexandre Blais, Ren-Shou Huang, Andreas Wallraff, S.~M. Girvin, and R.~J.
  Schoelkopf.
\newblock Cavity quantum electrodynamics for superconducting electrical
  circuits: An architecture for quantum computation.
\newblock {\em Phys. Rev. A}, 69:062320, Jun 2004.
\newblock URL: \url{http://link.aps.org/doi/10.1103/PhysRevA.69.062320}, \href
  {http://dx.doi.org/10.1103/PhysRevA.69.062320}
  {\path{doi:10.1103/PhysRevA.69.062320}}.

\bibitem{Schoelkopf2008}
R.~J. Schoelkopf and S.~M. Girvin.
\newblock Wiring up quantum systems.
\newblock {\em Nature}, 451(7179):664--669, 02 2008.
\newblock URL: \url{http://dx.doi.org/10.1038/451664a}.

\bibitem{Houck2012}
Andrew~A. Houck, Hakan~E. Tureci, and Jens Koch.
\newblock On-chip quantum simulation with superconducting circuits.
\newblock {\em Nat Phys}, 8(4):292--299, 04 2012.
\newblock URL: \url{http://dx.doi.org/10.1038/nphys2251}.

\bibitem{Cohen1989}
Claude Cohen-Tannoudji, Jacques Dupont-Roc, and Gilbert Grynberg.
\newblock {\em Photons and Atoms: Introduction to Quantum Electrodynamics}.
\newblock Wiley-VCH Verlag GmbH, 2007.
\newblock URL: \url{http://dx.doi.org/10.1002/9783527618422}, \href
  {http://dx.doi.org/10.1002/9783527618422} {\path{doi:10.1002/9783527618422}}.

\bibitem{Sorensen2000}
Anders S\o{}rensen and Klaus M\o{}lmer.
\newblock Entanglement and quantum computation with ions in thermal motion.
\newblock {\em Phys. Rev. A}, 62:022311, Jul 2000.
\newblock URL: \url{http://link.aps.org/doi/10.1103/PhysRevA.62.022311}, \href
  {http://dx.doi.org/10.1103/PhysRevA.62.022311}
  {\path{doi:10.1103/PhysRevA.62.022311}}.

\bibitem{Marzari2012}
Nicola Marzari, Arash~A. Mostofi, Jonathan~R. Yates, Ivo Souza, and David
  Vanderbilt.
\newblock Maximally localized wannier functions: Theory and applications.
\newblock {\em Rev. Mod. Phys.}, 84:1419, Oct 2012.
\newblock URL: \url{http://link.aps.org/doi/10.1103/RevModPhys.84.1419}, \href
  {http://dx.doi.org/10.1103/RevModPhys.84.1419}
  {\path{doi:10.1103/RevModPhys.84.1419}}.

\bibitem{Billangeon2015}
P.-M. Billangeon, J.~S. Tsai, and Y.~Nakamura.
\newblock Circuit-qed-based scalable architectures for quantum information
  processing with superconducting qubits.
\newblock {\em Phys. Rev. B}, 91:094517, Mar 2015.
\newblock URL: \url{http://link.aps.org/doi/10.1103/PhysRevB.91.094517}, \href
  {http://dx.doi.org/10.1103/PhysRevB.91.094517}
  {\path{doi:10.1103/PhysRevB.91.094517}}.

\bibitem{Richer2016}
Susanne Richer and David DiVincenzo.
\newblock Circuit design implementing longitudinal coupling: A scalable scheme
  for superconducting qubits.
\newblock {\em Phys. Rev. B}, 93:134501, Apr 2016.
\newblock URL: \url{http://link.aps.org/doi/10.1103/PhysRevB.93.134501}, \href
  {http://dx.doi.org/10.1103/PhysRevB.93.134501}
  {\path{doi:10.1103/PhysRevB.93.134501}}.

\bibitem{Ruskov2016}
Rusko Ruskov and Charles Tahan.
\newblock Spin-cavity longitudinal coupling for two-qubit gates and
  measurement.
\newblock {\em Contributed talk, APS 2016, Baltimore}, 2016.
\newblock URL: \url{http://meetings.aps.org/link/BAPS.2016.MAR.E45.7}.

\bibitem{Jaynes1963}
E.T. Jaynes and F.W. Cummings.
\newblock Comparison of quantum and semiclassical radiation theories with
  application to the beam maser.
\newblock {\em Proceedings of the IEEE}, 51(1):89, Jan 1963.
\newblock \href {http://dx.doi.org/10.1109/PROC.1963.1664}
  {\path{doi:10.1109/PROC.1963.1664}}.

\bibitem{Cummings1965}
F.~W. Cummings.
\newblock Stimulated emission of radiation in a single mode.
\newblock {\em Phys. Rev.}, 140:A1051--A1056, Nov 1965.
\newblock URL: \url{http://link.aps.org/doi/10.1103/PhysRev.140.A1051}, \href
  {http://dx.doi.org/10.1103/PhysRev.140.A1051}
  {\path{doi:10.1103/PhysRev.140.A1051}}.

\bibitem{Shore1993}
B.~W. {Shore} and P.~L. {Knight}.
\newblock {The Jaynes-Cummings Model}.
\newblock {\em Journal of Modern Optics}, 40:1195--1238, July 1993.
\newblock \href {http://dx.doi.org/10.1080/09500349314551321}
  {\path{doi:10.1080/09500349314551321}}.

\bibitem{Wu2007}
Ying Wu and Xiaoxue Yang.
\newblock Strong-coupling theory of periodically driven two-level systems.
\newblock {\em Phys. Rev. Lett.}, 98:013601, Jan 2007.
\newblock URL: \url{http://link.aps.org/doi/10.1103/PhysRevLett.98.013601},
  \href {http://dx.doi.org/10.1103/PhysRevLett.98.013601}
  {\path{doi:10.1103/PhysRevLett.98.013601}}.

\bibitem{Blais2007}
Alexandre Blais, Jay Gambetta, A.~Wallraff, D.~I. Schuster, S.~M. Girvin, M.~H.
  Devoret, and R.~J. Schoelkopf.
\newblock Quantum-information processing with circuit quantum electrodynamics.
\newblock {\em Phys. Rev. A}, 75:032329, Mar 2007.
\newblock URL: \url{http://link.aps.org/doi/10.1103/PhysRevA.75.032329}, \href
  {http://dx.doi.org/10.1103/PhysRevA.75.032329}
  {\path{doi:10.1103/PhysRevA.75.032329}}.

\bibitem{Schuch2003}
Norbert Schuch and Jens Siewert.
\newblock Natural two-qubit gate for quantum computation using the
  $\mathrm{XY}$ interaction.
\newblock {\em Phys. Rev. A}, 67:032301, Mar 2003.
\newblock URL: \url{http://link.aps.org/doi/10.1103/PhysRevA.67.032301}, \href
  {http://dx.doi.org/10.1103/PhysRevA.67.032301}
  {\path{doi:10.1103/PhysRevA.67.032301}}.

\bibitem{Tanamoto2008}
Tetsufumi Tanamoto, Koji Maruyama, Yu-xi Liu, Xuedong Hu, and Franco Nori.
\newblock Efficient purification protocols using $\mathrm{iSWAP}$ gates in
  solid-state qubits.
\newblock {\em Phys. Rev. A}, 78:062313, Dec 2008.
\newblock URL: \url{http://link.aps.org/doi/10.1103/PhysRevA.78.062313}, \href
  {http://dx.doi.org/10.1103/PhysRevA.78.062313}
  {\path{doi:10.1103/PhysRevA.78.062313}}.

\bibitem{Childs2000}
Andrew~M. Childs and Isaac~L. Chuang.
\newblock Universal quantum computation with two-level trapped ions.
\newblock {\em Phys. Rev. A}, 63:012306, Dec 2000.
\newblock URL: \url{http://link.aps.org/doi/10.1103/PhysRevA.63.012306}, \href
  {http://dx.doi.org/10.1103/PhysRevA.63.012306}
  {\path{doi:10.1103/PhysRevA.63.012306}}.

\bibitem{Wallraff2007}
A.~Wallraff, D.~I. Schuster, A.~Blais, J.~M. Gambetta, J.~Schreier, L.~Frunzio,
  M.~H. Devoret, S.~M. Girvin, and R.~J. Schoelkopf.
\newblock Sideband transitions and two-tone spectroscopy of a superconducting
  qubit strongly coupled to an on-chip cavity.
\newblock {\em Phys. Rev. Lett.}, 99:050501, Jul 2007.
\newblock URL: \url{http://link.aps.org/doi/10.1103/PhysRevLett.99.050501},
  \href {http://dx.doi.org/10.1103/PhysRevLett.99.050501}
  {\path{doi:10.1103/PhysRevLett.99.050501}}.

\bibitem{Leek2009}
P.~J. Leek, S.~Filipp, P.~Maurer, M.~Baur, R.~Bianchetti, J.~M. Fink,
  M.~G\"oppl, L.~Steffen, and A.~Wallraff.
\newblock Using sideband transitions for two-qubit operations in
  superconducting circuits.
\newblock {\em Phys. Rev. B}, 79:180511, May 2009.
\newblock URL: \url{http://link.aps.org/doi/10.1103/PhysRevB.79.180511}, \href
  {http://dx.doi.org/10.1103/PhysRevB.79.180511}
  {\path{doi:10.1103/PhysRevB.79.180511}}.

\bibitem{Chow2013}
Jerry~M Chow, Jay~M Gambetta, Andrew~W Cross, Seth~T Merkel, Chad Rigetti, and
  M~Steffen.
\newblock Microwave-activated conditional-phase gate for superconducting
  qubits.
\newblock {\em New Journal of Physics}, 15(11):115012, 2013.
\newblock URL: \url{http://stacks.iop.org/1367-2630/15/i=11/a=115012}.

\bibitem{Sapmaz2006}
Sami Sapmaz, Pablo Jarillo-Herrero, Leo~P Kouwenhoven, and Herre S~J van~der
  Zant.
\newblock Quantum dots in carbon nanotubes.
\newblock {\em Semiconductor Science and Technology}, 21(11):S52, 2006.
\newblock URL: \url{http://stacks.iop.org/0268-1242/21/i=11/a=S08}.

\bibitem{Zanardi1997}
Paolo Zanardi and Mario Rasetti.
\newblock Error avoiding quantum codes.
\newblock {\em Modern Physics Letters B}, 11(25):1085--1093, 1997.
\newblock URL:
  \url{http://www.worldscientific.com/doi/abs/10.1142/S0217984997001304}, \href
  {http://arxiv.org/abs/http://www.worldscientific.com/doi/pdf/10.1142/S0217984997001304}
  {\path{arXiv:http://www.worldscientific.com/doi/pdf/10.1142/S0217984997001304}},
  \href {http://dx.doi.org/10.1142/S0217984997001304}
  {\path{doi:10.1142/S0217984997001304}}.

\bibitem{Lidar1998}
D.~A. Lidar, I.~L. Chuang, and K.~B. Whaley.
\newblock Decoherence-free subspaces for quantum computation.
\newblock {\em Phys. Rev. Lett.}, 81:2594--2597, Sep 1998.
\newblock URL: \url{http://link.aps.org/doi/10.1103/PhysRevLett.81.2594}, \href
  {http://dx.doi.org/10.1103/PhysRevLett.81.2594}
  {\path{doi:10.1103/PhysRevLett.81.2594}}.

\bibitem{Viola2000}
Lorenza Viola, Emanuel Knill, and Seth Lloyd.
\newblock Dynamical generation of noiseless quantum subsystems.
\newblock {\em Phys. Rev. Lett.}, 85:3520--3523, Oct 2000.
\newblock URL: \url{http://link.aps.org/doi/10.1103/PhysRevLett.85.3520}, \href
  {http://dx.doi.org/10.1103/PhysRevLett.85.3520}
  {\path{doi:10.1103/PhysRevLett.85.3520}}.

\bibitem{Overhauser1953}
Albert~W. Overhauser.
\newblock Polarization of nuclei in metals.
\newblock {\em Phys. Rev.}, 92:411--415, Oct 1953.
\newblock URL: \url{http://link.aps.org/doi/10.1103/PhysRev.92.411}, \href
  {http://dx.doi.org/10.1103/PhysRev.92.411}
  {\path{doi:10.1103/PhysRev.92.411}}.

\bibitem{Abragam1961}
A.~Abragam.
\newblock {\em The Principles of Nuclear Magnetism}.
\newblock International series of monographs on physics. Clarendon Press, 1961.
\newblock URL: \url{https://books.google.de/books?id=9M8U\_JK7K54C}.

\bibitem{Slichter2010}
Charles~P. Slichter.
\newblock {\em Principles of Magnetic Resonance (Springer Series in Solid-State
  Sciences)}.
\newblock Springer, 2010.
\newblock URL:
  \url{https://www.amazon.com/Principles-Magnetic-Resonance-Springer-Solid-State/dp/3642080693%3FSubscriptionId%3D0JYN1NVW651KCA56C102%26tag%3Dtechkie-20%26linkCode%3Dxm2%26camp%3D2025%26creative%3D165953%26creativeASIN%3D3642080693}.

\bibitem{Hung2013}
Jo-Tzu Hung, \L{}ukasz Cywi\ifmmode~\acute{n}\else \'{n}\fi{}ski, Xuedong Hu,
  and S.~Das~Sarma.
\newblock Hyperfine interaction induced dephasing of coupled spin qubits in
  semiconductor double quantum dots.
\newblock {\em Phys. Rev. B}, 88:085314, Aug 2013.
\newblock URL: \url{http://link.aps.org/doi/10.1103/PhysRevB.88.085314}, \href
  {http://dx.doi.org/10.1103/PhysRevB.88.085314}
  {\path{doi:10.1103/PhysRevB.88.085314}}.

\bibitem{Delbecq2016}
M.~R. Delbecq, T.~Nakajima, P.~Stano, T.~Otsuka, S.~Amaha, J.~Yoneda,
  K.~Takeda, G.~Allison, A.~Ludwig, A.~D. Wieck, and S.~Tarucha.
\newblock Quantum dephasing in a gated gaas triple quantum dot due to
  nonergodic noise.
\newblock {\em Phys. Rev. Lett.}, 116:046802, Jan 2016.
\newblock URL: \url{http://link.aps.org/doi/10.1103/PhysRevLett.116.046802},
  \href {http://dx.doi.org/10.1103/PhysRevLett.116.046802}
  {\path{doi:10.1103/PhysRevLett.116.046802}}.

\bibitem{Falci2004}
G.~Falci, A.~D'Arrigo, A.~Mastellone, and E.~Paladino.
\newblock Dynamical suppression of telegraph and $1∕f$ noise due to quantum
  bistable fluctuators.
\newblock {\em Phys. Rev. A}, 70:040101, Oct 2004.
\newblock URL: \url{http://link.aps.org/doi/10.1103/PhysRevA.70.040101}, \href
  {http://dx.doi.org/10.1103/PhysRevA.70.040101}
  {\path{doi:10.1103/PhysRevA.70.040101}}.

\bibitem{Faoro2004}
Lara Faoro and Lorenza Viola.
\newblock Dynamical suppression of $1/f$ noise processes in qubit systems.
\newblock {\em Phys. Rev. Lett.}, 92:117905, Mar 2004.
\newblock URL: \url{http://link.aps.org/doi/10.1103/PhysRevLett.92.117905},
  \href {http://dx.doi.org/10.1103/PhysRevLett.92.117905}
  {\path{doi:10.1103/PhysRevLett.92.117905}}.

\bibitem{Rebentrost2009}
P.~Rebentrost, I.~Serban, T.~Schulte-Herbr\"uggen, and F.~K. Wilhelm.
\newblock Optimal control of a qubit coupled to a non-markovian environment.
\newblock {\em Phys. Rev. Lett.}, 102:090401, Mar 2009.
\newblock URL: \url{http://link.aps.org/doi/10.1103/PhysRevLett.102.090401},
  \href {http://dx.doi.org/10.1103/PhysRevLett.102.090401}
  {\path{doi:10.1103/PhysRevLett.102.090401}}.

\bibitem{Du2009}
Jiangfeng Du, Xing Rong, Nan Zhao, Ya~Wang, Jiahui Yang, and R.~B. Liu.
\newblock Preserving electron spin coherence in solids by optimal dynamical
  decoupling.
\newblock {\em Nature}, 461(7268):1265--1268, 10 2009.
\newblock URL: \url{http://dx.doi.org/10.1038/nature08470}.

\bibitem{Green2013}
Todd~J Green, Jarrah Sastrawan, Hermann Uys, and Michael~J Biercuk.
\newblock Arbitrary quantum control of qubits in the presence of universal
  noise.
\newblock {\em New Journal of Physics}, 15(9):095004, 2013.
\newblock URL: \url{http://stacks.iop.org/1367-2630/15/i=9/a=095004}.

\bibitem{Szankowski2016}
Piotr Sza\ifmmode~\acute{n}\else \'{n}\fi{}kowski, Marek Trippenbach, and
  \L{}ukasz Cywi\ifmmode~\acute{n}\else \'{n}\fi{}ski.
\newblock Spectroscopy of cross correlations of environmental noises with two
  qubits.
\newblock {\em Phys. Rev. A}, 94:012109, Jul 2016.
\newblock URL: \url{http://link.aps.org/doi/10.1103/PhysRevA.94.012109}, \href
  {http://dx.doi.org/10.1103/PhysRevA.94.012109}
  {\path{doi:10.1103/PhysRevA.94.012109}}.

\bibitem{Carr1954}
H.~Y. Carr and E.~M. Purcell.
\newblock Effects of diffusion on free precession in nuclear magnetic resonance
  experiments.
\newblock {\em Phys. Rev.}, 94:630--638, May 1954.
\newblock URL: \url{http://link.aps.org/doi/10.1103/PhysRev.94.630}, \href
  {http://dx.doi.org/10.1103/PhysRev.94.630}
  {\path{doi:10.1103/PhysRev.94.630}}.

\bibitem{Meiboom1958}
S.~Meiboom and D.~Gill.
\newblock Modified spinâecho method for measuring nuclear relaxation
  times.
\newblock {\em Review of Scientific Instruments}, 29(8):688--691, 1958.
\newblock URL:
  \url{http://scitation.aip.org/content/aip/journal/rsi/29/8/10.1063/1.1716296},
  \href {http://dx.doi.org/http://dx.doi.org/10.1063/1.1716296}
  {\path{doi:http://dx.doi.org/10.1063/1.1716296}}.

\bibitem{Uhrig2007}
G\"otz~S. Uhrig.
\newblock Keeping a quantum bit alive by optimized $\ensuremath{\pi}$-pulse
  sequences.
\newblock {\em Phys. Rev. Lett.}, 98:100504, Mar 2007.
\newblock URL: \url{http://link.aps.org/doi/10.1103/PhysRevLett.98.100504},
  \href {http://dx.doi.org/10.1103/PhysRevLett.98.100504}
  {\path{doi:10.1103/PhysRevLett.98.100504}}.

\bibitem{Uhrig2008}
G\"otz~S. Uhrig.
\newblock Exact results on dynamical decoupling by Ï pulses in quantum
  information processes.
\newblock {\em New Journal of Physics}, 10(8):083024, 2008.
\newblock URL: \url{http://stacks.iop.org/1367-2630/10/i=8/a=083024}.

\bibitem{Uhrig2009}
G\"otz~S. Uhrig.
\newblock Concatenated control sequences based on optimized dynamic decoupling.
\newblock {\em Phys. Rev. Lett.}, 102:120502, Mar 2009.
\newblock URL: \url{http://link.aps.org/doi/10.1103/PhysRevLett.102.120502},
  \href {http://dx.doi.org/10.1103/PhysRevLett.102.120502}
  {\path{doi:10.1103/PhysRevLett.102.120502}}.

\bibitem{Uys2009}
Hermann Uys, Michael~J. Biercuk, and John~J. Bollinger.
\newblock Optimized noise filtration through dynamical decoupling.
\newblock {\em Phys. Rev. Lett.}, 103:040501, Jul 2009.
\newblock URL: \url{http://link.aps.org/doi/10.1103/PhysRevLett.103.040501},
  \href {http://dx.doi.org/10.1103/PhysRevLett.103.040501}
  {\path{doi:10.1103/PhysRevLett.103.040501}}.

\bibitem{Johnson1928}
J.~B. Johnson.
\newblock Thermal agitation of electricity in conductors.
\newblock {\em Phys. Rev.}, 32:97, Jul 1928.
\newblock URL: \url{http://link.aps.org/doi/10.1103/PhysRev.32.97}, \href
  {http://dx.doi.org/10.1103/PhysRev.32.97} {\path{doi:10.1103/PhysRev.32.97}}.

\bibitem{Nyquist1928}
H.~Nyquist.
\newblock Thermal agitation of electric charge in conductors.
\newblock {\em Phys. Rev.}, 32:110, Jul 1928.
\newblock URL: \url{http://link.aps.org/doi/10.1103/PhysRev.32.110}, \href
  {http://dx.doi.org/10.1103/PhysRev.32.110}
  {\path{doi:10.1103/PhysRev.32.110}}.

\bibitem{Langsjoen2012}
Luke~S. Langsjoen, Amrit Poudel, Maxim~G. Vavilov, and Robert Joynt.
\newblock Qubit relaxation from evanescent-wave johnson noise.
\newblock {\em Phys. Rev. A}, 86:010301, Jul 2012.
\newblock URL: \url{http://link.aps.org/doi/10.1103/PhysRevA.86.010301}, \href
  {http://dx.doi.org/10.1103/PhysRevA.86.010301}
  {\path{doi:10.1103/PhysRevA.86.010301}}.

\bibitem{Poudel2013}
Amrit Poudel, Luke~S. Langsjoen, Maxim~G. Vavilov, and Robert Joynt.
\newblock Relaxation in quantum dots due to evanescent-wave johnson noise.
\newblock {\em Phys. Rev. B}, 87:045301, Jan 2013.
\newblock URL: \url{http://link.aps.org/doi/10.1103/PhysRevB.87.045301}, \href
  {http://dx.doi.org/10.1103/PhysRevB.87.045301}
  {\path{doi:10.1103/PhysRevB.87.045301}}.

\bibitem{Schottky1918}
W.~Schottky.
\newblock {\"U}ber spontane stromschwankungen in verschiedenen
  elektrizitätsleitern.
\newblock {\em Annalen der Physik}, 362(23):541--567, 1918.
\newblock URL: \url{http://dx.doi.org/10.1002/andp.19183622304}, \href
  {http://dx.doi.org/10.1002/andp.19183622304}
  {\path{doi:10.1002/andp.19183622304}}.

\bibitem{Ramon2015}
Guy Ramon.
\newblock Non-gaussian signatures and collective effects in charge noise
  affecting a dynamically decoupled qubit.
\newblock {\em Phys. Rev. B}, 92:155422, Oct 2015.
\newblock URL: \url{http://link.aps.org/doi/10.1103/PhysRevB.92.155422}, \href
  {http://dx.doi.org/10.1103/PhysRevB.92.155422}
  {\path{doi:10.1103/PhysRevB.92.155422}}.

\bibitem{Cywinski2014}
\L{}ukasz Cywi\ifmmode~\acute{n}\else \'{n}\fi{}ski.
\newblock Dynamical-decoupling noise spectroscopy at an optimal working point
  of a qubit.
\newblock {\em Phys. Rev. A}, 90:042307, Oct 2014.
\newblock URL: \url{http://link.aps.org/doi/10.1103/PhysRevA.90.042307}, \href
  {http://dx.doi.org/10.1103/PhysRevA.90.042307}
  {\path{doi:10.1103/PhysRevA.90.042307}}.

\bibitem{Friesen2016}
M.~{Friesen}, M.~A. {Eriksson}, and S.~N. {Coppersmith}.
\newblock {A decoherence-free subspace for charge: the quadrupole qubit}.
\newblock {\em ArXiv e-prints}, May 2016.
\newblock \href {http://arxiv.org/abs/1605.01797} {\path{arXiv:1605.01797}}.

\bibitem{Li1990}
Yuan~P. Li, D.~C. Tsui, J.~J. Heremans, J.~A. Simmons, and G.~W. Weimann.
\newblock Low frequency noise in transport through quantum point contacts.
\newblock {\em Applied Physics Letters}, 57(8):774--776, 1990.
\newblock URL:
  \url{http://scitation.aip.org/content/aip/journal/apl/57/8/10.1063/1.104094},
  \href {http://dx.doi.org/http://dx.doi.org/10.1063/1.104094}
  {\path{doi:http://dx.doi.org/10.1063/1.104094}}.

\bibitem{Dekker1991}
C.~Dekker, A.~J. Scholten, F.~Liefrink, R.~Eppenga, H.~van Houten, and C.~T.
  Foxon.
\newblock Spontaneous resistance switching and low-frequency noise in quantum
  point contacts.
\newblock {\em Phys. Rev. Lett.}, 66:2148--2151, Apr 1991.
\newblock URL: \url{http://link.aps.org/doi/10.1103/PhysRevLett.66.2148}, \href
  {http://dx.doi.org/10.1103/PhysRevLett.66.2148}
  {\path{doi:10.1103/PhysRevLett.66.2148}}.

\bibitem{Sakamoto1995}
T.~Sakamoto, Y.~Nakamura, and K.~Nakamura.
\newblock Distributions of single carrier traps in
  ${\mathrm{g}\mathrm{a}\mathrm{a}\mathrm{s}/\mathrm{a}\mathrm{l}}_{x}{\mathrm{ga}}_{1-x}\mathrm{As}$
  heterostructures.
\newblock {\em Applied Physics Letters}, 67(15):2220--2222, 1995.
\newblock URL:
  \url{http://scitation.aip.org/content/aip/journal/apl/67/15/10.1063/1.115109},
  \href {http://dx.doi.org/http://dx.doi.org/10.1063/1.115109}
  {\path{doi:http://dx.doi.org/10.1063/1.115109}}.

\bibitem{Kurdak1997}
C.~Kurdak, C.-J. Chen, D.~C. Tsui, S.~Parihar, S.~Lyon, and G.~W. Weimann.
\newblock Resistance fluctuations in
  ${\mathrm{g}\mathrm{a}\mathrm{a}\mathrm{s}/\mathrm{a}\mathrm{l}}_{x}{\mathrm{ga}}_{1-x}\mathrm{As}$
  quantum point contact and hall bar structures.
\newblock {\em Phys. Rev. B}, 56:9813--9818, Oct 1997.
\newblock URL: \url{http://link.aps.org/doi/10.1103/PhysRevB.56.9813}, \href
  {http://dx.doi.org/10.1103/PhysRevB.56.9813}
  {\path{doi:10.1103/PhysRevB.56.9813}}.

\bibitem{Hayashi2003}
T.~Hayashi, T.~Fujisawa, H.~D. Cheong, Y.~H. Jeong, and Y.~Hirayama.
\newblock Coherent manipulation of electronic states in a double quantum dot.
\newblock {\em Phys. Rev. Lett.}, 91:226804, Nov 2003.
\newblock URL: \url{http://link.aps.org/doi/10.1103/PhysRevLett.91.226804},
  \href {http://dx.doi.org/10.1103/PhysRevLett.91.226804}
  {\path{doi:10.1103/PhysRevLett.91.226804}}.

\bibitem{Buizert2008}
Christo Buizert, Frank H.~L. Koppens, Michel Pioro-Ladri\`ere, Hans-Peter
  Tranitz, Ivo~T. Vink, Seigo Tarucha, Werner Wegscheider, and Lieven M.~K.
  Vandersypen.
\newblock $insitu$ reduction of charge noise in
  $\mathrm{GaAs}/{\mathrm{al}}_{x}{\mathrm{ga}}_{1-x}\mathrm{As}$
  schottky-gated devices.
\newblock {\em Phys. Rev. Lett.}, 101:226603, Nov 2008.
\newblock URL: \url{http://link.aps.org/doi/10.1103/PhysRevLett.101.226603},
  \href {http://dx.doi.org/10.1103/PhysRevLett.101.226603}
  {\path{doi:10.1103/PhysRevLett.101.226603}}.

\bibitem{Petersson2010}
K.~Petersson, J.~Petta, H.~Lu, and A.~Gossard.
\newblock Quantum coherence in a one-electron semiconductor charge qubit.
\newblock {\em Phys. Rev. Lett.}, 105:246804, Dec 2010.
\newblock URL: \url{http://link.aps.org/doi/10.1103/PhysRevLett.105.246804},
  \href {http://dx.doi.org/10.1103/PhysRevLett.105.246804}
  {\path{doi:10.1103/PhysRevLett.105.246804}}.

\bibitem{Takeda2013}
K.~Takeda, T.~Obata, Y.~Fukuoka, W.~M. Akhtar, J.~Kamioka, T.~Kodera, S.~Oda,
  and S.~Tarucha.
\newblock Characterization and suppression of low-frequency noise in si/sige
  quantum point contacts and quantum dots.
\newblock {\em Applied Physics Letters}, 102(12), 2013.
\newblock URL:
  \url{http://scitation.aip.org/content/aip/journal/apl/102/12/10.1063/1.4799287;jsessionid=hasOC2L975pxcFxRJqQTzhuz.x-aip-live-06},
  \href {http://dx.doi.org/http://dx.doi.org/10.1063/1.4799287}
  {\path{doi:http://dx.doi.org/10.1063/1.4799287}}.

\bibitem{Freeman2016}
Blake~M. Freeman, Joshua~S. Schoenfield, and HongWen Jiang.
\newblock Comparison of low frequency charge noise in identically patterned
  si/sio2 and si/sige quantum dots.
\newblock {\em Applied Physics Letters}, 108(25), 2016.
\newblock URL:
  \url{http://scitation.aip.org/content/aip/journal/apl/108/25/10.1063/1.4954700},
  \href {http://dx.doi.org/http://dx.doi.org/10.1063/1.4954700}
  {\path{doi:http://dx.doi.org/10.1063/1.4954700}}.

\bibitem{Dial2013}
O.~E. Dial, M.~D. Shulman, S.~P. Harvey, H.~Bluhm, V.~Umansky, and A.~Yacoby.
\newblock Charge noise spectroscopy using coherent exchange oscillations in a
  singlet-triplet qubit.
\newblock {\em Phys. Rev. Lett.}, 110:146804, Apr 2013.
\newblock URL: \url{http://link.aps.org/doi/10.1103/PhysRevLett.110.146804},
  \href {http://dx.doi.org/10.1103/PhysRevLett.110.146804}
  {\path{doi:10.1103/PhysRevLett.110.146804}}.

\bibitem{Hu2011}
Xuedong Hu.
\newblock Two-spin dephasing by electron-phonon interaction in semiconductor
  double quantum dots.
\newblock {\em Phys. Rev. B}, 83:165322, Apr 2011.
\newblock URL: \url{http://link.aps.org/doi/10.1103/PhysRevB.83.165322}, \href
  {http://dx.doi.org/10.1103/PhysRevB.83.165322}
  {\path{doi:10.1103/PhysRevB.83.165322}}.

\bibitem{Gamble2012}
John~King Gamble, Mark Friesen, S.~N. Coppersmith, and Xuedong Hu.
\newblock Two-electron dephasing in single si and gaas quantum dots.
\newblock {\em Phys. Rev. B}, 86:035302, Jul 2012.
\newblock URL: \url{http://link.aps.org/doi/10.1103/PhysRevB.86.035302}, \href
  {http://dx.doi.org/10.1103/PhysRevB.86.035302}
  {\path{doi:10.1103/PhysRevB.86.035302}}.

\bibitem{Huang2014}
Peihao Huang and Xuedong Hu.
\newblock Electron spin relaxation due to charge noise.
\newblock {\em Phys. Rev. B}, 89:195302, May 2014.
\newblock URL: \url{http://link.aps.org/doi/10.1103/PhysRevB.89.195302}, \href
  {http://dx.doi.org/10.1103/PhysRevB.89.195302}
  {\path{doi:10.1103/PhysRevB.89.195302}}.

\bibitem{Culcer2009}
Dimitrie Culcer, \L{}ukasz Cywi\ifmmode~\acute{n}\else \'{n}\fi{}ski, Qiuzi Li,
  Xuedong Hu, and S.~Das~Sarma.
\newblock Realizing singlet-triplet qubits in multivalley si quantum dots.
\newblock {\em Phys. Rev. B}, 80:205302, Nov 2009.
\newblock URL: \url{http://link.aps.org/doi/10.1103/PhysRevB.80.205302}, \href
  {http://dx.doi.org/10.1103/PhysRevB.80.205302}
  {\path{doi:10.1103/PhysRevB.80.205302}}.

\bibitem{Culcer2010}
Dimitrie Culcer, \L{}ukasz Cywi\ifmmode~\acute{n}\else \'{n}\fi{}ski, Qiuzi Li,
  Xuedong Hu, and S.~Das~Sarma.
\newblock Quantum dot spin qubits in silicon: Multivalley physics.
\newblock {\em Phys. Rev. B}, 82:155312, Oct 2010.
\newblock URL: \url{http://link.aps.org/doi/10.1103/PhysRevB.82.155312}, \href
  {http://dx.doi.org/10.1103/PhysRevB.82.155312}
  {\path{doi:10.1103/PhysRevB.82.155312}}.

\bibitem{Rohling2012}
Niklas Rohling and Guido Burkard.
\newblock Universal quantum computing with spin and valley states.
\newblock {\em New Journal of Physics}, 14(8):083008, 2012.
\newblock URL: \url{http://stacks.iop.org/1367-2630/14/i=8/a=083008}.

\bibitem{Rohling2014}
Niklas Rohling, Maximilian Russ, and Guido Burkard.
\newblock Hybrid spin and valley quantum computing with singlet-triplet qubits.
\newblock {\em Phys. Rev. Lett.}, 113:176801, Oct 2014.
\newblock URL: \url{http://link.aps.org/doi/10.1103/PhysRevLett.113.176801},
  \href {http://dx.doi.org/10.1103/PhysRevLett.113.176801}
  {\path{doi:10.1103/PhysRevLett.113.176801}}.

\bibitem{Veldhorst2015b}
M.~Veldhorst, R.~Ruskov, C.~H. Yang, J.~C.~C. Hwang, F.~E. Hudson, M.~E.
  Flatt\'e, C.~Tahan, K.~M. Itoh, A.~Morello, and A.~S. Dzurak.
\newblock Spin-orbit coupling and operation of multivalley spin qubits.
\newblock {\em Phys. Rev. B}, 92:201401, Nov 2015.
\newblock URL: \url{http://link.aps.org/doi/10.1103/PhysRevB.92.201401}, \href
  {http://dx.doi.org/10.1103/PhysRevB.92.201401}
  {\path{doi:10.1103/PhysRevB.92.201401}}.

\bibitem{Rancic2016}
Marko~J. Ran\ifmmode \check{c}\else \v{c}\fi{}i\ifmmode~\acute{c}\else
  \'{c}\fi{} and Guido Burkard.
\newblock Electric dipole spin resonance in systems with a valley-dependent $g$
  factor.
\newblock {\em Phys. Rev. B}, 93:205433, May 2016.
\newblock URL: \url{http://link.aps.org/doi/10.1103/PhysRevB.93.205433}, \href
  {http://dx.doi.org/10.1103/PhysRevB.93.205433}
  {\path{doi:10.1103/PhysRevB.93.205433}}.

\bibitem{Boross2016}
P\'eter Boross, G\'abor Sz\'echenyi, Dimitrie Culcer, and Andr\'as P\'alyi.
\newblock Control of valley dynamics in silicon quantum dots in the presence of
  an interface step.
\newblock {\em Phys. Rev. B}, 94:035438, Jul 2016.
\newblock URL: \url{http://link.aps.org/doi/10.1103/PhysRevB.94.035438}, \href
  {http://dx.doi.org/10.1103/PhysRevB.94.035438}
  {\path{doi:10.1103/PhysRevB.94.035438}}.

\bibitem{Zimmerman2016}
N.~{Zimmerman}, P.~{Huang}, and D.~{Culcer}.
\newblock {Valley Phase, Interface Roughness and Voltage Control, and Coherent
  Manipulation in Si Quantum Dots}.
\newblock {\em ArXiv e-prints}, August 2016.
\newblock \href {http://arxiv.org/abs/1608.06881} {\path{arXiv:1608.06881}}.

\bibitem{Gamble2016}
J.~{King Gamble}, P.~{Harvey-Collard}, N.~T. {Jacobson}, A.~D. {Baczewski},
  E.~{Nielsen}, L.~{Maurer}, I.~{Monta{\~n}o}, M.~{Rudolph}, M.~S. {Carroll},
  C.~H. {Yang}, A.~{Rossi}, A.~S. {Dzurak}, and R.~P. {Muller}.
\newblock Valley splitting of single-electron si mos quantum dots.
\newblock {\em ArXiv e-prints}, page arXiv:1610.03388, October 2016.

\bibitem{Saraiva2011}
A.~L. Saraiva, M.~J. Calder√≥n, Rodrigo~B. Capaz, Xuedong Hu, S.~{Das
  Sarma}, and Belita Koiller.
\newblock Intervalley coupling for interface-bound electrons in silicon: An
  effective mass study.
\newblock {\em Phys. Rev. B}, 84:155320, Oct 2011.
\newblock URL: \url{http://link.aps.org/doi/10.1103/PhysRevB.84.155320}, \href
  {http://dx.doi.org/10.1103/PhysRevB.84.155320}
  {\path{doi:10.1103/PhysRevB.84.155320}}.

\bibitem{Lim2011}
W~H Lim, C~H Yang, F~A Zwanenburg, and A~S Dzurak.
\newblock Spin filling of valley--orbit states in a silicon quantum dot.
\newblock {\em Nanotechnology}, 22(33):335704, 2011.
\newblock URL: \url{http://stacks.iop.org/0957-4484/22/i=33/a=335704}.

\bibitem{Wu2012}
Yue Wu and Dimitrie Culcer.
\newblock Coherent electrical rotations of valley states in si quantum dots
  using the phase of the valley-orbit coupling.
\newblock {\em Phys. Rev. B}, 86:035321, Jul 2012.
\newblock URL: \url{http://link.aps.org/doi/10.1103/PhysRevB.86.035321}, \href
  {http://dx.doi.org/10.1103/PhysRevB.86.035321}
  {\path{doi:10.1103/PhysRevB.86.035321}}.

\bibitem{Culcer2012}
Dimitrie Culcer, A.~L. Saraiva, Belita Koiller, Xuedong Hu, and S.~Das~Sarma.
\newblock Valley-based noise-resistant quantum computation using si quantum
  dots.
\newblock {\em Phys. Rev. Lett.}, 108:126804, Mar 2012.
\newblock URL: \url{http://link.aps.org/doi/10.1103/PhysRevLett.108.126804},
  \href {http://dx.doi.org/10.1103/PhysRevLett.108.126804}
  {\path{doi:10.1103/PhysRevLett.108.126804}}.

\bibitem{Hao2014}
Xiaojie Hao, Rusko Ruskov, Ming Xiao, Charles Tahan, and HongWen Jiang.
\newblock Electron spin resonance and spin--valley physics in a silicon double
  quantum dot.
\newblock {\em Nature Communications}, 5:3860 EP --, 05 2014.
\newblock URL: \url{http://dx.doi.org/10.1038/ncomms4860}.

\end{thebibliography}

\end{document}